\documentclass[aps,prd,12pt,citeautoscript,superscriptaddress,showpacs,scrartcl]{revtex4-1}
\usepackage[utf8x]{inputenc}

\usepackage{xcolor}
\usepackage[colorlinks=true, pdfstartview=FitV,linkcolor=blue, citecolor=blue, urlcolor=blue]{hyperref}

\usepackage[bottom]{footmisc}

\usepackage{graphicx}

\makeatletter
\newcommand*\bigcdot{\mathpalette\bigcdot@{.5}}
\newcommand*\bigcdot@[2]{\mathbin{\vcenter{\hbox{\scalebox{#2}{$\m@th#1\bullet$}}}}}
\makeatother

\newcommand{\be}{\begin{equation}}
\newcommand{\ee}{\end{equation}}
\newcommand{\bea}{\begin{eqnarray}}
\newcommand{\eea}{\end{eqnarray}}
\newcommand{\x}{\bar{r}}
\newcommand{\nn}{\nonumber}

\newcommand{\bl}{\textcolor{black}}
\newcommand{\bbl}{\textcolor{black}}

\usepackage{natbib}
\usepackage{graphicx}
\usepackage{amsmath}
\usepackage{amssymb}
\usepackage{mathtools}
\usepackage{tensor}
\usepackage{accents}
\usepackage{cancel}
\usepackage{titlesec}
\usepackage{subfigure}
\usepackage[a4paper]{geometry}
\usepackage{psfrag}
\usepackage{latexsym,array,theorem,mathrsfs,subfigure,bm,float}
\usepackage{amsfonts}
\usepackage{amssymb}

\renewcommand{\theequation}{\arabic{section}.\arabic{equation}}

\newcommand\numberthis{\addtocounter{equation}{1}\tag{\theequation}}
\titleformat{\paragraph}
{\normalfont\normalsize\bfseries}{\theparagraph}{1em}{}
\titlespacing*{\paragraph}
{0pt}{3.25ex plus 1ex minus .2ex}{1.5ex plus .2ex}

\begin{document}

\title{Asymptotically flat hairy black holes in massive bigravity}

\author{Romain Gervalle}
\email{romain.gervalle@univ-tours.fr}
\affiliation{
Institut Denis Poisson, UMR - CNRS 7013, \\ 
Universit\'{e} de Tours, Parc de Grandmont, 37200 Tours, France}

\author{Mikhail~S.~Volkov}
\email{volkov@lmpt.univ-tours.fr}
\affiliation{
Institut Denis Poisson, UMR - CNRS 7013, \\ 
Universit\'{e} de Tours, Parc de Grandmont, 37200 Tours, France}
\affiliation{
Institute for Theoretical and Mathematical Physics, \\ 
Lomonosov Moscow State University,
Leninskie Gory, GSP-1, 119991 Moscow, Russia}
\affiliation{
Department of General Relativity and Gravitation, Institute of Physics,\\
Kazan Federal University, Kremlevskaya street 18, 420008 Kazan, Russia
}

\ 

\vspace{1 cm}

\begin{abstract}

\vspace{1 cm}

We study asymptotically flat black holes with massive graviton hair within the ghost-free
bigravity theory.  There have been contradictory  statements in the literature about their existence  --
such solutions were   reported some time ago, but later a different group claimed 
 the Schwarzschild solution to be the only asymptotically flat black hole in the theory. 
As a result, the controversy emerged. We have analyzed the issue ourselves and have been able to 
construct such solutions  within a carefully designed numerical scheme. 
We find that for given parameter values there can be one or two asymptotically flat hairy black hole solutions in addition 
to the Schwarzschild solution. We analyze their perturbative stability and find that they can be stable or unstable, 
depending on the parameter values. \bbl{The masses of stable hairy black holes that would be physically relevant 
range  form stellar values up to values typical for supermassive black holes.
One of their two metrics is extremely close to Schwarzschild, while all their  ``hair"  is hidden in the second metric 
that is not coupled to matter and   not directly seen.  If the massive bigravity theory indeed describes physics, 
the  hair of such black holes should manifest themselves
in violent processes like black hole mergers  and should be visible in the structure of the signals detected by LIGO/VIRGO. }

\end{abstract}


\maketitle

\section{INTRODUCTION}

Theories with massive gravitons provide a natural modification of the general relativity (GR) in the infrared regime 
and can be used to explain the current acceleration of our Universe \cite{1538-3881-116-3-1009,0004-637X-517-2-565}. 
Such theories have a long  history pioneered  by the work of  Fierz and Pauli \cite{Fierz:1939ix} and marked by subsequent discoveries of 
many interesting features, such as the vDVZ discontinuity \cite{vanDam:1970vg,Zakharov:1970cc}, 
the Vainshtein mechanism \cite{Vainshtein:1972sx},  the Boulware-Deser ghost \cite{Boulware:1973my}, 
culminating in the discovery of the ghost-free massive gravity  \cite{deRham:2010kj} and ghost-free bigravity  \cite{Hassan2012}
theories. 

The ghost-free bigravity theory is the most interesting physically. 
It contains two dynamical metrics, usually called $g_{\mu\nu}$ and $f_{\mu\nu}$, describing together two gravitons, 
one of which is massive and the other is massless. 
The theory admits self-accelerating cosmological solutions  \cite{Volkov:2011an, vonStrauss:2011mq, Comelli:2011zm} 
whose properties can agree with the observations 
\cite{Akrami:2015qga, Mortsell:2015exa, Aoki:2015xqa, Luben:2018ekw, Hogas:2019ywm},
with the $\Lambda$ term  mimicked by the graviton mass. The theory also admits solutions describing 
 black holes \cite{Volkov2012}, wormholes \cite{Sushkov:2015fma}, and other interesting 
solutions (see \cite{Volkov:2013roa} for a review). In what follows  we shall be discussing black holes.

The bigravity black holes can be either  ``bald" or ``hairy". The bald black holes 
are described by the known GR metrics. Such solutions were first discovered long ago  
\cite{Isham:1977rj, Gurses:1979qb, Gurses:1981an}  within the 
old bigravity theory inspired by physics of strong interactions \cite{Isham:1971gm}.  In the simplest case, their two metrics are both 
Schwarzschild-(anti-)de Sitter
and can be conveniently represented  in the Eddington-Finkelstein coordinates as \cite{Babichev:2014oua,Volkov:2014ooa} 
\be                                     \label{b1}
g_{\mu\nu}dx^\mu dx^\nu=-\Sigma_g\, dv^2+2dv dr +r^2 d\Omega^2,~~f_{\mu\nu}dx^\mu dx^\nu=C^2\left(-\Sigma_f\, dv^2+2dv dr +r^2 d\Omega^2\right),
\ee
with $\Sigma_g=1-2M_g/r+\Lambda_g/(3r^2)$ and  $\Sigma_f=1-2M_f/r+\Lambda_f/(3r^2)$, where values of constants 
$C,\Lambda_g,\Lambda_f$ are fixed by the field equations. 
Passing to the Schwarzschild coordinates, one can diagonalize one of the two metrics, but not both of them simultaneously. 
Such solutions have been much studied  \cite{Berezhiani:2008nr}, they exist also within the ghost-free bigravity \cite{Comelli:2011wq},
and they  admit the charged \cite{Babichev:2014fka} and spinning \cite{Babichev:2014tfa} generalizations.  These solutions also admit the 
massive gravity limit where $M_f=\Lambda_f=0$ hence the f metric is flat while the g metric remains nontrivial, and this yields    all possible
static black holes in the ghost-free massive gravity theory (it seems there can be also time-dependent black holes in this theory  \cite{Rosen:2018lki}).

Next, it was noticed in \cite{Volkov2012} that if the parameters of the potential are suitably adjusted, 
then the ghost-free bigravity reduces to the vacuum GR when  the two metrics coincide, $g_{\mu\nu}=f_{\mu\nu}$. 
Therefore, all vacuum black holes, as for example the Schwarzschild solution 
\bbl{(to be called  bald  to distinguish it from the ``hairy  Schwarzschild"  to be described below)},
\be                                    \label{b2}
g_{\mu\nu}dx^\mu dx^\nu=f_{\mu\nu}dx^\mu dx^\nu=-\left(1-\frac{2M}{r}\right)\, dt^2+
\frac{dr^2}{1-{2M}/{r} }+r^2 d\Omega^2,
\ee
or its spinning  generalization can be imbedded into the ghost-free bigravity. A $\Lambda$ term can be included by assuming 
the two metrics to be proportional to each other  \cite{Volkov2012,Volkov:2013roa,Volkov:2014ooa}. 
Such solutions are different from solutions of type  \eqref{b1}, for example  they do not admit the massive gravity limit. 
In addition, the solution \eqref{b1} is linearly stable \cite{Babichev:2015zub}, 
whereas   \eqref{b2} is unstable (for a small  $M$) with respect to fluctuations which do not respect the condition  $g_{\mu\nu}=f_{\mu\nu}$ 
\cite{Babichev:2013una}. 

These facts essentially exhaust the available knowledge of the  bald  black holes in the bigravity theory. 
At the same time,  more  general   hairy  black holes not described by the classical GR metrics can exist as well. 
The first example of hairy black holes in physics 
was found long ago  \cite{Volkov:1989fi}, 
followed by many other examples (see 
\cite{Volkov:1998cc,Volkov:2016ehx} for a review), so that nowadays hairy black holes are  considered as something 
usual. One may therefore wonder if they  exist in the ghost-free bigravity theory as well.

A  systematic  analysis of hairy black holes in the ghost-free massive bigravity 
has been carried out for the first  time  by one of the authors \cite{Volkov2012}, but none of the  solutions found were asymptotically flat. 
In that  analysis both metrics were  assumed to be 
static and spherically symmetric. If they are not simultaneously diagonal, then the most general solution is 
given by \eqref{b1}. If they are simultaneously diagonal, then one of the solutions is given by \eqref{b2}, 
but other  more general black hole solutions exist as well. 

Such solutions possess an event horizon --
a hypersurface that   is null simultaneously 
with respect to $g_{\mu\nu}$ and $f_{\mu\nu}$. Therefore,  both metrics share the horizon 
\cite{Deffayet:2011rh, Banados:2011hk}, but  its radius $r_H^g$  measured by $g_{\mu\nu}$ can be different from the 
radius $r_H^f$ measured by $f_{\mu\nu}$. One can set  $r_H\equiv r^g_H$ to unit value via rescaling the system 
(rescaling at the same time the graviton mass),  
but the ratio $u=r_H^g/r_H^f$ is  scale invariant. Choosing a value of 
 $u$ completely determines the boundary conditions at the horizon, which  allows one to integrate the 
 equations starting from the horizon toward large values of the radial coordinate $r$. As a result, the set of all 
 black hole solutions can be labeled by just  one parameter $u$, and integrating the equations for different values of $u$ 
 gives all possible black holes.

 Choosing  $u=1$ yields the 
 Schwarzschild solution \eqref{b2}. For $u\neq 1$ one finds more general  black holes supporting a massive graviton ``hair" 
 outside the horizon, but in the asymptotic region their two geometries do not become flat  \cite{Volkov2012}.  
 The latter property is generic,
 and trying different values of $u$ always gives either solutions with a curvature singularity somewhere outside 
 the horizon, or solutions which exist for all values of $r$ but show nonflat asymptotics.  

At the same time, these facts do not completely exclude a possibility of some other asymptotically flat black hole solutions  
different from  \eqref{b2},
which would correspond to some  special  values of $u$ different from $u=1$. 
 However, even if they exist, one does not find such solutions  by a brute force   via trying many different values of $u$, and the reason is the following. 
 The field equations reduce to three coupled first order 
 ordinary differential equations 
 (ODEs) \cite{Volkov2012}, whose 
 {\it local} at large $r$ solution  has  schematically the following structure when it is linearized around flat space
 ($A,B,C$ being integration constants):
 \be                    \label{large}
 \frac{A}{r}+B e^{-r}+C e^{+r}. 
 \ee
 Here $r={\rm mr}$  is the dimensionless radial coordinate, with  m and r being  the graviton mass and  dimensionful radial coordinate 
(we  assume the graviton mass to have the dimension of  inverse length,
 so that this is rather the inverse Compton wavelength ${\rm m}c/\hbar$).  
 The Newtonian mode $A/r$ in \eqref{large} arises due to the 
 massless graviton present in the theory, while the decaying mode $B e^{-r}$ and the growing mode 
 $C e^{+r}$ are due to the massive graviton. Now, when integrating from the horizon, the growing mode $C e^{+r}$  
 will be inevitably present in the numerical solution at large $r$ and will drive the solution away from flat space. 
 This is why one does not find asymptotically flat solutions in this way. 
 
 To get them, one should suppress the growing mode  by setting $C=0$, hence the local 
  solutions at large $r$ will comprise a two-parameter set labeled by $A$ and $B$. 
   The next step is to numerically extend this local solution toward small $r$, 
 extending at the same time the local solution at the horizon labeled by $u$ toward large $r$, 
 until the two solutions meet at some intermediate point. 
 For these solutions to agree, three (the number of the ODEs) matching conditions should be satisfied 
 via adjusting the three parameters $u,A,B$. In practice, this can be done 
  within the numerical multiple-shooting method  \cite{Press:2007:NRE:1403886}. 
  Once $u,A,B$ are adjusted, this yields global asymptotically
  flat solutions. 
  
  The difficulty, however, is that the numerical scheme requires some input values for $u,A,B$, which 
 should be close to the ``true values", otherwise the iterations do not converge. 
 It was {\it a priory}  unclear how to  choose these input values, whereas  choosing them randomly 
 does not give the convergence.  Some additional information was needed to properly choose
  these input values, but at the time of writing 
  the article \cite{Volkov2012} such  information was not available. 
 As a result,   the conclusion of  that work was that 
  asymptotically flat hairy black holes may exist, but they should be parametrically isolated form the 
  Schwarzschild  solution \eqref{b2}.  
  
 It is interesting  that by adding an extra matter source to obtain  not a black hole but a regular object like a star, 
 asymptotically flat solutions can be easily constructed, as was shown first  in \cite{Volkov2012}  and later in 
 \cite{Enander:2015kda, Aoki:2016eov}. The black hole case is more difficult.

 Fortunately, the additional information was later obtained within the analysis of  perturbations of the  Schwarzschild solution \eqref{b2}
 \cite{Babichev:2013una,Brito:2013wya}. Denoting $g^S_{\mu\nu}$ the  Schwarzschild metric, 
 the two perturbed metrics  are $g_{\mu\nu}=g^S_{\mu\nu}+\delta g_{\mu\nu}$ and 
 $f_{\mu\nu}=g^S_{\mu\nu}+\delta f_{\mu\nu}$.  Linearizing the field equations with respect to $\delta g_{\mu\nu}$ and
 $\delta f_{\mu\nu}$, one finds  that perturbations grow in time and hence the background Schwarzschild black hole is 
 unstable if $r_H\equiv {\rm mr_H}\leq  0.86$. On the other hand, for $r_H>0.86$ the perturbations are bounded in time 
 so that the background is stable \cite{Babichev:2013una}. Curiously, the mathematical structure of the perturbation equations is
 identical \cite{Babichev:2013una} to that previously discovered  by Gregory and Laflamm (GL) in their analysis of black strings  in $D=5$ GR \cite{Gregory:1993vy}. 
 We shall therefore refer to the Schwarzschild solution with $r_H=0.86$ as GL point. 
 
 This change of stability at the GL point suggests that for    $r_H$ close to  $0.86$  there could be two different asymptotically flat solutions:
 the Schwarzschild solution \eqref{b2} and  also some other  solution which 
 can be  approximated by the zero perturbation mode that exists at the GL point. 
 This new solution is different from  Schwarzschild although close to it, hence it describes an asymptotically flat hairy black hole. 
 To get this solution within the numerical scheme outlined above, one should  choose the input parameters $u,A,B$ to be close 
 the GL point, $u\approx 1$, $r_H\approx 0.86$, $A\approx -r_H/2$, $B\approx 0$, and it  is this essential piece of information that was missing 
when writing Ref.\cite{Volkov2012}. 
 As soon as the solution is obtained, one can change the value of $r_H$ iteratively, 
 thus obtaining  ``fully fledged" hairy black holes which may deviate considerably from the parent Schwarzschild  solution. 
 
 Remarkably, 
 this program was accomplished  by the Portuguese group  \cite{Brito:2013xaa} via explicitly constructing 
asymptotically flat hairy black holes in the theory in the region {\it below} the GL point, 
  for $r_H< 0.86$.
However, some time later spherically symmetric  bigravity solutions were  analyzed  by the Swedish group
\cite{Torsello:2017cmz}, and it was claimed that the Schwarzschild solution \eqref{b2} represents  
the unique  asymptotically flat black hole in the theory. 
As a result, a controversy  emerged and it was  unclear  if asymptotically flat hairy black holes  exist or not. 

We have therefore reconsidered the issue ourselves  and below are our results. 
In brief, we were able  to construct  asymptotically flat hairy black holes in the theory, 
thereby confirming the finding of \cite{Brito:2013xaa}. We apply a very carefully designed  numerical scheme  
to exclude any ambiguities 
and to take into account the arguments of \cite{Torsello:2017cmz}. 
In fact, these arguments correctly point to some drawbacks of the numerical analysis commonly present in 
many publications. From the mathematical viewpoint, one has to solve a nonlinear boundary value problem 
where the boundaries are singular points of the 
differential equations (horizon and infinity). Since it is difficult to approach such points numerically, 
various approximations are used in practice,
which may give reasonable results in some cases but inevitably increase the numerical errors and lead to a numerical instability. 
Only very rarely does one find in the literature a correct treatment of the problem (apart from  the relaxation approach), as for example in 
\cite{Breitenlohner:1991aa,BFM,Breitenlohner:1993es}.  We therefore pay  special attention to the details of our numerical scheme 
and describe them in a very explicit way.  From the methodological viewpoint,  our paper gives an example of how one should properly 
tackle a nonlinear boundary value problem with singular endpoints.

We cross-check our results with  two different numerical codes written independently  by two of us. 
Our results strongly suggest that the hairy solutions exist and are indeed asymptotically flat and regular. 
We discover many new features of these solutions,  for example we obtain 
hairy black holes  also {\it above} the GL point for $r_H>0.86$, and we study for the first time the perturbative stability of the solutions. 
We were able to identify regions in the parameter space which correspond to stable solutions, and we 
determined subsets of these regions which agree with the constraints imposed by the cosmological observations. 
We find that the viable hairy black holes should be described by the g metric that is very close to Schwarzschild,
but their f metric is different. \bbl{ Therefore, if the bigravity theory indeed describes physics,
the  astrophysical black holes should hide  the  hair  in their f-metric. 
We find masses of such  black holes to range from $\sim 0.2\,{\rm M}_\odot$ 
to  $\sim 0.3\times 10^6\,{\rm M}_\odot$. }

We have also attentively considered  the arguments of Ref. \cite{Torsello:2017cmz}.
In brief, this work seems to agree that the hairy solutions exist but judges them physically unacceptable. 
\bl{We analyze the arguments and we think  some of them  are interesting and should be taken into consideration, 
but none of them is decisive, so that they should rather be viewed as a conjecture. 
To understand its origin, we notice that the  numerical procedure 
adopted in that work is not suitable for suppressing the growing at infinity mode, which  generates artificial numerical singularities. 
This must be the  reason why 
the solutions  were judged   unacceptable in that work.} However, no singularities appear within the properly 
chosen numerical scheme, and we specially adapt our scheme to be able to cope with 
the arguments of  \cite{Torsello:2017cmz}.
We shall postpone a more detailed  discussion  of 
Ref. \cite{Torsello:2017cmz} until the  end of this text to be able to make a comparison with our results.

The rest of the text is organized as follows.  In Sec. \ref{sec2} we introduce the massive bigravity 
theory of Hassan and Rosen \cite{Hassan2012}. 
The field equations, their reduction to the static and spherically symmetric sector, and the simplest solutions are 
described in Sec. \ref{sec3}--\ref{simp}. In Sec. \ref{BC},\ref{BI} we describe in detail our analysis 
of boundary conditions at the horizon and at infinity, and then summarize in Sec. \ref{hairybh}  the structure of 
our numerical procedure. In  Sec. \ref{NR} we show our 
solutions for asymptotically flat hairy black holes and also describe 
the duality  relation  yielding the solutions above the GL point. 
After that, we discuss in Sec. \ref{pert} the perturbations of the hairy backgrounds 
and the analysis of the negative perturbation modes. Our discussion culminates in Sec. \ref{par} 
where we describe various limits and identify regions in the parameter space where the solutions exist and where they are stable. 
In Sec. \ref{CR} we give a brief summary of our results and discuss the arguments of Ref.\cite{Torsello:2017cmz}. 
The two appendixes contain the description of 
 the desingularization of the equations at the horizon, as well as the complete set of the field equations 
in the time-dependent case.


\section{THE GHOST-FREE BIGRAVITY \label{sec2}}
\setcounter{equation}{0}


The theory is  defined on a four-dimensional spacetime manifold endowed with 
two Lorentzian metrics ${\rm g}_{\mu\nu}$ and ${\rm f}_{\mu\nu}$ with the signature
$(-,+,+,+)$.   
The action is  \cite{Hassan2012}
\bea                                      \label{1}
&&S[{\rm g},{\rm f}]
=\frac{1}{2{\boldsymbol\kappa}_1}\int \, 
R({\rm g})\sqrt{-{\rm g}}\, d^4{\rm x}
+\frac{1}{2{\boldsymbol\kappa}_2}\,\int R({\rm f})\sqrt{-{\rm f}}\, d^4{\rm x}
-\frac{{\rm m}^2}{\boldsymbol\kappa }\int {\cal U}\sqrt{-{\rm g}} \,d^4{\rm x}
\,,~~~~~~~~~~
\eea
where ${\boldsymbol\kappa}_1$ and ${\boldsymbol\kappa}_2$ are the gravitational couplings,
${\boldsymbol\kappa}$ is a parameter with 
the same dimension, and 
${\rm m}$ is a mass parameter.
 The interaction between the two metrics 
is expressed by  
 a scalar function 
of the tensor (the hat denotes matrices) 
\be                             \label{gam}
\hat{\gamma}=\sqrt{\hat{{\rm g}}^{-1}\hat{{\rm f}}}. 
\ee
Here the matrix square root is understood in the sense that 
$\hat{\gamma}^2=\hat{{\rm g}}^{-1}\hat{{\rm f}}$, which can be written in components as 
\be                     \label{gamgam}
(\gamma^2)^\mu_{~\nu}\equiv \gamma^\mu_{~\alpha}\gamma^\alpha_{~\nu}=
{\rm g}^{\mu\alpha}
{\rm f}_{\alpha\nu}.
\ee 
If $\lambda_a$ ($a=1,2,3,4$) are the eigenvalues of $\gamma^\mu_{~\nu}$
then the interaction potential is 
\be                             \label{2}
{\cal U}=\sum_{n=0}^4 b_k\,{\cal U}_k,
\ee
where $b_k$ are dimensionless parameters while  
${\cal U}_k$ are defined 
by the 
relations
\bea                        \label{4}
{\cal U}_0&=&1,~~~~
{\cal U}_1=
\sum_{a}\lambda_a=[\gamma],\nonumber \\
{\cal U}_2&=&
\sum_{a<b}\lambda_a\lambda_b
=\frac{1}{2!}([\gamma]^2-[\gamma^2]),\nonumber \\
{\cal U}_3&=&
\sum_{a<b<c}\lambda_c\lambda_b\lambda_c
=
\frac{1}{3!}([\gamma]^3-3[\gamma][\gamma^2]+2[\gamma^3]),\nonumber \\
{\cal U}_4&=&
\lambda_1\lambda_2\lambda_3\lambda_4
=\det(\hat{\gamma})
\,. \nonumber 
\eea
Here 
$[\gamma]={\rm tr}(\hat{\gamma})\equiv \gamma^\mu_{~\mu}$ and 
$[\gamma^k]={\rm tr}(\hat{\gamma}^k)\equiv (\gamma^k)^\mu_{~\mu}$. 
The two metrics 
actually enter the action in a completely symmetric way, 
since the action is invariant under 
\be                                  \label{sym}
{\rm g}_{\mu\nu}\leftrightarrow {\rm f}_{\mu\nu},~~~~
{\boldsymbol\kappa}_1 \leftrightarrow {\boldsymbol\kappa}_2,~~~~
b_k \leftrightarrow b_{4-k}\,.
\ee 
The action is also invariant under rescalings
${\boldsymbol\kappa}\to \pm \lambda^2 {\boldsymbol\kappa}$, $b_k\to \pm b_k,$
${\rm m}\to\lambda\,{\rm m}$,
and this allows one to 
impose, without any loss of generality, the normalization 
condition 
${\boldsymbol\kappa}={\boldsymbol\kappa}_1+{\boldsymbol\kappa}_2$.
Varying the action with respect to 
the two metrics 
gives two sets of Einstein equations,
\bea                                  \label{Enst0}
G_{\mu\nu}({\rm g})&=&{\rm m}^2\,\kappa_1\, T_{\mu\nu},~~~~~~~~~~~~
G_{\mu\nu}({\rm f})= {\rm m}^2\,\kappa_2\, {\cal T}_{\mu\nu}, 
\eea
where $\kappa_1\equiv {\boldsymbol\kappa}_1/{\boldsymbol\kappa}$ and 
$\kappa_2\equiv {\boldsymbol\kappa}_2/{\boldsymbol\kappa}$, and the 
normalization of $\boldsymbol\kappa$ implies   
that $\kappa_1+\kappa_2=1$. 
The source terms in \eqref{Enst0} 
are obtained by varying the interaction potential ${\cal U}$, 
\bea                        \label{T}
&&
T^{\mu}_{~\nu}={\rm g}^{\mu\alpha}T_{\alpha\nu}=
\,\tau^\mu_{~\nu}-{\cal U}\,\delta^\mu_\nu,~~~~~
{\cal T}^{\mu}_{~\nu}=
{\rm f}^{\mu\alpha}{\cal T}_{\alpha\nu}=
-\frac{\sqrt{-{\rm g}}}{\sqrt{-{\rm f}}}\,\tau^\mu_{~\nu}\,,
\eea
where ${\rm f}^{\mu\alpha}$ is the inverse of ${\rm f}_{\mu\alpha}$ and 
\bea                                \label{tau1}
\tau^\mu_{~\nu}&=&
\{b_1\,{\cal U}_0+b_2\,{\cal U}_1+b_3\,{\cal U}_2
+b_4\,{\cal U}_3\}\gamma^\mu_{~\nu} \nonumber \\
&-&\{b_2\,{\cal U}_0+b_3\,{\cal U}_1+b_4\,{\cal U}_2\}(\gamma^2)^\mu_{~\nu} \nonumber  \\
&+&\{b_3\,{\cal U}_0+b_4\,{\cal U}_1\}(\gamma^3)^\mu_{~\nu} 
-b_4\,{\cal U}_0\,(\gamma^4)^\mu_{~\nu}\,. 
\eea
There is an identity relation following from the diffeomorphism invariance of the interaction term in the action,
\be               \label{id}
\sqrt{-g}\stackrel{(g)}{\nabla}_\mu T^{\mu}_{~\nu}+ \sqrt{-f}\stackrel{(f)}{\nabla}_\mu {\cal T}^{\mu}_{~\nu}\equiv 0\,,
\ee
where $\stackrel{(g)}{\nabla}_\rho$ and 
$\stackrel{(f)}{\nabla}_\rho$ are the covariant 
derivatives with respect to g$_{\mu\nu}$ and f$_{\mu\nu}$.

Equations \eqref{Enst0} describe two interacting gravitons, one massive and one massless. This
can be easily seen in the flat space limit. Setting ${\rm g}_{\mu\nu}={\rm f}_{\mu\nu}=\eta_{\mu\nu}$ (the Minkowski metric),
 Eqs.\eqref{Enst0} reduce to  
\bea                                  \label{Enst000}
0&=&-{\rm m}^2\,\kappa_1\,(P_0+P_1)\, \eta_{\mu\nu},~~~~~~~~~~~~
0= -{\rm m}^2\,\kappa_2\,(P_1+P_2)\, \eta_{\mu\nu}, 
\eea
with $P_m\equiv b_m+2b_{m+1}+b_{m+2}$. Therefore, the flat space will be a solution if only 
the parameters $b_k$ fulfil the conditions 
$P_1=-P_0=-P_2$.
Assuming this to be the case, let us set 
${\rm g}_{\mu\nu}=\eta_{\mu\nu}+\delta {\rm g}_{\mu\nu}$ and 
${\rm f}_{\mu\nu}=\eta_{\mu\nu}+\delta {\rm f}_{\mu\nu}$
where the deviations $\delta {\rm g}_{\mu\nu}$ and $\delta {\rm f}_{\mu\nu}$ are small. 
Linearizing the equations \eqref{Enst0} with respect to the deviations yields 
\bea
&&\hat{{\cal E}}_{\mu\nu}^{\alpha\beta}h^{(0)}_{\alpha\beta}=0,~~~\label{m=0}\\
&&\hat{{\cal E}}_{\mu\nu}^{\alpha\beta}h_{\alpha\beta}+\frac{\rm m^2_{\rm FP}}{2} \label{FP}\,
(h_{\mu\nu}-\eta_{\mu\nu}h)=0,~
\eea
where $\hat{{\cal E}}_{\mu\nu}^{\alpha\beta}$ denotes the linear part of the Einstein operator,
and where 
$h_{\mu\nu}^{(0)}=\kappa_1\delta {\rm f}_{\mu\nu}+\kappa_2\delta {\rm g}_{\mu\nu}$
and $h_{\mu\nu}=\delta {\rm f}_{\mu\nu}-\delta {\rm g}_{\mu\nu}$
with $h=\eta^{\alpha\beta}h_{\alpha\beta}$. The $h_{\mu\nu}^{(0)}$ equations are the 
linearized Einstein equations describing 
a massless graviton with two dynamical polarizations. 
The $h_{\mu\nu}$ field fulfills the Fierz-Pauli equations for massive gravitons with five polarizations and 
with the mass
\be                                \label{FP-mass}
{\rm m^2_{\rm FP}}=P_1\, {\rm m}^2.
\ee
Therefore, one will have ${\rm m}_{\rm FP}={\rm m}$ if 
\be            \label{P1}
P_1=1.
\ee 
This condition can be solved together with the conditions 
$P_0=P_2=-1$ implied by  \eqref{Enst000}
to express the five $b_k$ in terms of two independent  parameters, sometimes called $c_3$
and $c_4$, 
\be                   \label{bbb}        
{b}_0=4c_3+c_4-6,~~
{b}_1=3-3c_3-c_4,~~
{b}_2=2c_3+c_4-1,~~
{b}_3=-(c_3+c_4),~~
{b}_4=c_4.
\ee
At the same time, the theory has exactly 7 propagating degrees of freedom also away from the flat space limit 
and for arbitrary $b_k$ (see  \cite{Hassan:2011ea,Alexandrov:2013rxa,Soloviev:2020zht} for its Hamiltonian 
formulation). 

Let us finally pass from the dimensionful spacetime coordinates ${\rm x}^\mu$ 
to the dimensionless ones, 
\be        \label{conf}
x^\mu = {\rm m x}^\mu. 
\ee
This  is equivalent to the conformal rescaling of the metrics, 
\be
{\rm g}_{\mu\nu}=\frac{1}{{\rm m}^2}\,g_{\mu\nu},~~~~~
{\rm f}_{\mu\nu}=\frac{1}{{\rm m}^2}\,f_{\mu\nu},
\ee
after which the field equations \eqref{Enst0} reduce to 
\bea                                  \label{Enst-eq}
G^\mu_{~\nu}({g})&=&\kappa_1\, T^\mu_{~\nu},~~~~~~~~~~~~
G^\mu_{~\nu}({f})= \kappa_2\, {\cal T}^\mu_{~\nu}, 
\eea
where $T^\mu_{~\nu}$ and ${\cal T}^\mu_{~\nu}$ are 
still given by \eqref{T},\eqref{tau1} with $\hat{\gamma}=\sqrt{\hat{{g}}^{-1}\hat{{f}}}$. 
The Bianchi identities for these equations imply that 
\be                                   \label{T1} 
\stackrel{(g)}{\nabla}_\rho T^\rho_\lambda=0\,,~~~~~
\stackrel{(f)}{\nabla}_\rho {\cal T}^\rho_\lambda=0\,,~~~~~
\ee
which is consistent with \eqref{id}. 
All fields and coordinates are now dimensionless and no trace of the 
mass parameter m is left in the equations. However, one has to remember that the unity of length
corresponds to the dimensionful  $1/{\rm m}$, which is 
the physical length scale. 

In what follows we shall be analyzing equations \eqref{Enst-eq} without 
making any assumptions about values of $\kappa_1$, $\kappa_2$ and $b_k$. However, 
when integrating the equations numerically, we shall assume that $\kappa_1+\kappa_2=1$ 
and choose $b_k$ according to \eqref{bbb}.  Therefore, our solutions depend on three parameters of the theory, 
$c_3,c_4$ and $\eta$,  where 
\be                \label{eta}
\kappa_1=\cos^2\eta,~~~~~~\kappa_2=\sin^2\eta.
\ee
\bbl{We shall assume in what follows that if the theory is extended to include an extra matter variables
denoted by $\Psi$, then the action \eqref{1} becomes $S[{\rm g},{\rm f}]\to S[{\rm g},{\rm f}]+S_{\rm mat}[{\rm g},\Psi]$, so that 
the matter couples only to the g metric. The g-geometry is therefore physically measurable as test particles follow its geodesics. 
The f-geometry is not directly coupled to matter, hence it cannot be directly seen and remains hidden.}

\section{SPHERICAL SYMMETRY  \label{sec3}}
\setcounter{equation}{0}
Let us introduce coordinates $(x^0,x^1,x^2,x^3)=(t,r,\vartheta,\varphi)$ 
and choose both metrics to be static, spherically symmetric,  and diagonal, 
\bea                             \label{ansatz0}
ds_g^2&=&g_{\mu\nu}dx^\mu dx^\nu=-Q^2dt^2+\frac{dr^2}{\Delta^2}+R^2d\Omega^2\,, \nonumber \\
ds_f^2&=&f_{\mu\nu}dx^\mu dx^\nu=-q^2 dt^2+\frac{dr^2}{W^2}+U^2d\Omega^2,
\eea
where $d\Omega^2=d\vartheta^2+\sin^2\vartheta d\varphi^2$ while 
$Q,\Delta,R,q,W,U$ are functions of the radial coordinate $r={\rm mr}$.
In fact, this is not the most general form of the spherically symmetric fields, since one could also 
include the off-diagonal metric element $f_{01}$ as shown by Eq.\eqref{anz} in Appendix \ref{time}. 
However, in the {\it static} case this would imply that  \eqref{b1} is 
the only possible solution   \cite{Volkov2012} (the situation changes in the time-dependent case). 
Therefore, we choose the static metrics to be both diagonal, which leads to nontrivial solutions.

The tensor $\gamma^\mu_{~\nu}$ in \eqref{gam} then  reads 
\be 
\gamma^\mu_{~\nu}={\rm diag}\left[
\frac{q}{Q},\frac{\Delta}{W},\frac{U}{R},\frac{U}{R}
\right],
\ee
and one obtains from  \eqref{T} 
\bea 
T^\mu_{~\nu}&=&{\rm diag}\left[
T^0_{~0},T^1_{~1},T^2_{~2},T^2_{~2}
\right],\nonumber \\
{\cal T}^\mu_{~\nu}&=&{\rm diag}\left[
{\cal T}^0_{~0},{\cal T}^1_{~1},{\cal T}^2_{~2},{\cal T}^2_{~2}
\right],
\eea
where 
\bea
T^0_{~0}&=&-{\cal P}_0-{\cal P}_1\,\frac{\Delta}{W}, \nonumber \\
T^1_{~1}&=&-{\cal P}_0-{\cal P}_1\,\frac{q}{Q},~~\nonumber \\
T^2_{~2}&=&-{\cal D}_0-{\cal D}_1\left(\frac{q}{Q}
+\frac{\Delta}{W}\right)-{\cal D}_2\,
\frac{q\Delta}{QW},\nonumber \\
{\bf u}^2{\cal T}^0_{~0}&=&-{\cal P}_2-{\cal P}_1\,\frac{W}{\Delta}, \nonumber \\
{\bf u}^2{\cal T}^1_{~1}&=&-{\cal P}_2-{\cal P}_1\,\frac{Q}{q},~\nonumber \\
{\bf u}{\cal T}^2_{~2}&=&-{\cal D}_3-{\cal D}_2\left(\frac{Q}{q}
+\frac{W}{\Delta}\right)-{\cal D}_1\,
\frac{QW}{q\Delta}. 
\eea
Here ${\bf u}={U}/{R}$ and 
\bea 
{\cal P}_m&=&b_m+2b_{m+1}{\bf u}+b_{m+2}{\bf u}^2\,, \nonumber \\ 
{\cal D}_m&=&b_m+b_{m+1}{\bf u}\,,~~~~  ~~~~~~~            ~~~(m=0,1,2). 
\label{e5} 
\eea
The independent field equations are 
\bea                                  \label{Ein}
G^0_0(g)&=&\kappa_1\, T^{0}_{~0}, ~~~~
G^1_1(g)=\kappa_1\, T^{1}_{~1}, ~~~~\nonumber \\
G^0_0(f)&=&\kappa_2\, {\cal T}^{0}_{~0},~~~~
G^1_1(f)=\kappa_2\, {\cal T}^{1}_{~1},~
\eea
plus the conservation condition 
$
\stackrel{(g)}{\nabla}_\mu T^\mu_\nu=0\,,
$
which has only one nontrivial component, 
\be                                \label{CONS}
\stackrel{(g)}{\nabla}_\mu T^\mu_{~1}=
\left(T^1_{~1}\right)^\prime
+\left.\left.\frac{Q^\prime}{Q}\right(T^1_{~1}-T^0_{~0}\right)
+2\left.\left.\frac{R^\prime}{R}
\right(T^1_{~1}-T^2_{~2}\right)=0,
\ee
where the prime denotes differentiation with respect to $r$. 
The
conservation condition for the second energy-momentum tensor 
also has only one nontrivial component,  
\be                                \label{CONSf}
\stackrel{(f)}{\nabla}_\mu {\cal T}^\mu_{~1}=
\left({\cal T}^1_{~1}\right)^\prime
+\left.\left.\frac{q^\prime}{q}\right({\cal T}^1_{~1}-{\cal T}^0_{~0}\right)
+2\left.\left.\frac{U^\prime}{U}
\right({\cal T}^1_{~1}-{\cal T}^2_{~2}\right)=0,
\ee
but this  follows from \eqref{CONS} due to the identity relation \eqref{id}. 
As a result, there are  5 independent equations in \eqref{Ein}, \eqref{CONS},
which is enough to determine the 6 field amplitudes $Q,\Delta,R,q,W,U$, 
 because  
 the freedom of 
reparametrization  of the radial coordinate
$
r\to \tilde{r}(r)
$
allows one to fix one of the amplitudes. 

\section{FIELD EQUATIONS  \label{sec4}}
\setcounter{equation}{0}

Let us introduce new functions 
\be                                 \label{NY}
N=\Delta R^\prime\,,~~~~Y=WU^\prime\,,
\ee
in terms of which the 
 two metrics read 
\bea                             \label{ansatz000}
ds_g^2&=&-Q^2dt^2+\frac{dR^2}{N^2}+R^2d\Omega^2\,, \nonumber \\
ds_f^2&=&-q^2 dt^2+\frac{dU^2}{Y^2}+U^2d\Omega^2.
\eea
The advantage of this parametrization 
is that 
the second derivatives disappear from the Einstein tensor and 
the four Einstein equations \eqref{Ein} become 
\bea
N^\prime&=&-\frac{\kappa_1}{2}\frac{R}{NY}\left(R^\prime Y{\cal P}_0
+U^\prime N {\cal P}_1
 \right)+\frac{(1-N^2)R^\prime}{2RN}\,,  \label{e1} \\
Y^\prime&=&-\frac{\kappa_2}{2}\frac{ R^2}{UNY}
\left(R^\prime Y {\cal P}_1+U^\prime N {\cal P}_2
 \right)+\frac{(1-Y^2)U^\prime}{2UY}\,, \label{e2}  \\
Q^\prime&=&-\left(
\kappa_1(Q{\cal P}_0+q{\cal P}_1)+\frac{Q(N^2-1)}{R^2}
\right) \frac{RR^\prime}{2N^2}\,, \label{e3}  \\
q^\prime&=&-\left(
\kappa_2(Q{\cal P}_1+q{\cal P}_2)+\frac{q(Y^2-1)}{R^2}
\right) \frac{R^2U^\prime}{2Y^2U}\,. \label{e4} 
\eea
The conservation condition \eqref{CONS} reads 
\bea                                \label{CONSg}
\stackrel{(g)}{\nabla}_\mu T^\mu_{~1}&=&
\frac{U^\prime}{R}\left(1-\frac{N}{Y}\right)\left(d{\cal P}_0+
\frac{q}{Q}\,d{\cal P}_1\right)
+\left(
\frac{q^\prime}{Q}-\frac{NQ^\prime U^\prime}{YQR^\prime}
\right){\cal P}_1=0,
\eea
and using Eqs.\eqref{e3} and \eqref{e4}, this 
reduces to 
\bea                                \label{CONSgg}
R^2Q\stackrel{(g)}{\nabla}_\mu T^\mu_{~1}&= &
\frac{U^\prime}{Y}\,{\bf C}=0\,,
\eea
where
\bea                             \label{C}
{\bf C}&=&\left(
\kappa_2\,\frac{R^4{\cal P}_1^2}{2UY}
-\kappa_1\,\frac{R^3\,{\cal P}_0{\cal P}_1}{2N}
-\frac{(N^2-1)\,R{\cal P}_1}{2N}+(N-Y)Rd{\cal P}_0
\right)Q \nonumber \\
&+&\left(
\kappa_2\,\frac{R^4{\cal P}_1{\cal P}_2}{2UY}
-\kappa_1\,\frac{R^3\,{\cal P}_1^2 }{2N}
+\frac{(Y^2-1)\,R^2{\cal P}_1}{2UY}+(N-Y)Rd{\cal P}_1
\right)q\,,
\eea
with
\be  
d{\cal P}_m=
2\,(b_{m+1}+b_{m+2}{\bf u})~~~(m=0,1).          \label{e5a} 
\ee
The conservation condition \eqref{CONSf} becomes 
\be                                \label{CONSff}
-U^2q\stackrel{(f)}{\nabla}_\mu {\cal T}^\mu_{~1}=
\frac{R^\prime }{N}\,{\bf C}=0\,.
\ee
The two conditions \eqref{CONSgg} and \eqref{CONSff} will be fulfilled if
$U^\prime=R^\prime=0$, in which case 
both metrics are degenerate. 
If the metrics are not degenerate, then conditions \eqref{CONSgg} and \eqref{CONSff} 
reduce to the algebraic constraint 
\be                          \label{CC}
{\bf C}=0. 
\ee 
This constraint can be resolved with respect to $q$ to give
\be                \label{q}
q=\Sigma(R,U,N,Y)\,Q,
\ee
where $\Sigma(N,Y,R,U)$ is the (negative) ratio of the coefficients in front of $Q$ and $q$ in \eqref{C}. 

As a result, we obtain four differential equations \eqref{e1}--\eqref{e4} 
plus one algebraic constraint \eqref{CC}. The  same equations 
can be obtained by inserting the metrics \eqref{ansatz000} directly to the 
action \eqref{1}, which gives 
\bea                                      \label{1a}
&&S
=\frac{4\pi}{{\rm m}^2\boldsymbol\kappa}\int L\, dt dr
\,,
\eea
where, dropping a total derivative, 
\bea 
L&=&\frac{1}{\kappa_1}\left(
\frac{(1-N^2)\, R^\prime}{N}-2RN^\prime
\right)Q
+\frac{1}{\kappa_2}\left(
\frac{(1-Y^2)\,U^\prime }{Y}-2UY^\prime
\right)q \nonumber \\
&-&\frac{QR^2 R^\prime }{N}\,{\cal P}_0
-\left(
\frac{QR^2 U^\prime}{Y}+\frac{qR^2R^\prime}{N}
\right){\cal P}_1
-\frac{qR^2 U^\prime }{Y}\,{\cal P}_2\,.
\eea
Varying $L$ with respect to $N,Y,Q,q$ gives 
Eqs.\eqref{e1}--\eqref{e4}, while varying it with respect to $R,U$ 
reproduces conditions  \eqref{CONSgg} and \eqref{CONSff}.  
The  
equations and the Lagrangian $L$ are invariant under the interchange symmetry 
\eqref{sym}, which now reads 
\be                                 \label{change}
\kappa_1\leftrightarrow\kappa_2,~~Q\leftrightarrow q,~~~
N\leftrightarrow Y,~~~R \leftrightarrow U,~~~ b_m\leftrightarrow b_{4-m}\,.
\ee
Equations \eqref{e1}--\eqref{e4} contain $R^\prime$ and $U^\prime$ which are not yet known.
One of these two amplitudes can be fixed by imposing a gauge condition, but the other one should be 
determined dynamically. We need therefore one more condition, and the only way to get it is to 
differentiate the constraint. Since the constraint should be stable, 
this gives the secondary constraint:
\be 
{\bf C}^\prime=
\frac{\partial \bf C}{\partial N}\,N^\prime+
\frac{\partial \bf C}{\partial Y}\,Y^\prime+
\frac{\partial \bf C}{\partial Q}\,Q^\prime+
\frac{\partial \bf C}{\partial q}\,q^\prime+
\frac{\partial \bf C}{\partial R}\,R^\prime+
\frac{\partial \bf C}{\partial U}\,U^\prime=0.
\ee 
Expressing here  the derivatives $N^\prime$, $Y^\prime$, $Q^\prime$, $q^\prime$  by  
Eqs.\eqref{e1}--\eqref{e4} and using  the relation \eqref{q}, this condition reduces to 
\be                          \label{C1}
{\bf C}^\prime={\cal A}(R,U,N,Y)\, R^\prime+{\cal B}(R,U,N,Y)\, U^\prime=0,
\ee
where the functions ${\cal A}(R,U,N,Y)$ and ${\cal B}(R,U,N,Y)$ are rather complicated and 
we do not show them explicitly. When the radial coordinate change, both $R^\prime$ and $U^\prime$ change,
\be                         \label{rr}
r\to \tilde{r}(r),~~~~~~R^\prime\to \tilde{R}^\prime=R^\prime \,\frac{dr}{d\tilde{r}},
~~~~~~U^\prime\to \tilde{U}^\prime=U^\prime \,\frac{dr}{d\tilde{r}},
\ee
but the relation \eqref{C1} between $R^\prime$ and $U^\prime$ remains the same. The secondary constraint can be 
resolved with respect to $U^\prime$, 
\be               \label{U1}
U^\prime =-\frac{{\cal A}(R,U,N,Y)}{{\cal B}(R,U,N,Y) }\,R^\prime\equiv {\cal D}_U(R,U,N,Y)\,R^\prime\,.
\ee
We can now use the gauge symmetry \eqref{rr} to impose the coordinate condition 
\be
R^\prime=1~~~~\Rightarrow~~~~~R=r,
\ee            
and then \eqref{U1} reduces to 
\be            \label{UU1}
U^\prime = {\cal D}_U(r,U,N,Y)\,.
\ee
Now, $U^\prime$ appears in the right-hand sides of Eqs.\eqref{e1} and \eqref{e2}, and replacing it there by 
the value \eqref{UU1}, these two equations together with \eqref{UU1} form a closed system of 
three equations 
\bea        \label{eqs}
N^\prime&=&{\cal D}_N(r,U,N,Y),\nonumber \\
Y^\prime&=&{\cal D}_Y(r,U,N,Y), \nonumber \\
U^\prime&=&{\cal D}_U(r,U,N,Y). 
\eea
The amplitudes $Q,q$ are determined as follows. Injecting \eqref{q} to \eqref{e3} yields 
the equation 
\be                \label{Q1}
Q^\prime=-\frac{r}{2N^2}\left(
\kappa_1({\cal P}_0+\Sigma(r,U,N,Y){\cal P}_1)+\frac{N^2-1}{r^2}
\right) \, Q\equiv {\cal F}(r,U,N,Y) Q,
\ee
which determines $Q$, and when its solution is known, $q$ is determined algebraically from \eqref{q}. 

\bbl{Solutions of \eqref{eqs},\eqref{Q1} and \eqref{q} are automatically compatible with \eqref{e1}--\eqref{e4}
and  with the constraint \eqref{CC}. For example,  the algebraic solution for $q$ given by \eqref{q} is compatible 
with its differential equation \eqref{e4}  because the latter contains $U^\prime$ not defined by 
\eqref{e1}--\eqref{e4}. To determine   $U^\prime$ one needs to differentiate the constraint \eqref{CC} whose 
algebraic solution is \eqref{q}  and to use  \eqref{e1}--\eqref{e4}. This completes the procedure in a consistent way. 
}

In what follows we shall mainly focus on the three coupled equations  \eqref{eqs} determining $N,Y,U$.
As soon as their solution is obtained, 
the amplitudes $Q,q$ are determined  from \eqref{Q1},\eqref{q}.

\section{ANALYTICAL SOLUTIONS \label{simp}} 
\setcounter{equation}{0}
Some simple solutions of the equations can be 
obtained analytically \cite{Volkov2012,Hassan:2012wr},
for which it is  convenient to use the equations in the form \eqref{e1}--\eqref{e4}.

\subsection{Proportional backgrounds} 

Choosing 
the two metrics to be conformally related \cite{Volkov2012,Hassan:2012wr},
\be                           \label{prop1}
ds_f^2=C^2 ds_g^2\,,
\ee
with a constant $C$, the solution is given by 
\be                           \label{prop2}
Q^2=N^2=Y^2= 1-\frac{2M}{r}-\frac{\Lambda(C)}{3}\,r^2\,,~~~~R=r,~~~q=C Q,~~~U=C R,
\ee 
which describes two proportional Schwarzschild-(anti-)de Sitter geometries. The constant $C$ and the cosmological 
constant $\Lambda(C)$ are determined by 
\be                                 \label{sgm}
\kappa_1({\cal P}_0+C{\cal P}_1)=
\frac{\kappa_2}{C}({\cal P}_1+C{\cal P}_2)\equiv \Lambda(C).
\ee 
Since ${\cal P}_m$ defined by \eqref{e5} are polynomials in ${\bf u}=U/R=C$, this yields 
an algebraic equation for $C$ that  can have up to four real roots. If the parameters $b_k$ are chosen according to \eqref{bbb},
then one of the roots is $C=1$, in which case $\Lambda=0$. 

The value of the dimensionful cosmological constant ${\bm \Lambda}$ should agree with the observation, hence one should have 
\be
{\bm \Lambda}={\rm m}^2 \Lambda\sim 1/R_{\rm Hub}^2
\ee
where $R_{\rm Hub}$  is the Hubble radius of our Universe. One way to fulfill this relation is to assume that the graviton mass is 
extremely small such that the Compton length is of the order of the Hubble radius,
\be                           \label{Hubble}
1/{\rm m}\sim R_{\rm Hub}.
\ee
However, the relation can also be fulfilled by assuming that $\Lambda$ is very 
small, which is possible if there is a hierarchy between the two couplings: $\kappa_1\ll \kappa_2=1-\kappa_1\sim 1$. 
Equation \eqref{sgm} implies then that $\Lambda\sim \kappa_1$ and that $C$ should be very close to a root of ${\cal P}_1+C{\cal P}_2$. 
\bbl{The hierarchy between the two couplings
is in fact necessary to reconcile with the observations the perturbation spectrum  of the massive bigravity cosmology,
because it contains an instability in the scalar sector \cite{Comelli:2012db,Konnig:2014xva,Lagos:2014lca}. 
For this  one should assume that 
\cite{Akrami:2015qga, Mortsell:2015exa, Aoki:2015xqa, Luben:2018ekw, Hogas:2019ywm} 
\be                             \label{m0}
\frac{\kappa_1}{\kappa_2}\approx \kappa_1\,\bbl{ \leq}\, \left(\frac{{\rm M}_{\rm ew}}{{\rm M}_{\rm Pl}} \right)^2\sim 10^{-34}\ll 1,
\ee
where ${\rm M}_{\rm ew}\sim 100$~GeV is the electroweak energy scale and ${\rm M}_{\rm Pl}\sim 10^{19}$~GeV is the Planck mass. 
Here $10^{-34}$ is the {\it upper bound} for $\kappa_1$ imposing which shifts  the instability toward early times 
making it unobservable. However, $\kappa_1$ can also be less then this bound \cite{Akrami:2015qga}, hence 
\be
\kappa_1=\gamma^2\times 10^{-34}~~~~~~\mbox{with}~~~~~\gamma\in[0,1]. 
\ee 
As a result,
\be                              \label{m}
1/{\rm m}\sim \sqrt{\Lambda}\,R_{\rm Hub}=\sqrt{\kappa_1}\,R_{\rm Hub}\,=\gamma\times  \left(\frac{{\rm M}_{\rm ew}}{{\rm M}_{\rm Pl}} \right)\,R_{\rm Hub}\sim \gamma\times 10^6~\mbox{ km},
\ee
which is of the order of the  solar size if $\gamma\sim 1$. However, in what follows we shall not be always assuming $\kappa_1$ to be small and shall
present our results for arbitrary  $\kappa_1\in[0,1]$. }

\subsection{Deformed AdS background}

Choosing 
$U,q$ to be constant, 
\be                \label{Uq}
U=U_0,~~~~q=q_0, 
\ee
solves Eqs.\eqref{e4} and \eqref{CONSgg}, 
while Eqs.\eqref{e1}--\eqref{e3} then can be integrated in quadratures  \cite{Volkov2012}.
However, such solution is unacceptable, since the f metric degenerates if $U^\prime=0$. 
At the same time, there are other, more general solutions which approach \eqref{Uq} for $r\to\infty$, and for these 
solutions $U^\prime$ vanishes only asymptotically, hence they are acceptable. The leading at large $r$ terms of such 
solutions are 
 \bea                           \label{NQQQ}
N^2&=&
-\kappa_1\frac{b_0}{3}\,r^2
-\kappa_1 b_1U_0\, r  
+{\cal O}(1)
 \,, ~~~~~
Y=-\frac{\sqrt{3}\kappa_2 b_1}{4U_0\sqrt{-\kappa_1 b_0} }\,r^2+{\cal O}(r)
, \nonumber \\
Q&=&\frac{q_0 }{4U_0}\, r+{\cal O}(1), ~~~~
U=U_0+{\cal O}\left(\frac{1}{r}\right), ~~~~~
q=q_0+{\cal O}\left(\frac{1}{r}\right).
\eea
The g metric  approaches the AdS metric in the leading ${\cal O}(r^2)$ order, but the subleading terms 
do not have the AdS structure. 

It turns out that solutions of Eqs.\eqref{e1}--\eqref{e4}  generically approach 
for $r\to\infty$ either \eqref{prop2} or \eqref{NQQQ}  (or they show a curvature singularity at a finite $r$),
hence they are not asymptotically flat \cite{Volkov2012}.

\section{BOUNDARY CONDITIONS AT THE HORIZON  \label{BC}}
\setcounter{equation}{0}

Let us require the g metric to have a regular event horizon at some $r=r_H$
by demanding the metric components $g_{00}=Q^2$ and $g^{rr}=N^2$ to show simple zeroes at this point.
Therefore, we demand that close to this point one has 
$Q^2\sim N^2\sim r-r_H$ and we consider only the exterior region $r\geq r_H$ where $Q^2>0$ and $N^2>0$. 
Such a behavior is compatible with the field equations if only 
the f metric also  shows a regular horizon at the same place,
hence $q^2\sim Y^2\sim r-r_H$.  As a result, both metrics share a horizon at the same place $r=r_H$,
in agreement with  \cite{Deffayet:2011rh, Banados:2011hk}. 
However, the horizon radius measured by the g metric, $r_H$, can be different from the radius measured by the second metric, 
$U(r_H)$. We therefore introduce the parameter $u\equiv {\bf u}(r_H)=U(r_H)/r_H$. 

As a result, the local solutions close to the horizon are expected to  have the form 
\begin{align}        \label{l1}
    N^2&=\sum_{n\geq 1}a_n(r-r_H)^n,~~~~~~Y^2=\sum_{n\geq 1}b_n(r-r_H)^n,\nonumber  \\
    U&=u\,r_H+\sum_{n\geq 1}c_n(r-r_H)^n,
\end{align}
the two other amplitudes being 
\begin{align}
    \label{l2}
    Q^2&=\sum_{n\geq 1}d_n(r-r_H),~~~~~~~~
    q^{2}=\sum_{n\geq 1}e_n(r-r_H)^n.
\end{align}
The equations then allow one to recurrently determine the coefficients $a_n,b_n,c_n,d_n,e_n$. 
It turns out they all can be expressed in terms of $a_1$, which should fulfil a quadratic equation 
\be            \label{qqq}
{\cal A} a_1^2+{\cal B} a_1+{\cal C}=0~~~~~\Rightarrow~~~~
a_1=\frac{1}{2{\cal A}}\,\left(-{\cal B}+\sigma\sqrt{{\cal B}^2-4{\cal AC}}   \right),~~~~~~~\sigma=\pm 1,
\ee
where ${\cal A,B,C}$ are functions of $u,r_H$ and  of  the theory  parameters 
$b_k,\kappa_1,\kappa_2$. It turns out that one should choose $\sigma=+1$, since choosing $\sigma=-1$ always yields singular solutions. 
Therefore, for a chosen a value of the horizon size $r_H$, the local solutions 
\eqref{l1},\eqref{l2} comprise a set labeled by a continuous parameter $u$. 
These local solutions determine the boundary 
conditions at the horizon, and they can be numerically extended to the region $r>r_H$. 

The surface gravity for each metric is  \cite{Volkov2012} 
\be
\kappa_g^2=\lim_{r\to r_H} Q^2N^{\prime 2}=\frac14 d_1 a_1,~~~~~
\kappa_f^2=\lim_{r\to r_H}  q^2\left(\frac{Y}{U^\prime} \right)^{\prime 2}= \frac{e_1 b_1}{4\,c_1^2}, 
\ee
and using the values of the expansion coefficients determined by the equations yields the relation $\kappa_g=\kappa_f$, hence 
the two surface gravities coincide, as coincide the Hawking temperatures,
\be                 \label{TT}
T=\frac{\kappa_g}{2\pi}=\frac{\kappa_f}{2\pi}. 
\ee

{One has close to the horizon $N(r)\sim Y(r)\sim \sqrt{r-r_H}$ hence  the derivatives $N^\prime$ and $Y^\prime$ are not defined at the horizon. 
The usual practice would then be to start the numerical integration not at $r=r_H$ but at a nearby point $r=r_H+\epsilon$. However, 
although the dependence  on $\epsilon$ is expected to be small, still its presence  in the procedure may 
lead to numerical instabilities. This point was emphasized in \cite{Torsello:2017cmz}. This difficulty can be resolved as follows. 
Setting 
\be
N(r)={S(r)}\,\nu(r),~~~~~Y(r)=S(r)\,y(r)~~~~~\mbox{with}~~S(r)=\sqrt{1-\frac{r_H}{r}},
\ee  
the functions $\nu(r),y(r)$ and all their derivatives assume finite values at $r=r_H$. Making this change of variables in \eqref{eqs}  gives 
a ``desingularized" version of the equations that allows us to  start the numerical integration exactly at $r=r_H$.
This form of the equations is described in Appendix \ref{Des}. }

To recapitulate,  all black holes for a given $r_H$ can be   labeled by only one parameter $u$.  If $u=1$ then the two metrics 
coincide everywhere and the solution is Schwarzschild \eqref{b2}. If $u=C$ where $C$ is a root of the 
algebraic equation \eqref{sgm}, then the solutions is Schwarzschild-(anti-)de Sitter and is described by 
\eqref{prop1} and \eqref{prop2}. For other values of $u$ the numerical integration 
produces more general solutions which 
 describe hairy black holes and which  can be of the  following three qualitative types,
depending on their asymptotic behavior  \cite{Volkov2012}.


{\bf a)} Solutions extending up to arbitrarily  large values of $r$ and asymptotically approaching a proportional AdS 
background \eqref{prop1}, \eqref{prop2}. At large $r$ one has  $N=N_0\,(1+\delta N)$, 
$Y=Y_0\,(1+\delta Y)$, $U=U_0\,(1+\delta U)$ where $N_0,Y_0,U_0$ are given by \eqref{prop2}, while the 
deviations $\delta N,\delta Y,\delta U$ approach zero. In the linear approximation, the latter are described by 
\be
\delta N=\frac{A}{r^3}, ~~~~~~\delta U=B_1 e^{\lambda_1 r}+B_2 e^{\lambda_2 r},~~~~~~\delta Y={\cal O}(\delta U),
\ee
where $A,B_1,B_2$ are integration constants and real parts of $\lambda_1$ and $\lambda_2$ are {\it negative}. 
All of these three perturbation modes vanish for  $r\to\infty$, and since 
the number of equations \eqref{eqs} is also three, it follows that 
the  AdS background is an {\it attractor} at large $r$. 

{\bf b)} Solutions extending up to arbitrarily  large values of $r$ and asymptotically approaching a deformed AdS background 
\eqref{NQQQ}. The latter is also an attractor  at  large $r$. 

{\bf c)} Solutions extending only up to   $r=r_s<\infty$ where derivatives of some  metric functions diverge, 
which corresponds to a curvature singularity. 


This exhausts the possible types of {\it generic} solutions. If one integrates the equation for many 
different values of $u$, one always obtains solutions of the above three types and one does not find 
asymptotically flat solutions other than  Schwarzschild. For example, choosing $u=1+\epsilon$ 
 yields solutions which are almost Schwarzschild in a region close to the horizon, but 
for larger values of $r$ 
they deviate from the Schwarzschild  metric more and more \cite{Volkov2012}
(this means the Schwarzschild solution is Lyapunov unstable
\cite{Torsello:2017cmz}). 
All of this does not mean that 
the Schwarzschild  is the only asymptotically flat black hole solution. There may be others, but they are not 
parametrically close to the Schwarzschild   solution and should 
 correspond  to some   discrete
values of $u$ which are difficult to detect by a ``brute force" method.

\section{BOUNDARY CONDITIONS AT INFINITY \label{BI}}

\setcounter{equation}{0}

Let us suppose the solutions to approach flat space with $g_{\mu\nu}=f_{\mu\nu}=\eta_{\mu\nu}$ at large $r$
and set 
\be            \label{inf}
N=1+\delta N,~~~~Y=1+\delta Y,~~~~~U=r+\delta U. 
\ee
In fact, a more general possibility would be to require the g metric to approach  the flat Minkowski   metric 
${\rm diag}(-1,1,1,1)$ and the f metric  to approach  just a flat metric, 
as for example ${\rm diag}(-a^2,b^2,b^2,b^2)$  with constat $a,b$. This would lead to solutions whose Lorentz invariance 
is broken in the asymptotic region
\cite{Berezhiani:2008nr,Comelli:2011wq}. 
However, we shall not analyze this option.

Inserting \eqref{inf} to \eqref{eqs} yields 
\bea          \label{eqinf}
\delta N^\prime&=&-\frac{1}{r}(\kappa_2\,\delta N+\kappa_1\, \delta Y)-\kappa_1 \delta U+{\cal N}_N, \nonumber \\
\delta Y^\prime&=&-\frac{1}{r}(\kappa_2\,\delta N+\kappa_1\,\delta Y )+\kappa_2\, \delta U+{\cal N}_Y, \nonumber \\
\delta U^\prime&=&\left(1+\frac{2}{r^2} \right)\,\left(\delta Y-\delta N\right)+{\cal N}_U,
\eea
where ${\cal N}_N,{\cal N}_Y,{\cal N}_U$ are the nonlinear in $\delta N,\delta Y,\delta U$ 
parts of the right-hand sides ${\cal D}_N,{\cal D}_Y,{\cal D}_U$ in \eqref{eqs}. Neglecting the 
nonlinear terms, the solution of these equations is 
\bea                    \label{infXX}
\delta N=\frac{A}{r}+B\,\kappa_1\,\frac{1+r}{r}\,e^{-r}+C\,\kappa_1\,\frac{1-r}{r}\,e^{+r}, \nonumber \\
\delta Y=\frac{A}{r}-B\,\kappa_2\,\frac{1+r}{r}\,e^{-r}-C\,\kappa_2\,\frac{1-r}{r}\,e^{+r}, \nonumber \\
\delta U=B\,\frac{r^2+r+1}{r^2}\,e^{-r}+C\,\frac{r^2-r+1}{r^2}\,e^{+r},
\eea
where $A,B,C$ are integration constants. 
The part of this solution proportional to $A$
is the Newtonian mode describing the massless graviton subject to the 
linearized Einstein equations \eqref{m=0}. The other two modes proportional to $B$ and $C$ 
fulfill  the Fierz-Pauli equations \eqref{FP} and describe the massive graviton, hence they contain the 
Yukawa exponents (remember that $r={\rm mr}$). 

As one can see, among the three modes only two are stable for $r\to \infty$ while the third one 
diverges in this limit, hence {\it flat space is not an attractor}. This is why one cannot get 
asymptotically flat solutions by simply integrating from the horizon -- trying to approach flat 
space in this way, 
 the unstable mode $e^{+r}$ rapidly wins and drives the 
solution away from flat space. The only way to proceed is to suppress the unstable 
mode from the very beginning by requiring the solution at large $r$ to  be 
\bea                \label{nnn}
\delta N&=&\frac{A}{r}+B\,\kappa_1\,\frac{1+r}{r}\,e^{-r}+\ldots, \nonumber \\
\delta Y&=&\frac{A}{r}-B\,\kappa_2\,\frac{1+r}{r}\,e^{-r}+\ldots, \nonumber \\
\delta U&=&r+B\,\frac{r^2+r+1}{r^2}\,e^{-r}+\ldots, 
\eea
where the dots denote non-inear corrections. The usual practice would be  to neglect the dots and assume that 
the linear terms approximate the solution everywhere for $r>r_\star$, where $r_\star$  is some large value. 
However,  one can check that already the quadratic correction contains an additional factor of $\ln(r)$ and
 hence dominates the linear part for $r\to\infty$. 
Therefore, nonlinear corrections are important, but if all of them are taken into account, 
it is not obvious  that the solution will remain asymptotically flat. 

Fortunately, problems of this kind have  been studied -- see, e.g., \cite{BFM}. 
To take the nonlinear corrections into account, the procedure is as follows. 
Let us express $\delta N,\delta Y,\delta U$  in terms of three functions $Z_0,Z_{+},Z_{-}$:
\bea                   \label{ZZZ}
\delta N&=&Z_0+\kappa_1\,\frac{1+r}{r}\,Z_{+}+\kappa_1\,\frac{1-r}{r}\,Z_{-}\,, \nonumber  \\
\delta Y&=&Z_0-\kappa_2\,\frac{1+r}{r}\,Z_{+}-\kappa_2\,\frac{1-r}{r}\,Z_{-}\,,  \nonumber \\
\delta U&=&\frac{1+r+r^2}{r^2}\,Z_{+}+\frac{1-r+r^2}{r^2}\,Z_{-}\, . 
\eea
Equations \eqref{eqinf} then assume the form 
\bea                 \label{Z}
Z_0^\prime+\frac{Z_0}{r}&=&{\cal S}_{0}(r,Z_0,Z_\pm)\equiv \kappa_1\, {\cal N}_Y+ \kappa_2\, {\cal N}_N\,, \nonumber  \\
Z_{+}^\prime+Z_{+}&=&{\cal S}_{+}(r,Z_0,Z_\pm)\equiv \frac{r^2-r+1}{2r^2}\,\left({\cal N}_N-{\cal N}_Y \right) 
+\frac{r-1}{2r}\,{\cal N}_U\,,  \nonumber  \\
Z_{-}^\prime-Z_{-}&=&{\cal S}_{-}(r,Z_0,Z_\pm)\equiv \frac{r^2+r+1}{2r^2}\,\left({\cal N}_Y-{\cal N}_N \right) 
+\frac{r+1}{2r}\,{\cal N}_U\, . 
\eea
Terms on the left in these equations are linear in $Z_0,Z_\pm$, while those on the right are 
non-inear. Neglecting the nonlinear terms, the solution is $Z_0=1/r$, $Z_{+} =e^{-r}$, $Z_{-}=e^{+r}$, and if we
set 
\be              \label{0}
Z_0=\frac{A}{r},~~~~~Z_{+}=B\,e^{-r},~~~~~~Z_{-}=0,
\ee
this reproduces the linear part of \eqref{nnn}. Now, to take the nonlinear terms into account, 
one converts Eqs.\eqref{Z} into the equivalent set of integral equations,
\bea                         \label{integral}
Z_{0}(r)&=&\frac{A}{r}-\int_{r}^\infty \frac{\x}{r}\,{\cal S}_{0}(\x,Z_0(\x),Z_\pm(\x))\, d\x\,, \nonumber \\
Z_{+}(r)&=&B\,e^{-r}+\int_{r_\star}^r e^{\x-r}\,{\cal S}_{+}(\x,Z_0(\x),Z_\pm(\x))\, d\x\,,\nonumber   \\
Z_{-}(r)&=&-\int_{r}^\infty e^{r-\x}\,{\cal S}_{-}(\x,Z_0(\x),Z_\pm(\x))\, d\x\,,
\eea
where $r_\star$ is some large value. These equations determine the solution for $r>r_\star$, and they 
are solved by iterations. To start the iterations, one neglects the nonlinear terms, which gives 
the configuration \eqref{0}. The next step is to inject this configuration to the integrals, 
which gives the corrected configuration, and so on. 
In practice, one introduces  variables $x=r_\star/r$ and 
$\bar{x}=r_\star/\bar{r}$ assuming values in the interval $[0,1]$, and then one discretizes 
the interval  to compute the integrals. 

\begin{figure}
    \centering
    \includegraphics[scale=0.65]{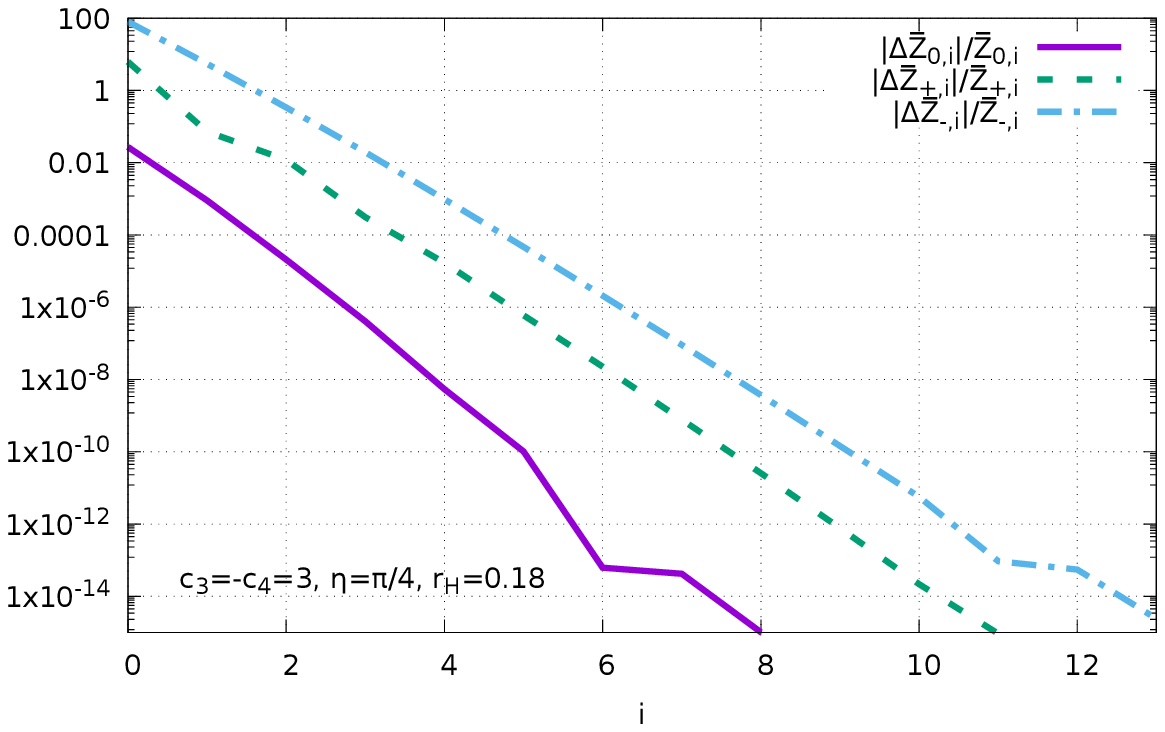}
    \includegraphics[scale=0.65]{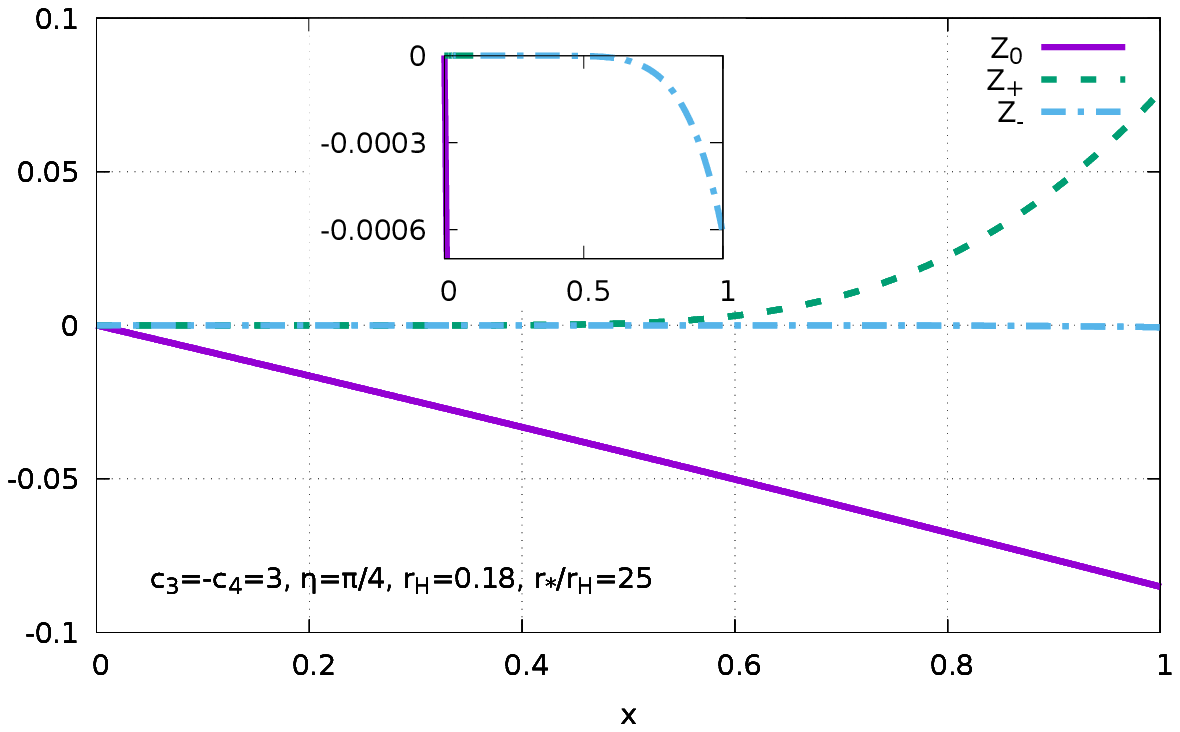}
   \caption{\bl{Left: convergence of the iterations of the integral equations \eqref{integral}. 
   Right: the amplitudes $Z_0$, $Z_\pm$ against $x=r_\star/r$,
  where  the insertion shows  a closeup of $Z_{-}$.}}
    \label{Fig0}
\end{figure}

To see the convergence of the iterations, we compute for each $Z$ and for each discretization  point 
the difference $\Delta Z_i=Z_{i+1}-Z_{i}$ of the results of the consecutive $(i+1)$th and  $i$th iterations, 
and then  we take the average $\overline{{\Delta {Z}}}_i$ over all discretization  points. 
Computing similarly  the average  $\bar{Z}_i$ of $Z_i$, 
the ratios $\overline{\Delta Z}_i/\bar{Z}_i$ 
decrease with $i$ exponentially fast, as seen on the left panel in Fig.\ref{Fig0},  hence the iterations converge. 
\bl{The solution of the integral equations is shown on the right panel in Fig.\ref{Fig0}: 
the amplitudes $Z_0$ and $Z_\pm$ against $x=r_\star/r$ (for $r_\star =25\, r_H$).  One can see that 
the amplitude $Z_{-}$ is always small but nonvanishing,  and that 
all the three amplitudes vanish for $x=0$,  hence the solution is indeed 
asymptotically flat. } 

This yields an asymptotically flat solution in the region $r>r_\star$. 
To extend this solution  to the region $r_H<r<r_\star$ 
one only needs its values at $r=r_\star$, 
\bea              \label{0a}
Z_0(r_\star)&=&\frac{A}{r_\star}-\int_{r_\star}^\infty \frac{\x}{r}\,{\cal S}_{0}(\x,Z_0(\x),Z_\pm(\x))\, d\x\,,
~~~~~Z_{+}(r_\star)=B\,e^{-r_\star},~~~~~\nn\\
Z_{-}(r_\star)&=&-\int_{r_\star}^\infty e^{r_\star-\x}\,{\cal S}_{-}(\x,Z_0,Z_\pm)\, d\x\,.
\eea

To recapitulate, the described above procedure yields the boundary values for the fields at a large $r_\star$ and makes sure that 
the solution for $r>r_\star$ exists and is indeed asymptotically flat. 
It is worth noting that the parameter $A$ determines the ADM mass, 
\be                \label{ADM}
M=-A. 
\ee

\section{NUMERICAL PROCEDURE  \label{hairybh}}

\setcounter{equation}{0}

Summarizing the above discussion, the asymptotically flat black holes are described by solutions 
of the three coupled first order ODEs \eqref{eqs} for the three functions $N(r),Y(r),U(r)$ 
[which determines also  $Q(r),q(r)$]
with the following boundary conditions.
At the horizon $r=r_H$ one has 
 \be
 N(r)=\sqrt{1-\frac{r_H}{r}}\,\nu(r), ~~~~~~Y(r)=\sqrt{1-\frac{r_H}{r}}\,y(r),
 \ee 
where the horizon values $\nu(r_H)\equiv \nu_H$ and $y(r_H)\equiv y_H$ are finite and determined by the 
 Eqs. \eqref{alg}, \eqref{yh} in Appendix \ref{Des}, while  $U(r_H)\equiv U_H\equiv u\,r_H$ can be arbitrary. 
 Therefore, all possible boundary conditions at the horizon are labeled by just one  free parameter $u$, 
 and choosing some value for it, the equations can be integrated directly from the horizon, as explained in Appendix \ref{Des},
 to the outer region $r>r_H$. 
 
 Far from the horizon, at $r=r_\star\gg r_H$, one has 
\bea                   \label{ZZZ1}
N(r_\star)&=&1+Z_0(r_\star)+\kappa_1\,B\,\frac{1+r_\star}{r_\star}\,e^{-r_\star}+\kappa_1\,\frac{1-r_\star}{r_\star}\,Z_{-}(r_\star),\, \nonumber  \\
Y(r_\star)&=&1+Z_0(r_\star)-\kappa_2\,B\,\frac{1+r_\star}{r_\star}\,e^{-r_\star}-\kappa_2\,\frac{1-r_\star}{r_\star}\,Z_{-}(r_\star),\,  \nonumber \\
U(r_\star)&=&r_\star+B\,\frac{1+r_\star+r_\star^2}{r_\star^2}\,e^{-r_\star}+\frac{1-r_\star+r_\star^2}{r^2_\star}\,Z_{-}(r_\star)\, ,
\eea
where $Z_0(r_\star)$ and $Z_{-}(r_\star)$ are functions of $A,B$ determined by \eqref{0a} via 
iterating the integral equations \eqref{integral}.  

As a result, we have the boundary conditions  at $r=r_H$ labeled by $u$ 
and the boundary conditions at $r=r_\star$  labeled by $A,B$. We use them 
to construct solutions in the region $r_H\leq r\leq r_\star$. To this end, we choose some value of $u$ and 
integrate numerically the equations starting from $r=r_H$ as far as 
some $r=r_0<r_\star$ and we obtain at this point  some  values which will depend on $r_H$ and $u$:
\be
N(r_0)\equiv N_{\rm hor}(r_H,u),~~~~~~~Y(r_0)\equiv Y_{\rm hor}(r_H,u),~~~~~~U(r_0)\equiv U_{\rm hor}(r_H,u). 
\ee
Then we choose $A,B$ and numerically extend the large $r$ data \eqref{ZZZ1} 
from $r=r_\star$ down to $r=r_0$,
thereby obtaining 
\be
N(r_0)\equiv N_{\rm inf}(A,B),~~~~~~~Y(r_0)\equiv Y_{\rm inf}(A,B),~~~~~~U(r_0)\equiv U_{\rm inf}(A,B). 
\ee
If the two sets of values agree, hence if 
\bea     \label{match}
\Delta N(r_H,u,A,B)&\equiv& N_{\rm hor}(r_H,u)-N_{\rm inf}(A,B)=0, ~~\nonumber \\
\Delta Y(r_H,u,A,B)&\equiv& Y_{\rm hor}(r_H,u)-Y_{\rm inf}(A,B)=0, ~~\nonumber \\
\Delta U(r_H,u,A,B)&\equiv& U_{\rm hor}(r_H,u)-U_{\rm inf}(A,B)=0, 
\eea
then the solution in the interval $r\in [r_H,r_0]$ 
merges smoothly with the solution in the interval $r\in [r_0,r_\star]$ to 
represent one single solution in the interval $r\in [r_H,r_\star]$. 
The extension  to the region $r>r_\star$ is then  provided by  the integral equations \eqref{integral}, finally yielding 
an asymptotically flat black hole solution in the region  $r\in [r_H,\infty)$. It is worth noting that these solutions 
will depend neither on $r_0$ nor on $r_\star$;  these values could be varied without affecting  the global solution
(which is a good consistency check). 

In some cases using just two zones $[r_H,r_0]$ and $[r_0,r_\star]$ produces too large numerical errors. 
To keep the numerical instability under control, 
one should  then integrate through many smaller zones  $[r_H,r_0]$, $[r_0,r_1],[r_1,r_2]\ldots [r_k,r_\star]$
and perform matchings at $r_0,r_1,\ldots r_k$ (see Sec.7.3.5 in \cite{stoer}). 
This yields numerically stable results. 

In the case of  just two zones, the problem reduces to solving the matching conditions \eqref{match} by adjusting the values 
$u,A,B$.  At least one solution to these three conditions certainly exists and corresponds to the 
Schwarzschild solution, for which 
\be                      \label{Sw}
u=1,~~~~~ A=-\frac{r_H}{2},~~~~~ B=0.
\ee
 Are there other solutions ? 
Since there are three matching conditions for the three variables, their solutions must 
constitute a {\it discrete} set of points $(u_k,A_k,B_k)$  in the 3-space spanned by  $u,A,B$. This implies 
 that different black hole solutions with the same $r_H$ are {\it parametrically isolated} from each other. This
creates a problem, since in order to solve numerically algebraic equations \eqref{match}, an input  configuration ${u},{A},{B}$ is needed
to start the numerical iterations within the Newton-Raphson procedure
 \cite{Press:2007:NRE:1403886}. 
 However, unless the input configuration 
is close to the solution, the numerical iterations do not converge, hence 
some additional information is necessary  to specify where to start the iterations. 
 
 As explained in the Introduction, the additional information is provided by the 
 stability analysis of the Schwarzschild solution \eqref{b2}  \cite{Babichev:2013una,Brito:2013wya}. In this analysis one 
 considers the two metrics of the form \eqref{ansatz000} with 
 \bea
 Q&=&{S}+\delta Q,~~~N={S}+\delta N,~~~R=r\nonumber \\
 q&=&{S}+\delta q,~~~Y={S}+\delta Y,~~~U=r+\delta U,~~~~~f_{01}=\delta \alpha, 
 \eea
 where $S=\sqrt{1-\frac{r_H}{r}}$ while the perturbations $\delta Q,\delta N,\delta q,\delta Y,\delta U,\delta\alpha$ are assumed
 to be small and depend on $t,r$.  It turns out that at the GL point, for  $r_H=0.86$, 
 the perturbation equations admit a 
 {\it static} solution (zero mode) 
 for which   $\delta Q,\delta N,\delta q,\delta Y,\delta U$ depend only on $r$ and are bounded everywhere in the region   $r\geq r_H$
 while $\delta\alpha=0$. 
 This  solution can be viewed as a perturbative approximation of a new solution that merges 
 with the Schwarzschild solution for  $r_H=0.86$.

This suggests that  to get new solutions of the matching conditions 
 \eqref{match}, one should choose the event horizon size to be close 
 $r_H=0.86$ and choose the input configuration ${u},{A},{B}$ to be close to \eqref{Sw}. 
 Then the numerical iterations should converge to values $u,A,B$ which are slightly different from  \eqref{Sw}
 and correspond to an almost Schwarzschild black hole slightly distorted by a massive hair. 
 Changing  then iteratively  the value of $r_H$ yields solutions which deviate considerably from the Schwarzschild 
 metric close to the horizon, but always approach flat metric in the asymptotic region. 

\begin{figure}
    \centering
    \includegraphics[scale=0.65]{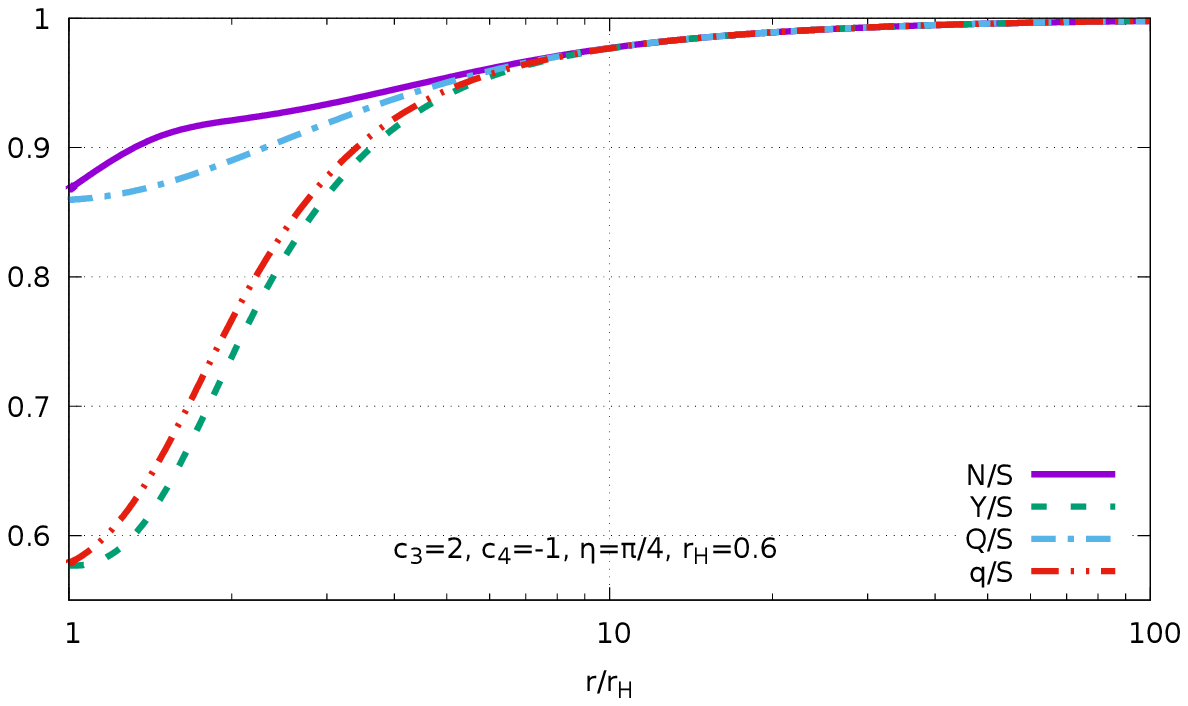}
    \includegraphics[scale=0.65]{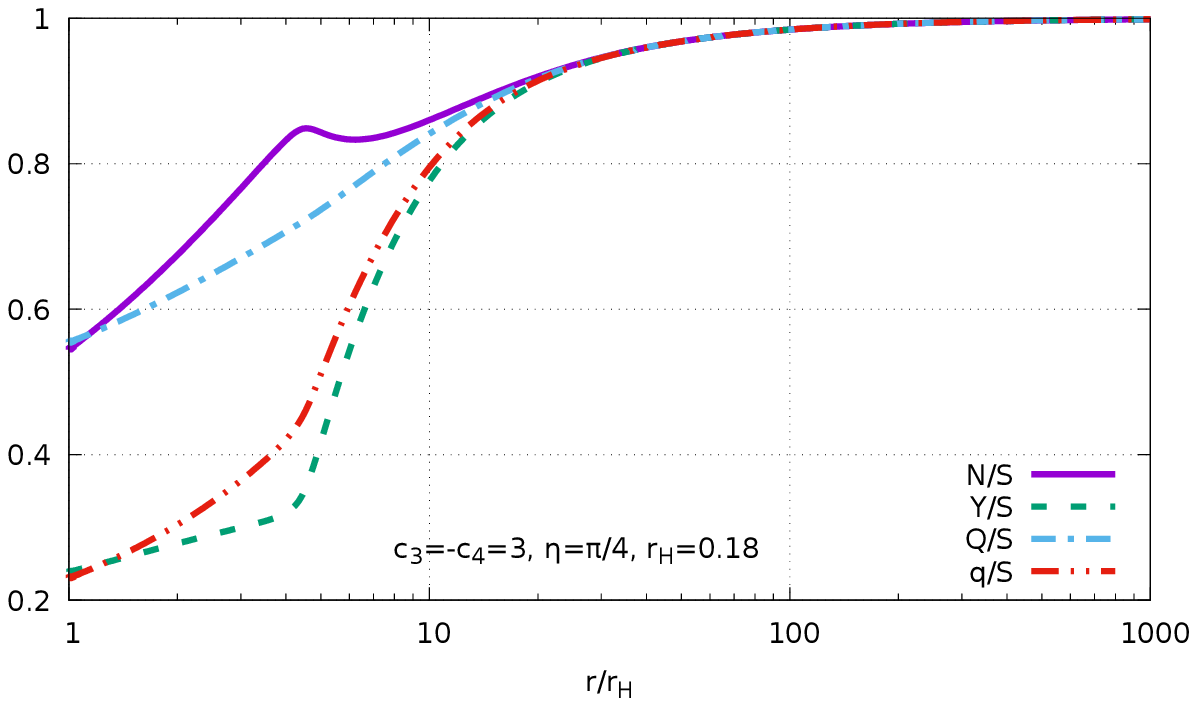}
        \centering
    \includegraphics[scale=0.65]{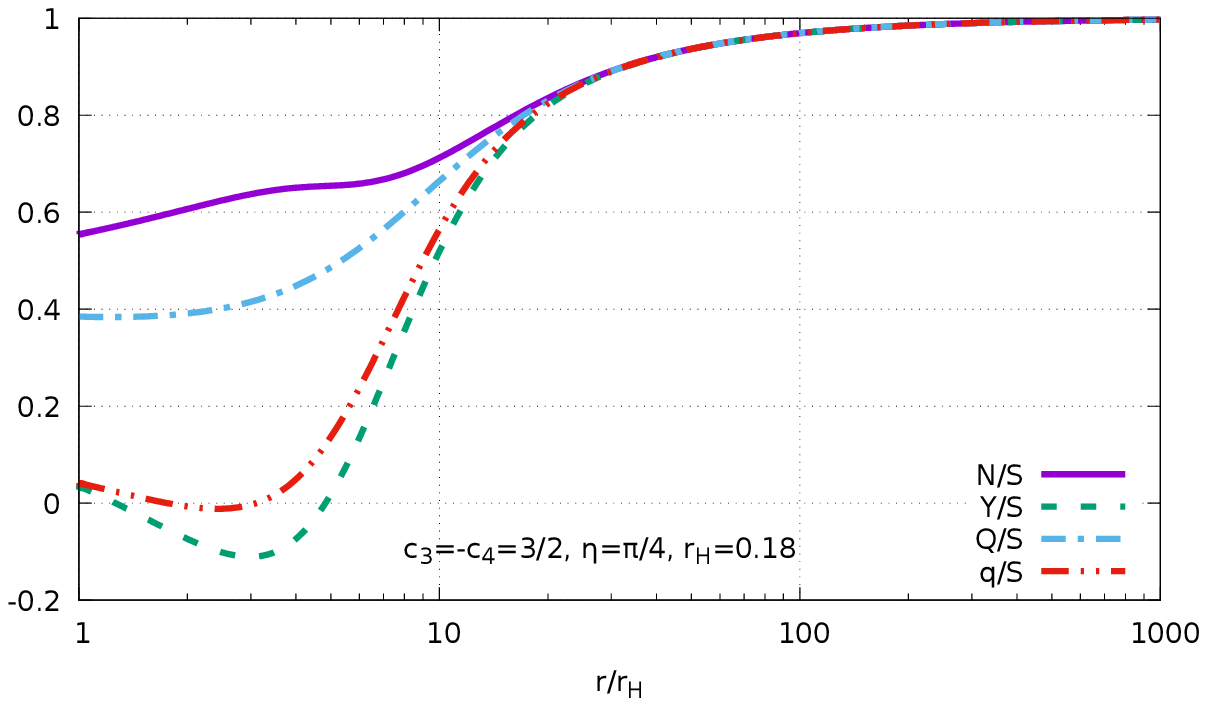}
    \includegraphics[scale=0.65]{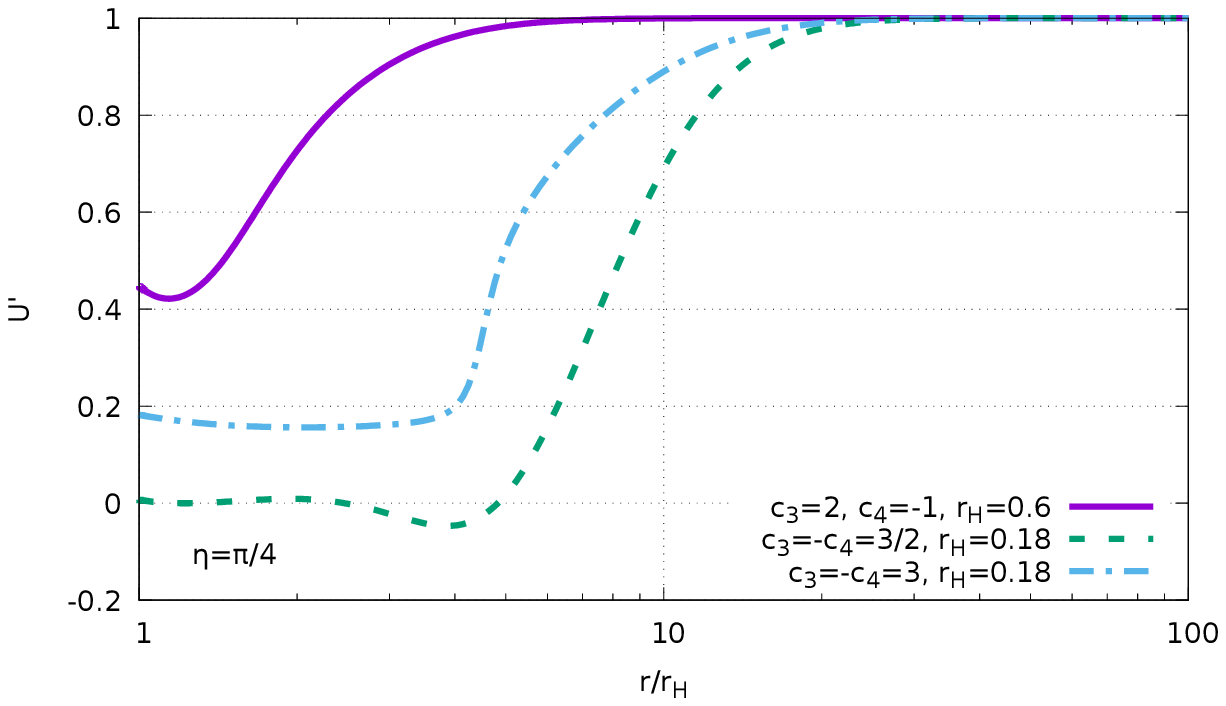}   
    \caption{Profiles of  $N/S$, $Y/S$, $Q/S$, $q/S$ with $S=\sqrt{1-r_H/r}$ and that of $U^\prime$ 
    for solutions with $\eta=\pi/4$  but  for various values of $r_H,c_3,c_4$. 
   Solution with $c_3=-c_4=3/2$  shown on the two lower panels is singular because the amplitudes $q,Y,U^\prime$ develop  zeros outside the horizon. 
      }
    \label{Fig1}
\end{figure}

  \section{ASYMPTOTICALLY FLAT HAIRY BLACK HOLES  \label{NR}}
  
  \setcounter{equation}{0}
 
 Applying the procedure outlined above, we were able to construct asymptotically flat hairy solutions. We confirm 
 the results of Ref. \cite{Brito:2013xaa} and obtain many new results.

First of all, we  find that for $r_H$ approaching from below the GL value, $r_H\approx 0.86$,  there are asymptotically flat hairy black hole
solutions for any $c_3,c_4,\eta$. They are very close to the
Schwarzschild solution: one has  $u=U_H/r_H\approx 1$ and the ADM mass $M\approx r_H/2$. 
However, for smaller values of $r_H$ the solutions deviate more and more from Schwarzschild.
To illustrate this, we plot in Fig.~\ref{Fig1} 
the functions $N/S$, $Q/S$, $Y/S$, $q/S$, and $U^\prime$. If these 
functions all equal to one, then the solution is Schwarzschild. As one can see, they indeed approach unity  far away 
from the horizon, but close to the horizon  they deviate considerably from unity, 
hence  the massive graviton hair is concentrated in this region.

\begin{figure}
     
       \centering
    \includegraphics[scale=0.65]{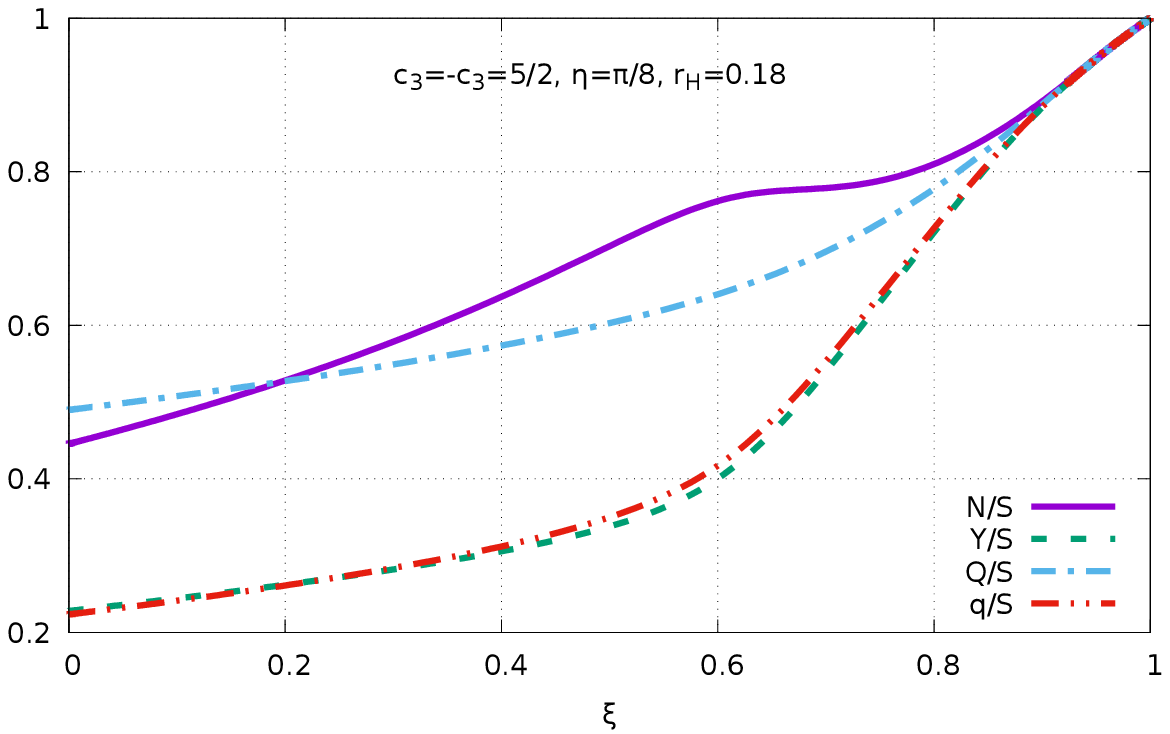}
    \includegraphics[scale=0.65]{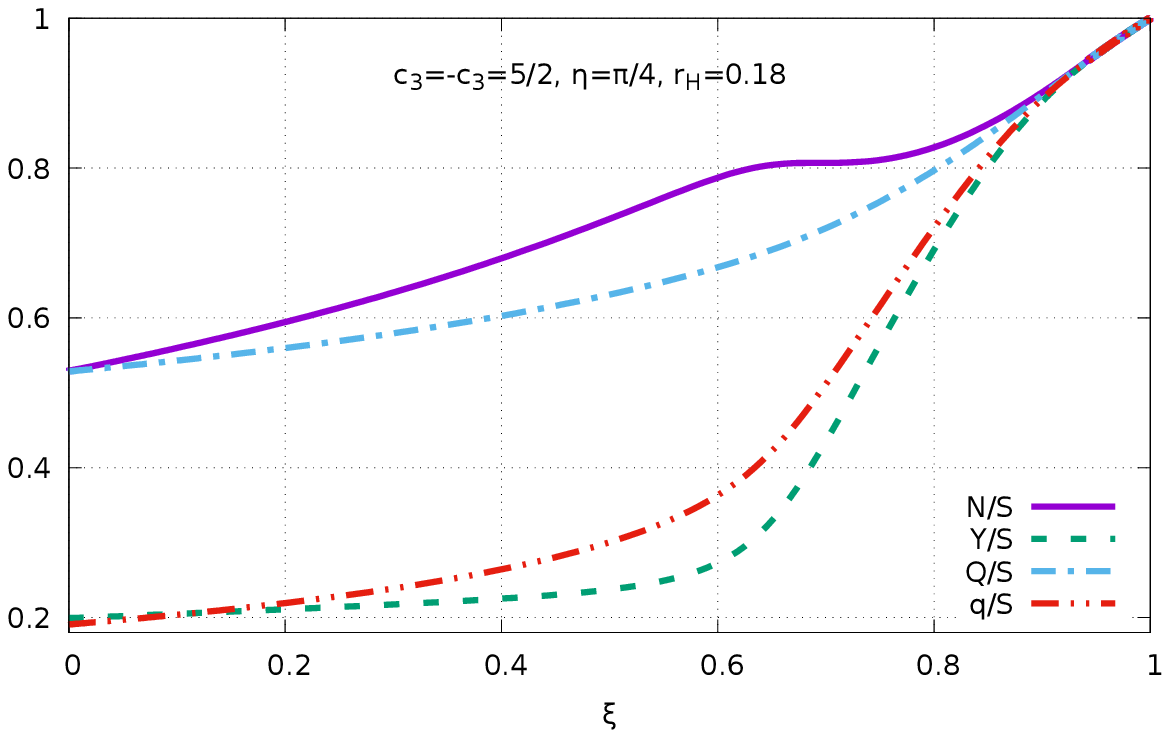}
     
         \centering
         
     \includegraphics[scale=0.65]{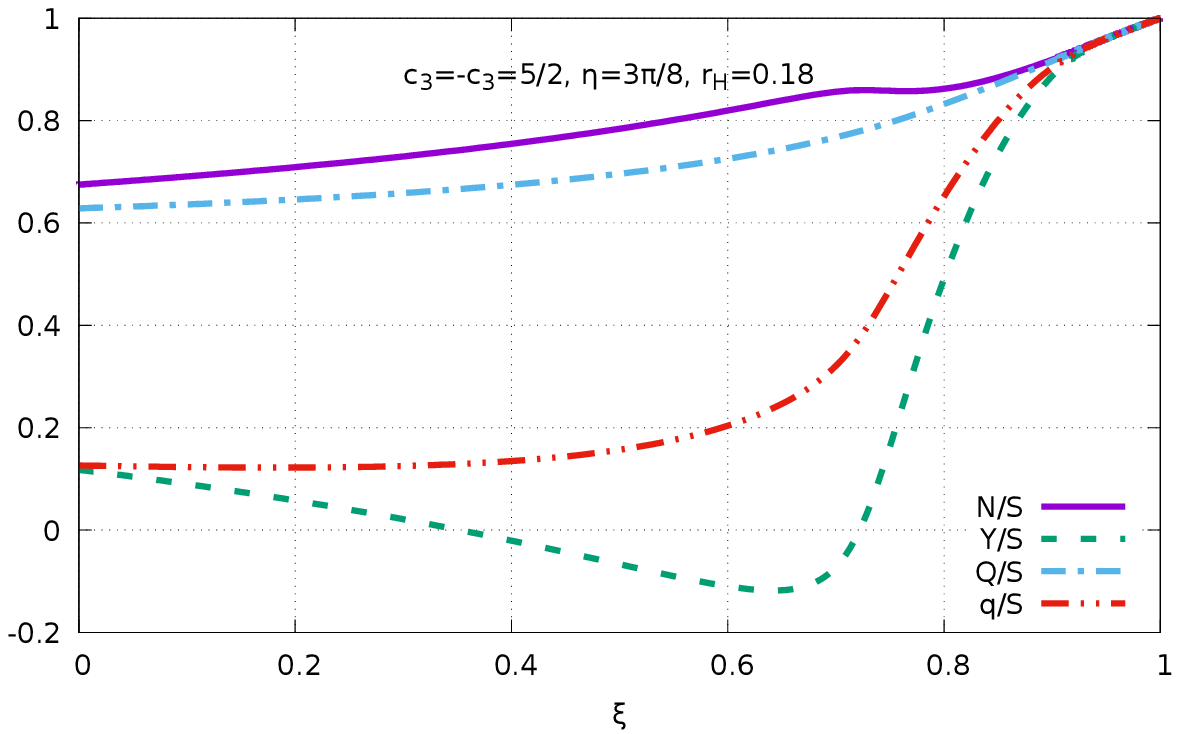}
    \includegraphics[scale=0.65]{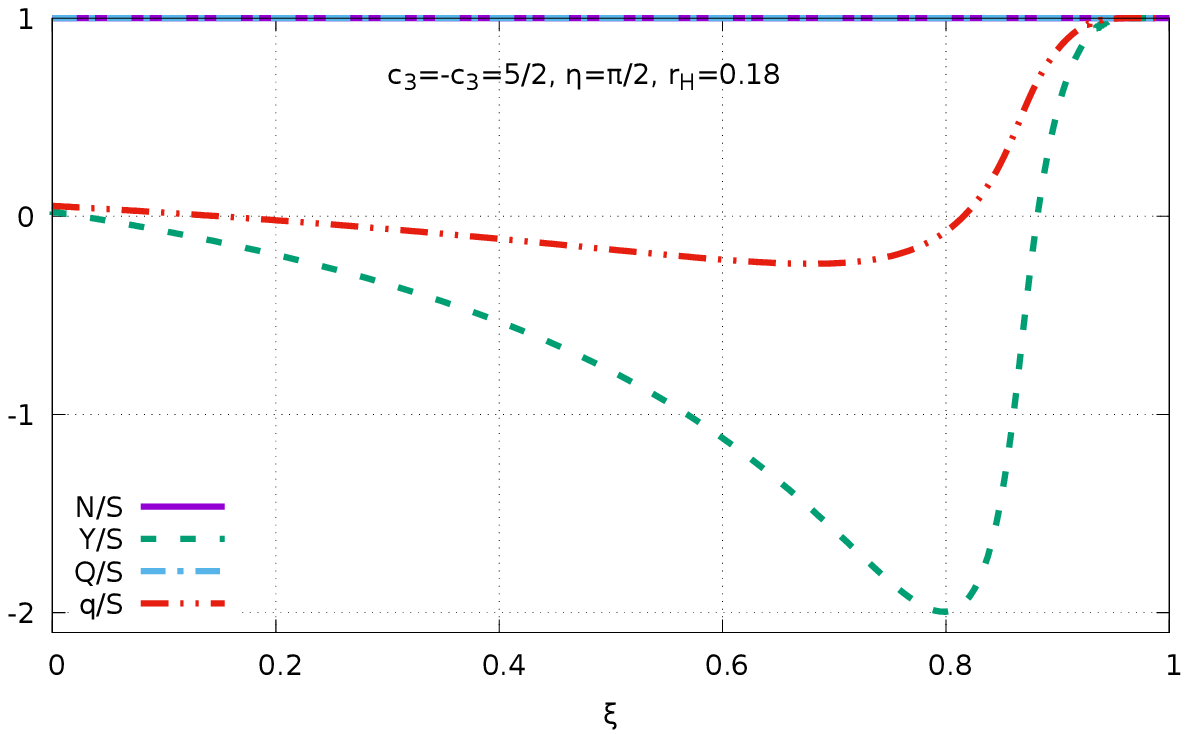}

    \caption{Profiles of  $N/S$, $Y/S$, $Q/S$, $q/S$ with $S=\sqrt{1-r_H/r}$ \bl{against $\xi=(r-r_H)/(r+r_H)$} 
    for  solutions with the same $c_3,c_4,r_H$ but for different values of $\eta$. 
 When $\eta$ approaches $\pi/2$ then  the amplitudes
 $Y,q$ start showing zeroes. For $\eta=\pi/2$ 
 the g metric is Schwarzschild with $N/S=Q/S=1$. 
      }
    \label{Fig-eta}
\end{figure}

Solutions are regular for $r_H$ close to $0.86$; however, for smaller $r_H$ and depending on values of $c_3,c_4,\eta$, 
the amplitudes $Y,q,U^\prime$ may show additional zeros outside the horizon, 
whereas $Q,N$ always remain positive. 
This implies that the f metric is singular, because the invariants of its Riemann tensor diverge where the zeros are located. 
\bbl{An example of this is shown  on the lower  two  panels in Fig.~\ref{Fig1}, and also on the lower  two  panels 
in Fig.~\ref{Fig-eta} where one can see  that the phenomenon occurs 
 when  $\eta$ approaches $\pi/2$. The fact that the f metric becomes singular does not invalidate the solutions because the f-geometry is not 
 directly measurable and its singularities are not seen, while the g metric, which can be probed by test particles, remains always regular. 
 We shall therefore keep  such solutions in our consideration. }

\bl{Solutions  in Fig.~\ref{Fig1} are shown up to large but still finite values  of the radial coordinate, $r/r_H\leq 100$ or $r/r_H\leq 1000$. 
What is shown is the combination 
of the solutions of  differential equations \eqref{eqs} in the region $r_H\leq r\leq r_\star$ and of 
 the integral equations \eqref{integral} for $r>r_\star$ where $r_\star/r_H=25$. At the same time, our procedure yields  
 solutions in the whole region  $r\in [r_H,\infty)$. 
Introducing the compactified radial variable
\be                  \label{comp}
\xi=\frac{r-r_H}{r+r_H}\in [0,1], 
\ee
we plot  in Fig.~\ref{Fig-eta}  the  amplitudes $N,Y,Q,q$ against $\xi$. 
As seen, the amplitudes approach unity as $\xi\to 1$ (same is true for $U^\prime$) hence the solutions are indeed asymptotically flat. 
The disadvantage of this parametrization  is that the slope of the functions does not vanish for $\xi\to 1$. Indeed, for large $r$ one has 
$N=1-M/r+\ldots$ and $\xi=1-r_H/r+\ldots$ hence at infinity $dN/d\xi=M/r_H$.  }

\begin{figure}
    \centering
    \includegraphics[scale=0.65]{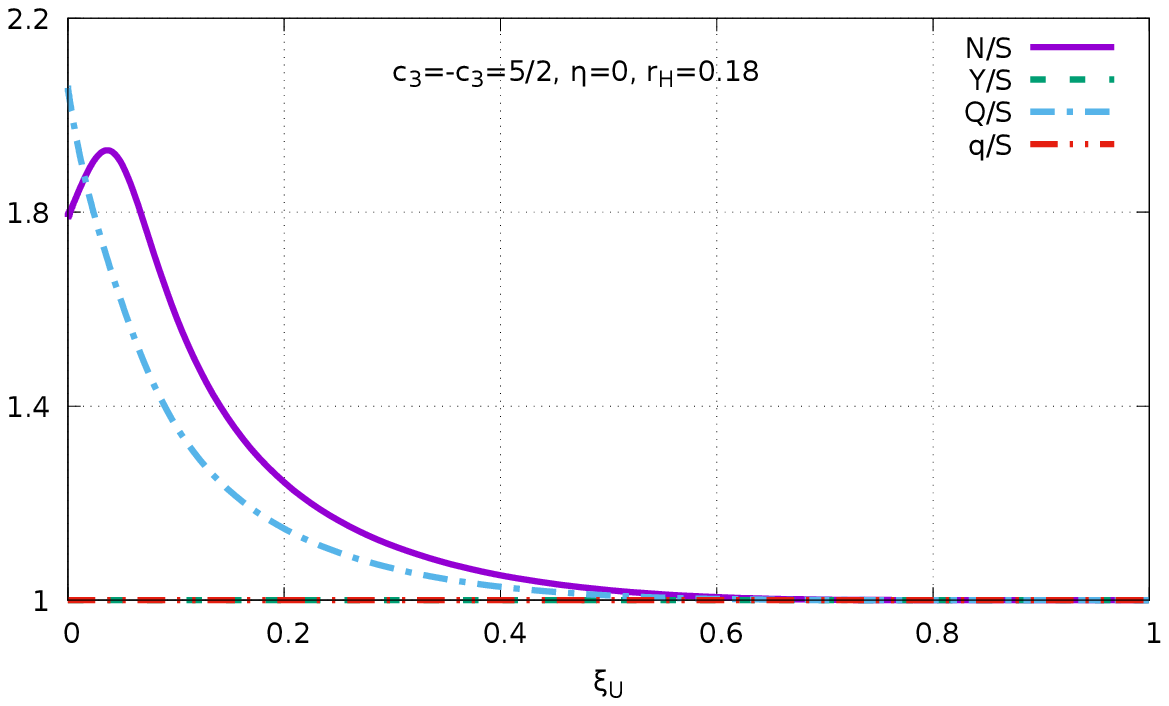}
      \includegraphics[scale=0.65]{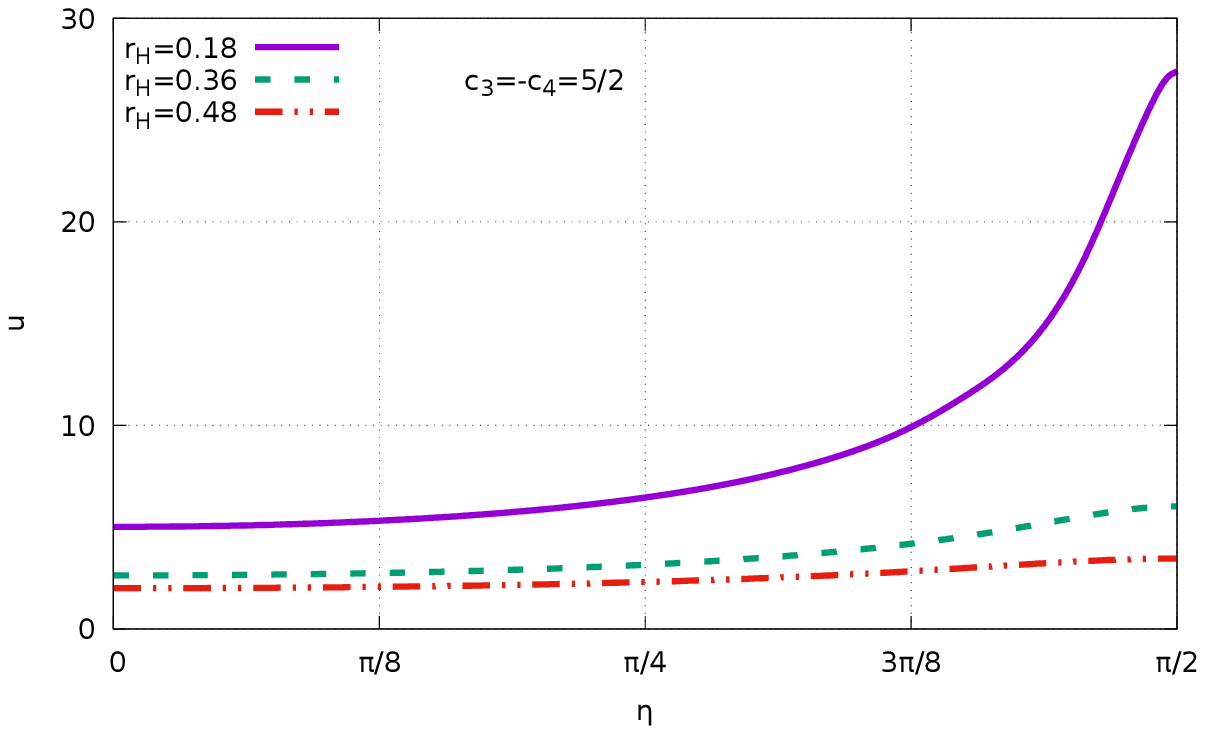}

         \centering
    \includegraphics[scale=0.65]{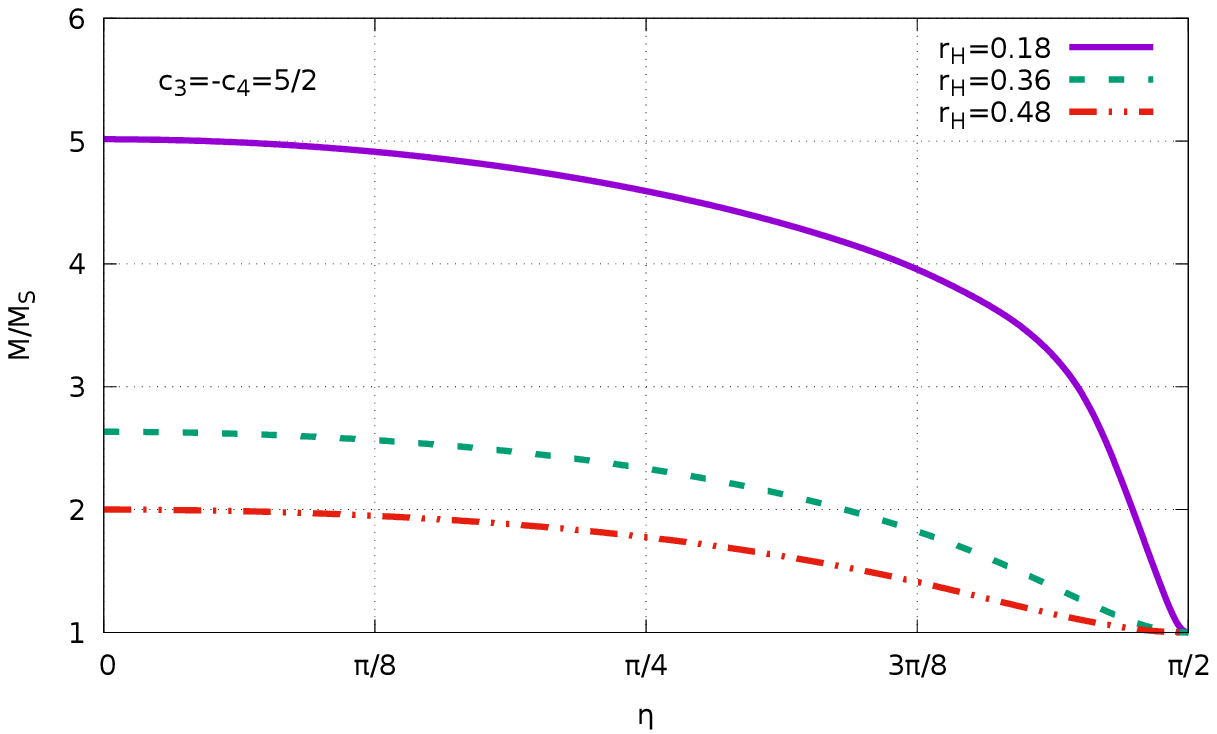}
        \includegraphics[scale=0.65]{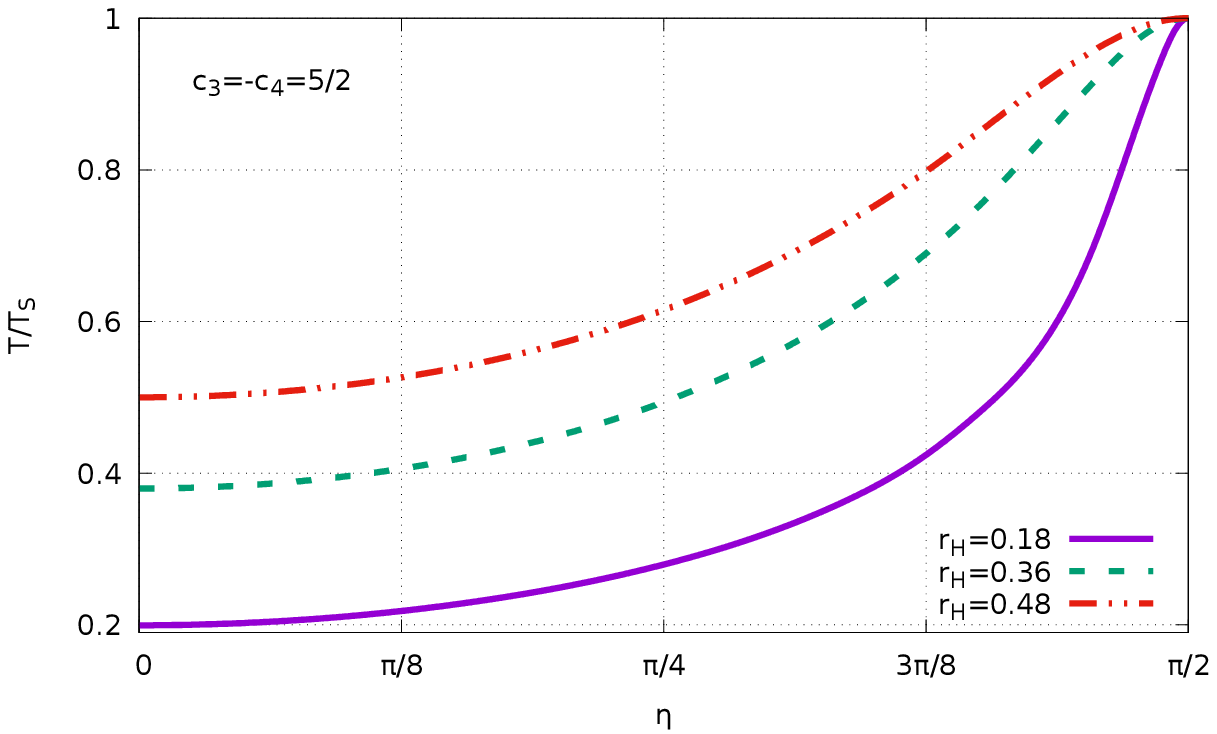}

       \caption{Upper left:   $N/S$, $Y/S$, $Q/S$, $q/S$ with $S=\sqrt{1-U_H/U}$   
       against the compact variable $\xi_U=(U-U_H)/(U+U_H)$ for  $\eta=0$. One has $Y/S=q/S=1$  hence 
       the f metric is Schwarzschild. The other three panels show $u=U_H/r_H$, 
       the ADM mass $M$, and the temperature $T$ against $\eta$.}
    \label{Fig-eta0}
\end{figure}

If $\eta=\pi/2$ then $\kappa_1=0$  
and the g metric becomes  Schwarzschild. The theory reduces then to the massive gravity for the 
dynamical f metric on a fixed Schwarzschild background. 
The solution for the f metric is shown on the lower right panel in Fig.~\ref{Fig-eta}. Similarly, for $\eta=0$ 
one has $\kappa_2=0$ and the f metric is Schwarzschild, while the g metric is a solution of the 
massive gravity on the Schwarzschild 
background shown on the  upper left panel in Fig.~\ref{Fig-eta0}.  One should emphasize that the radii  of the 
background Schwarzschild  black  holes for $\eta=0$ and for $\eta=\pi/2$ are not the same. 
For example, for $\eta=\pi/2$ the  Schwarzschild black hole has 
$r_H=0.18$ for the solution shown in Fig.~\ref{Fig-eta}, while for $\eta=0$ the event horizon size is determined
not by $r_H$ but 
by $U_H=ur_H$ where $u\approx 5$ (as seen in Fig.\ref{Fig-eta0})  hence this time the  Schwarzschild black hole
 is much larger.  As a result, solutions on these different backgrounds
look quite different -- the solution for the f metric  on the lower right panel in Fig.~\ref{Fig-eta} shows zeros hence it is singular, 
while the solution for the g metric on the upper left panel in  Fig.~\ref{Fig-eta0} is regular. 

\bbl{Solutions for $\eta=\pi/2$ will play an important role below. We shall call them 
``hairy Schwarzschild" because  their g metric is Schwarzschild but their f metric supports 
hair.}

Figure \ref{Fig-eta0} shows the $\eta$ dependence of  $u=U_H/r_H$ and of  the ADM mass $M$ expressed in units of the 
Schwarzschild mass $M_S=r_H/2$, as well as the temperature $T$ expressed in units of the Schwarzschild temperature $T_S=1/(4\pi r_H)$. 
As one can see, the dependence is rather strong for small $r_H$,
in particular for $u$. The decrease of the mass $M$ with $\eta$ can be understood by noting that  the mass is the same with respect to 
each metric (the same is true for the temperature). If $\eta=\pi/2$ then the g metric is Schwarzschild hence $M=M_S$
and $T=T_S$. 
If $\eta=0$ then the f metric is Schwarzschild with a larger radius $U_H=u r_H$, hence  the mass is larger, 
$M=U_H/2=u M_S$, while the temperature is smaller,  $T=T_S/u$. Therefore, if $\eta=0$ then 
$M/M_S=u$  so that, for example, $u\approx 5$ for $r_H=0.18$, as seen in Fig.~\ref{Fig-eta0}. 

\begin{figure}
    
     \centering   
      \includegraphics[scale=0.65]{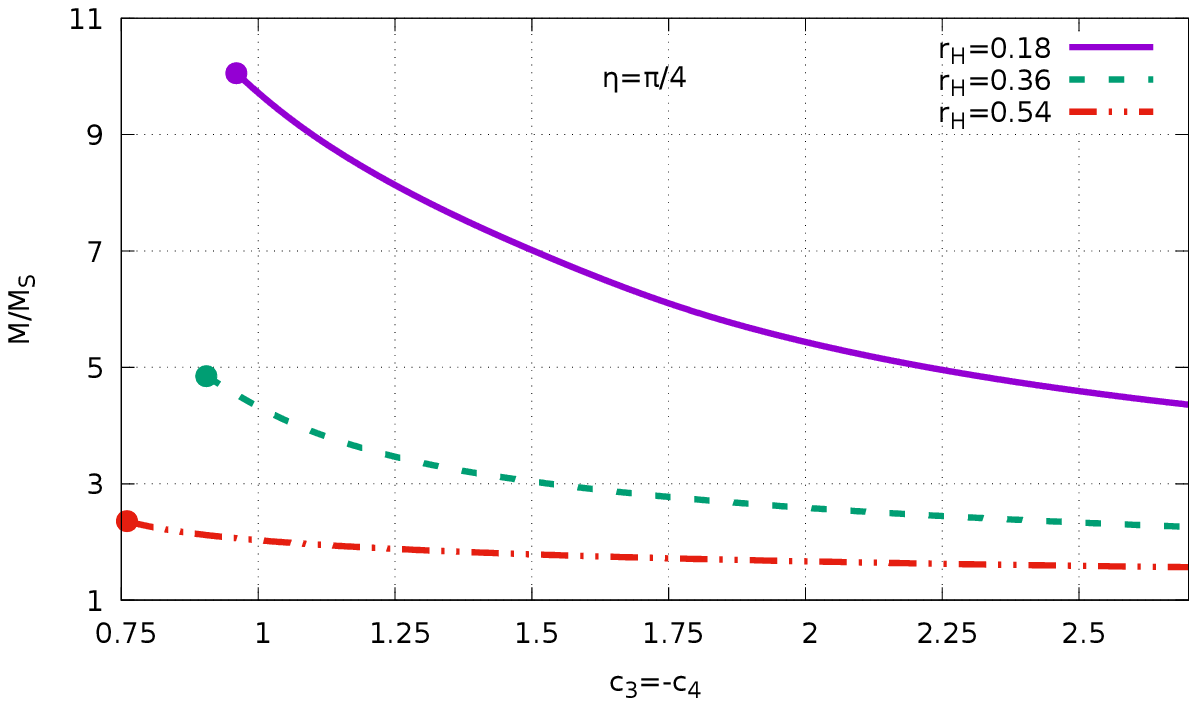}
    \includegraphics[scale=0.65]{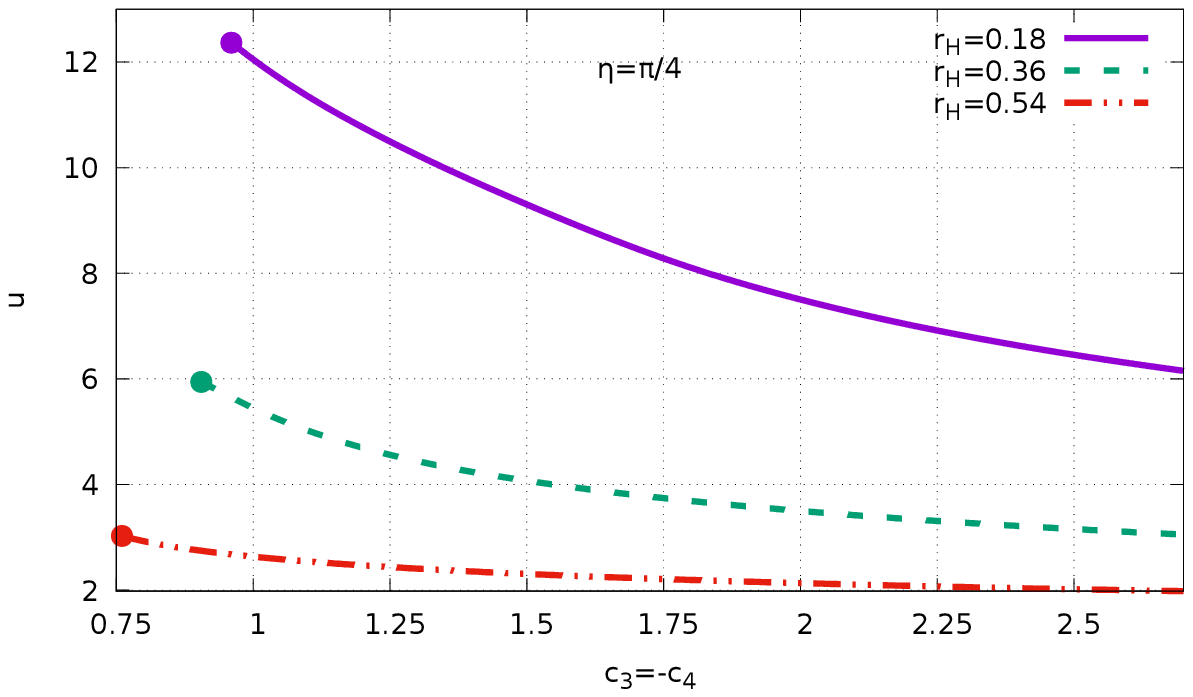}
    
   \caption{The $M/M_S$ (left) and $u=U_H/r_H$  (right) against $c_3=-c_4$. }   \label{Fig4}
\end{figure}

 Figure \ref{Fig4} shows the dependence of $u$ and $M$ on $c_3$ in the case where $c_3=-c_4$. One can see that the solutions 
 exist if only the value of  $c_3=-c_4$ is not too small. 
Similarly, not all hairy black holes exist for however small values of $r_H$. 
As was noticed in \cite{Brito:2013xaa}, small $r_H$ 
black holes exist if the coefficient $b_3$ in the potential \eqref{2} vanishes so that  the 
cubic part of the potential is absent. In view of \eqref{bbb}, this requires that 
$c_3=-c_4$, but this is not the only condition. Depending on the parameter values, one can 
distinguish  the  following two cases: 
\bea           \label{typeI}
\mbox{I}: ~~~~ c_3\neq -c_4~~\mbox{or}~~c_3= -c_4< 1, ~~~~~~~
\mbox{II}: ~~~c_3= -c_4\geq 1.
\eea
In case I asymptotically flat hairy black holes exist only if $0<r_H^{\rm min}\leq r_H<0.86$ hence  they cannot be 
arbitrarily small. In  case II
they exist for any  $0<r_H<0.86$, although their f metric may be singular for small $r_H$.
We shall see below in Sec. \ref{par}  what happens when $r_H$ approaches 
the lower bound.

\subsection{Duality relation}

The results described above in this section 
essentially reproduce those of Ref.~\cite{Brito:2013xaa}, the only important difference being  that we show 
solutions for different values of $\eta$, whereas Ref.~\cite{Brito:2013xaa} shows them only for $\eta=\pi/4$. However,
starting from this moment and in the following two Sections we  shall be describing new results. 

Reference \cite{Brito:2013xaa} finds  solutions only below the GL point, for  $r_H\leq 0.86$. At the same time, the 
consistency of the procedure requires 
that there should be asymptotically flat hairy black holes  also for $r_H>0.86$. This 
follows from the symmetry \eqref{change} of the equations, which now reads 
\be               \label{dual}
\eta\to\frac{\pi}{2}-\eta,~~~Q\leftrightarrow q,~~~N\leftrightarrow Y,~~~U\leftrightarrow r,
~~~c_3\to 3-c_3,~~~c_4\to 4c_3+c_4-6. 
\ee
More precisely, this means that if  for some  values of $\eta,c_3,c_4$ there is a solution 
\be             \label{QQ1}
Q(r), q(r), N(r), Y(r), U(r), 
\ee
then for $\tilde{\eta}=\pi/2-\eta$, $\tilde{c}_3=3-c_3$, $\tilde{c}_4= 4c_3+c_4-6$
there should be  the  ``dual" solution described by 
\be            \label{Q2}
\tilde{Q}(r)=q(w(r)),~~~\tilde{q}(r)=Q(w(r)),~~~\tilde{N}(r)=Y(w(r)),~~~\tilde{Y}(r)=N(w(r)),~~~\tilde{U}(r)=w(r),
\ee
where  $w(r)$ is the function inverse for $U(r)$, such that $U(w(r))=r$. 
This duality correspondence relates between themselves black holes of different size, because \eqref{QQ1}
has the horizon at $r=r_H$ while  the horizon of \eqref{Q2} is located  where $w(r)=r_H$, that is at $r=\tilde{r}_H=U(r_H)$. 
One has
\be
\tilde{u}=\frac{\tilde{U}(\tilde{r}_H)}{\tilde{r}_H}=\frac{r_H}{U(r_H)}=\frac{1}{u}. 
\ee
Now, for hairy solutions with $r_H<0.86$ one always has $U(r_H)>0.86$ and 
$u=U(r_H)/r_H>1$. 
It follows that their duals are characterized by $\tilde{r}_H>0.86$ and by $\tilde{u}_H<1$. 

\begin{figure}
    \centering
    \includegraphics[scale=0.65]{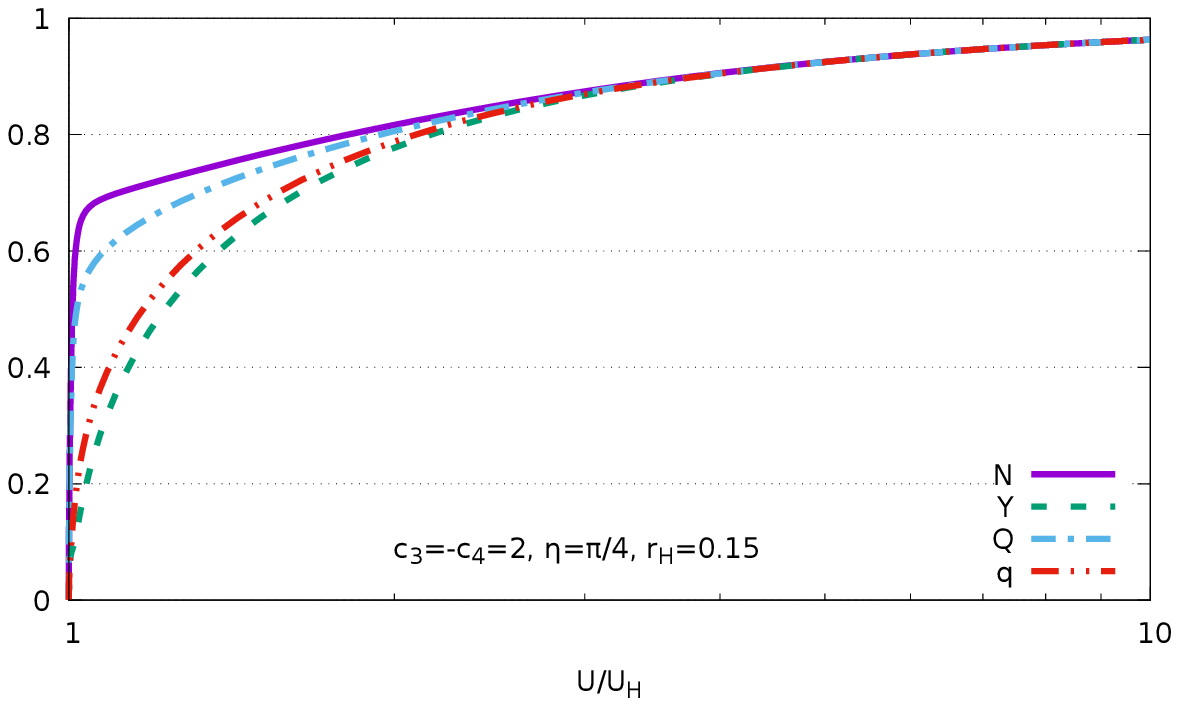}
    \includegraphics[scale=0.65]{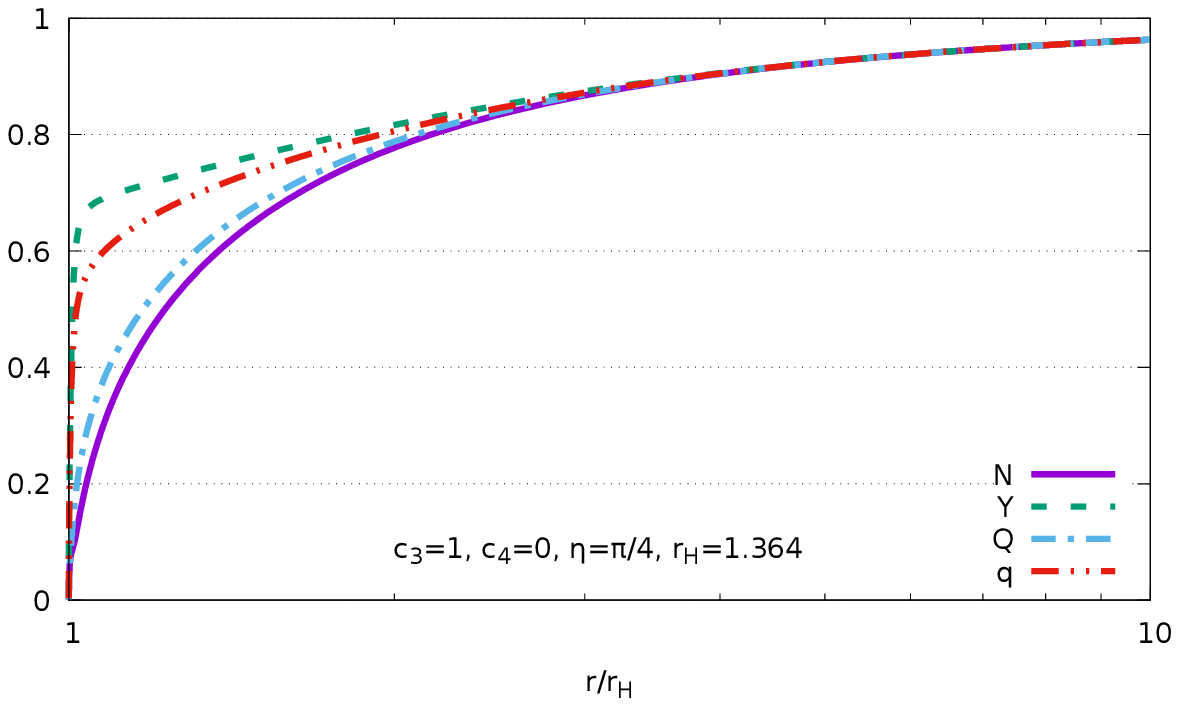}
   \caption{The solution with $c_3=-c_4=2$, $\eta=\pi/4$, $r_H=0.15$ (left) 
   and the dual solution with $c_3=1$, $c_4=0$, $\eta=\pi/4$, $r_H=1.364$ (right). 
   The curves on the two panels are 
   exactly the same, up to the interchange  $N\leftrightarrow Y$, $Q\leftrightarrow q$,  $r/r_H\leftrightarrow U/U_H$. }
    \label{Fig2}
\end{figure}

An explicit example of the duality relation is shown in Fig.~\ref{Fig2}, which presents on the left panel the solution for 
$c_3=-c_4=2$, $\eta=\pi/4$, $r_H=0.15$ for which $U(r_H)=1.364$, hence $u=1.364/0.15=2.42$. 
The duality implies that for $c_3=1$, $c_4=0$, $\eta=\pi/4$ there must be the dual solution with $r_H=1.364$ and 
$u=0.15/1.364=0.41$, which is indeed confirmed by our numerics. 
Plotting the first solution against $U/U_H$ and the second one against $r/r_H$, as shown in Fig. \ref{Fig2},
yields exactly the same curves, up  to the interchange $N\leftrightarrow Y$, $Q\leftrightarrow q$. 

It is unclear why solutions with $r_H>0.86$ were not found in \cite{Brito:2013xaa}. 

The duality is in fact a powerful tool for studying the solutions, because sometimes their properties may look puzzling
in one description  but become obvious within the dual description.

\section{STABILITY ANALYSIS\label{pert}}
\setcounter{equation}{0}

In this section we analyze  the stability of the hairy solutions by studying their perturbations within the ansatz 
described in Appendix \ref{time}, 
\begin{align*}  \label{anz0}
    ds^2_g&=-Q^2 dt^2+\frac{dr^2}{N^2}+r^2 d\Omega^2,\\ 
    ds^2_f&=-\big(q^2-\alpha^2Q^2N^2\big)dt^2-2\alpha\bigg(q+\frac{QNU^\prime}{Y}\bigg)dtdr+\bigg(\frac{U^{\prime 2}}{Y^2}-\alpha^2\bigg)dr^2+U^2 d\Omega^2,\numberthis{}
\end{align*}
where $Q$, $q$, $N$, $Y$, $\alpha$, $U$ are functions of $r$ and $t$. 
The full set of the field equations in this case is shown in  Appendix \ref{time}. 
If we set  $\alpha=0$ and assume  that nothing depends on time,
then we return back to  the static case studied above. Therefore, small deviations from the static solutions are described by \eqref{anz0} 
with 
\begin{align*}                      \label{ptb}
    Q(r,t)&=\accentset{(0)}Q(r)+\delta Q(r,t),\\
    q(r,t)&=\accentset{(0)}q(r)+\delta q(r,t),\\
    N(r,t)&=\accentset{(0)}N(r)+\delta N(r,t),\numberthis{}\\
    Y(r,t)&=\accentset{(0)}Y(r)+\delta Y(r,t),\\
    U(r,t)&=\accentset{(0)}U(r)+\delta U(r,t),\\
    \alpha(r,t)&=\delta\alpha(r,t),
\end{align*}
where the functions $\accentset{(0)}Q(r)$, $\accentset{(0)}q(r)$, $\accentset{(0)}N(r)$, $\accentset{(0)}Y(r)$, $\accentset{(0)}U(r)$ correspond to the background black hole solution while the perturbations $\delta Q, \delta q, \delta N, \delta Y, \delta U, \delta\alpha $ are small. 

We therefore inject \eqref{ptb} to Eqs.\eqref{Ein00} and \eqref{cons00} and linearize with respect to the 
perturbations. Linearizing  the  $G^0_{~1}(g)=\kappa_1 T^0_{~1}$ equation yields 
\be           \label{XXX}
\frac{2}{rNQ^2}\,\delta \dot{N}=\kappa_1 \frac{{\cal P}_1}{Q}\,\delta{\alpha}, 
\ee
where $N,Q,{\cal P}_1$ relate to the static background, and we do not write their over sign ``$(0)$'' for simplicity. 
\bbl{In the linear  perturbation theory one can  consistently separate the time variable by assuming 
the harmonic time dependence for all amplitudes (this would no longer be possible if non-inear corrections are taken into account), 
so that we choose}
\be
\delta N(t,r)=e^{i\omega t}\, \delta N(r)~~~~~\delta \alpha(t,r)=e^{i\omega t}\, \delta \alpha(r), 
\ee 
and similarly for $\delta Y,\delta Q,\delta q,\delta U$. 
Injecting to \eqref{XXX} yields the algebraic relation 
\be
\delta \alpha(r)=\frac{2i\omega}{rNQ{\cal P}_1}\,\delta N(r). 
\ee
Linearizing similarly the $G^0_{~1}(f)=\kappa_2 {\cal T}^0_{~1}$ equation yields a  linear relation 
between $\delta\alpha(r)$, $\delta Y(r)$, $\delta U(r)$.  Using these two algebraic relations one  finds  that  
the three equations $G^0_{~0}(g)=\kappa_1 T^0_{~0}$, $G^0_{~0}(f)=\kappa_2 {\cal T}^0_{~0}$ and  
$ \stackrel{(g)}{\nabla}_\mu T^\mu_{~0}=0$ yield upon the linearization three equivalent to each other relations. 
Therefore, among the 8 equations \eqref{Ein00}, \eqref{cons00} only 6 are independent (at least at the linearized level). 

Taking all of this into account and linearizing  similarly the remaining 3 equations 
$G^1_{~1}(g)=\kappa_1 T^1_{~1}$, $G^1_{~1}(f)=\kappa_2 {\cal T}^1_{~1}$ and $ \stackrel{(g)}{\nabla}_\mu T^\mu_{~1}=0$, 
one finds  that all 6 perturbation amplitudes 
$\delta Q(r)$, $\delta q(r)$, $\delta N(r)$, $\delta Y(r)$, $\delta U(r)$ and $\delta\alpha(r)$
can be expressed in terms of a single master amplitude $\Psi(r)$ subject to the Schrödinger-type equation,
\begin{equation}               \label{eqpert}
    \frac{d^2\Psi}{dr_*^2}+\big(\omega^2-V(r)\big)\Psi=0.
\end{equation}
\bbl{The master amplitude $\Psi(r)$  is a linear combination of $\delta N(r)$ and $\delta Y(r)$ with rather  complicated  coefficients whose explicit 
expression is not particularly illuminating, hence we do not show it explicitly. 
The potential $V(r)$ is also a complicated function of the background amplitudes that we do not show. }
The tortoise radial coordinate $r_\ast\in (-\infty,+\infty)$ is defined by the relation 
\begin{equation}
    dr_*=\frac{1}{{a(r)}}\, dr, 
\end{equation}
where the function $a(r)$ (also complicated) varies from $0$ to $1$ as $r$ changes from $r_H$ to $\infty$. 
The potential  $V$ always tends to zero at the horizon, for $r_*\to -\infty$, and it approaches unit value at infinity, for
 $r_*\to +\infty$. 
One should remember  that our dimensionless variables are related to the 
dimensionful ones via $r=$mr, $r_H={\rm mr}_H$, $V={\rm V/m^2}$, $\omega={\bm \omega}/{\rm m}$.

For the bald Schwarzschild background with 
$Q=q=N=Y=\sqrt{1-r_H/r}$ and $U=r$, one has $a(r)=Q^2(r)$ and the potential reduces to 
\begin{equation}
    \label{potsch}
    V(r)=\bigg(1-\frac{r_H}{r}\bigg)\bigg(1+\frac{r_H}{r^3}+6\,\frac{r_H(r_H-2r)+r^3(r-2r_H)}{(r_H+r^3)^2}\bigg),
\end{equation}
in agreement with Ref. \cite{Brito:2013wya}. 
In the flat space limit $r_H\to 0$ this reduces to $V(r)=1+6/r^2$, which is the potential of a massive particle of unit mass
(in units of the graviton mass) with spin $s=2$. 

Equation \eqref{eqpert} defines the eigenvalue problem on the line $r_\ast\in (-\infty,+\infty)$. Solutions of this problem with $\omega^2>0$ 
describe scattering states of gravitons. In addition, there can be bound states with purely imaginary 
frequency $\omega =i\sigma$ and 
hence with $\omega^2=-\sigma^2<0$. For such solutions the wave function $\Psi$ is everywhere bounded and square-integrable,
because one has 
$
e^{+\sigma r_\ast}\leftarrow \Psi \rightarrow e^{-\sqrt{1+\sigma^2}\,r_\ast}
$
as $-\infty \leftarrow r_\ast \rightarrow +\infty$, respectively. Such bound state solutions grow in time as
$e^{i\omega t}=e^{\pm \sigma t}$. Therefore, they correspond to unstable modes of the 
background black holes.

\subsection{Computing the eigenfrequencies}
\label{eigenfrequencies}

Our aim is to investigate a potential existence 
of negative modes with $\omega^2<0$ in the spectrum of the eigenvalue problem \eqref{eqpert}. 
If such modes exist, then the background black holes are unstable. 
If they do not exist, then the black holes are stable with respect to spherically symmetric perturbations, which would 
strongly suggest that they should be stable with respect to all perturbations. Indeed, in most known cases the $S$-channel is 
usually the only place
where the instability can reside (of course, this should be proven case to case).

\begin{figure}
    \centering
    \includegraphics[scale=0.65]{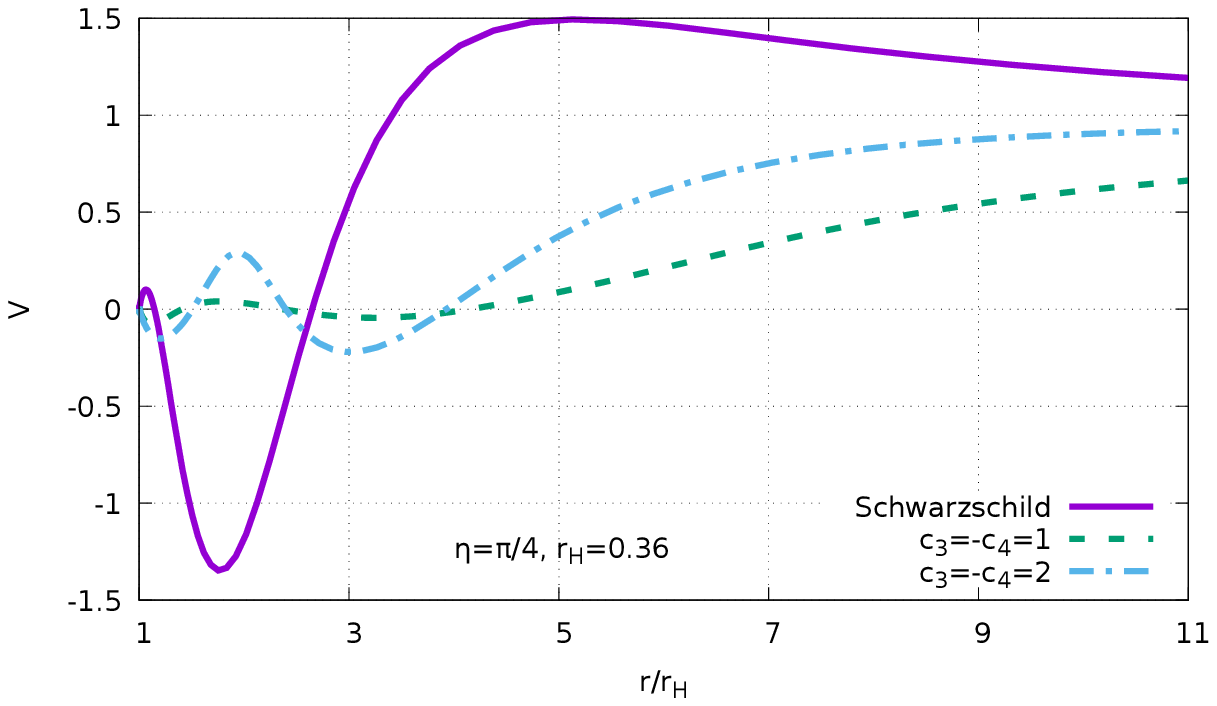}
    \includegraphics[scale=0.65]{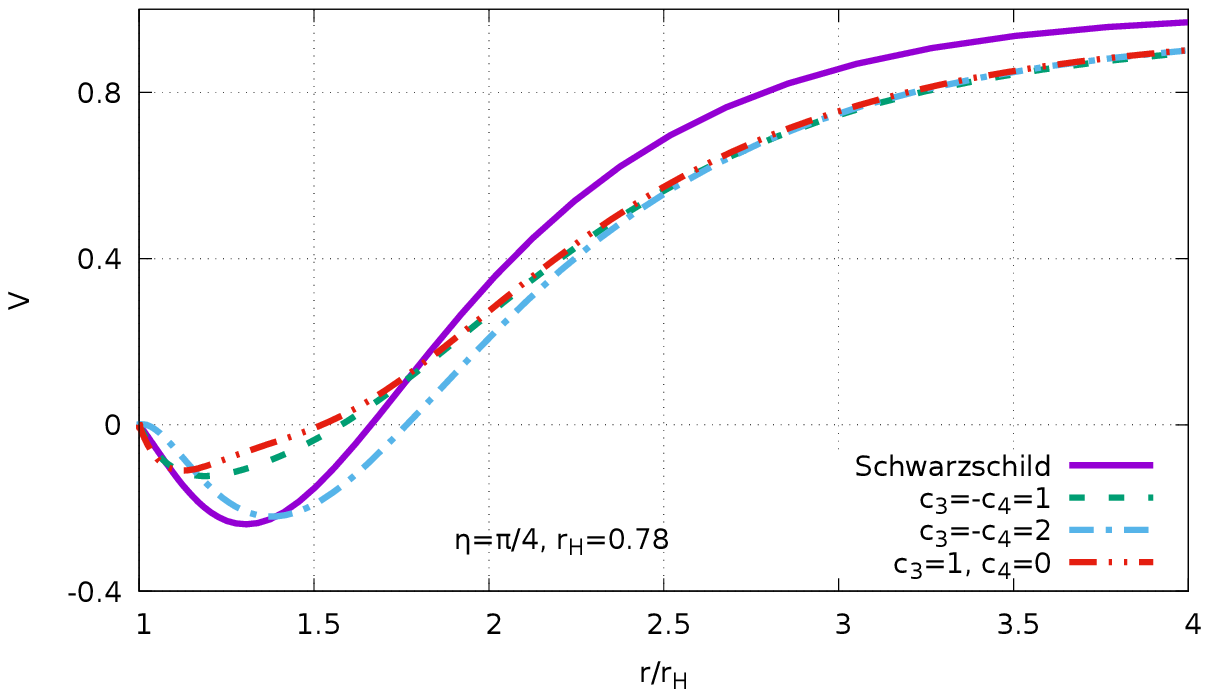}
    \caption{Potential $V(r)$ for $r_H=0.36$ (left) and for 
    $r_H=0.78$ (right) for different $c_3,c_4$ with $\eta=\pi/4$.}
    \label{potplots}
\end{figure}

The first thing to check  is the shape of the potential $V(r)$, because  if it 
is everywhere positive, then there are no bound states. We therefore show in Fig.~\ref{potplots} the potential $V(r)$ for the hairy 
backgrounds for several values of the event horizon size $r_H$ and for different $c_3,c_4$,  
and we also show $V(r)$  for the bald Schwarzschild solution with the same $r_H$ (it does not depend on $c_3,c_4$). 
We observe that in each case the potential vanishes at the horizon, then shows negative values in its vicinity,
and then approaches unity as $r\to\infty$. Since the potential is not positive definite, bound states {\it may} exist, 
but their existence is not yet guaranteed. 

We know that a bound state certainly exists for the bald Schwarzschild background with  
$r_H<0.86$ \cite{Babichev:2013una,Brito:2013wya}. When looking at the potentials for the hairy solution 
with $r_H=0.78$   in Fig.~\ref{potplots},    we notice that they are close to the Schwarzschild potential, hence 
a bound state could exist for these potentials as well.

In order to know whether bound states exist or not, we  use the well-known Jacobi criterion \cite{gelfand2000calculus} and construct 
the solution of the Schrödinger equation \eqref{eqpert} with $\omega=0$. If this solution $\Psi(r)$ crosses zero somewhere,
then there are bound states. We start in the asymptotic region where 
the tortoise coordinate $r_\ast$ becomes identical to the usual $r$, hence Eq.\eqref{eqpert} reduces simply to $\Psi^{\prime\prime}=\Psi$
so that the bounded solution is $\Psi=e^{-r}$. Then we extend this solution numerically toward small values of $r$,
and we find that, depending on values of $r_H,\eta,c_3,c_4$, it may indeed 
show a zero as $r$ approaches $r_H$. Therefore, there exists 
a bound state. 

\begin{figure}
    \centering
    \includegraphics[scale=0.65]{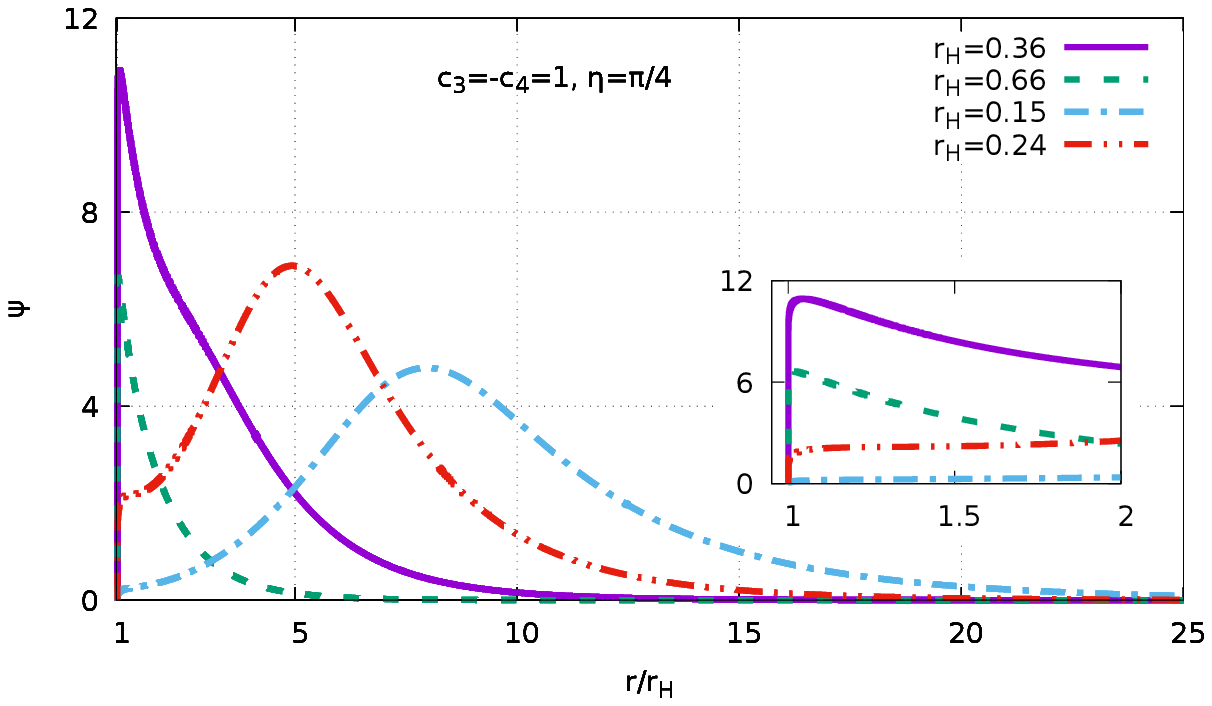}
    \includegraphics[scale=0.65]{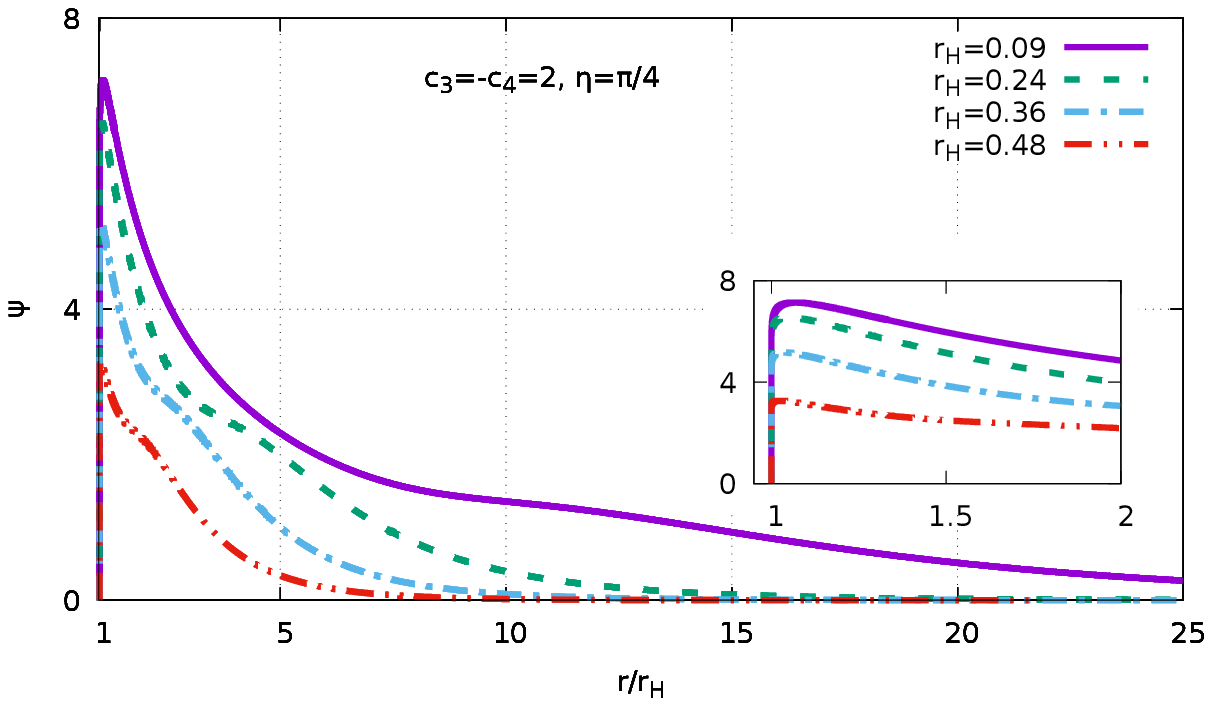}
    \caption{Negative mode eigenfunctions $\Psi(r)$ for $\eta=\pi/4$ and different  $r_H$, with  $c_3=-c_4=1$ (left) and $c_3=-c_4=2$ (right).
    They vanish at the horizon and at infinity.}
    \label{psi}
\end{figure}

The next step is to actually find the bound state by solving the eigenvalue problem \eqref{eqpert} with the potential $V(r)$ 
obtained by numerically solving the background equations. For this we set $\omega^2=-\sigma^2$ and determine 
the local solutions at infinity and close to the horizon, 
\be               \label{BBB}
B\,(r-r_H)^{\sigma r_H}\leftarrow \Psi(r) \rightarrow e^{-\sqrt{1+\sigma^2}r}~~~~\mbox{as}~~~~r_H\leftarrow r\to \infty,
\ee
where $B$ is an integration constant. Then we apply the multiple shooting method and numerically 
extend the horizon solution toward large $r$, extending at the same time the large $r$ solution toward small $r$. 
The two solutions meet at some intermediate point $r=r_0$, where the values of $\Psi(r_0)$ and $\Psi^\prime(r_0)$ should agree. 
This gives two conditions to be fulfilled by adjusting the two parameters $B$ and $\sigma$ in \eqref{BBB},
which finally yields the  bound state solution on the whole line (see \cite{Pani:2013pma,Berti:2009kk} 
for a review on the black hole  perturbation theory and the tools that can be used to solve the perturbation equation).

The eigenfunctions $\Psi$ against the ordinary radial coordinate $r$ are shown in Fig.~\ref{psi}. 
They vanish at the horizon, then show a maximum, sometimes  very close to the 
horizon, and then approach zero  for $r\to \infty$. 

As a result, we find the negative eigenvalues $\omega^2<0$ 
for all hairy black holes obtained in \cite{Brito:2013xaa}.
Therefore, all these solutions are unstable. It is worth emphasizing that all of them  correspond to 
the particular choice $\eta=\pi/4$, hence $\kappa_1=\kappa_2=1/{2}$. 
 In order to test our method, we have also computed the negative mode for the bald Schwarzschild solution as in  \cite{Brito:2013wya}.

\begin{figure}
    \centering
    \includegraphics[scale=0.9]{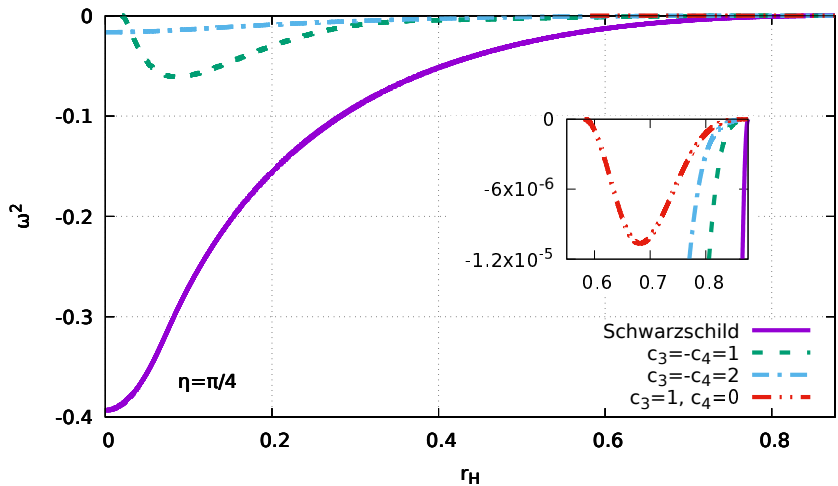}
    \caption{The negative mode eigenvalue $\omega^2(r_H)$ for the hairy and for the bald Schwarzschild black holes against $r_H$ for 
 different values of $c_3$, $c_4$. In all cases $\eta=\pi/4$. }
    \label{omega}
\end{figure}

As seen in Fig.~\ref{omega}, the absolute value of the negative mode eigenvalue for the Schwarzschild  solution is always larger than that 
for the hairy solutions. Therefore, the instability growth rate for the hairy black holes is not as large as for the Schwarzschild  solution. 
In all cases, since one has $\omega={\bm\omega}/{\rm m}$ where ${\bm\omega}$ is the dimensionful physical frequency, 
the instability growth time is $1/{\bm\omega}=1/(\omega {\rm m})$. If we assume the graviton mass ${\rm m}$ to be very small and given by 
\eqref{Hubble}, then the instability growth time will be cosmologically large, hence the instability will not play any role. 
However, as we shall see below, it is preferable to assume that $1/{\rm m}\leq 10^6$~km according to \eqref{m}, in which 
case the instability growth time will be  less than $10^3$ seconds, hence the instability is dangerous and should be 
avoided.

As seen in Fig.\ref{omega},  the eigenvalue $\omega^2(r_H)<0$ approaches zero when $r_H\to 0.86$, 
therefore all hairy black holes become then stable.
However, they are no longer hairy in this limit, because  they ``lose their hair" and merge with the bald  Schwarzschild solution. 
Near  $r_H=0.86$ all solutions  are close to each other and $\omega^2$ is close to zero 
for any $c_3,c_4,\eta$, while for smaller $r_H$ the backgrounds and  $\omega^2$ become parameter dependent. 
\bl{The eigenvalue $\omega^2(r_H)<0$ may approach  zero also for type I solutions 
for a 
small $r_H\neq 0$ 
when they  cease to exist. }
For example, for $c_3=1$, $c_4=0$,  the hairy solution disappears at $r_H\sim 0.58$, and at the same time 
the eigenvalue $\omega^2$ approaches zero, as seen in the insertion in  Fig.~\ref{omega}.


The instability of hairy black holes is in fact a somewhat puzzling phenomenon, since it is unclear what they may decay into. 
Since the hairy solutions with $r_H<0.86$ are more energetic than 
the bald Schwarzschild solution, they probably may approach the latter via absorbing and/or radiating away their hair during their decay.  
However, the bald Schwarzschild solution is also unstable for $r_H<0.86$ and should decay into something.

The perturbative instability of the Schwarzschild solution in the massive bigravity theory 
is mathematically equivalent \cite{Babichev:2013una} to the Gregory-Laflamme instability of the vacuum black string in D=5 
\cite{Gregory:1993vy}. It is known that the nonlinear development of the latter leads
 to the formation of an infinite string of ``black hole beads"
 in $D=5$, but the event horizon topology does not change \cite{Lehner:2011wc}.  
 This fact being established within the $D=5$ vacuum GR, 
 a similar scenario is not possible in the  $D=4$ bigravity theory, 
 hence the fate of the bigravity black holes should be different.  One possibility is that
 the black hole radiates away all of its energy within the S-channel (some radiative solutions are known explicitly 
 \cite{Kocic:2017hve, Hogas:2019cpg}),
 but it is unclear  what happens to the horizon, whether  it disappears 
 or not.  In  GR the horizon cannot disappear via a classical process \cite{Hawking:1973uf}, but 
in the bimetric theory the situation might be different.

\begin{figure}
    \centering
    \includegraphics[scale=0.9]{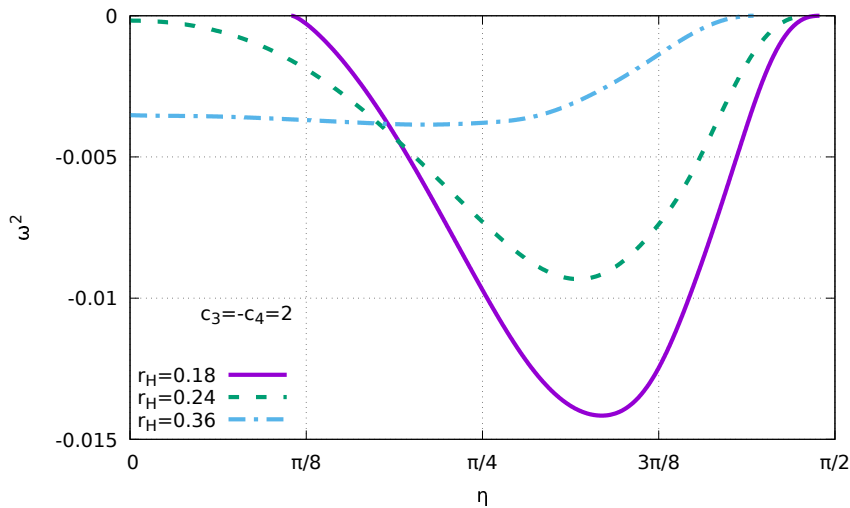}
    \caption{The negative mode eigenvalue $\omega^2(\eta)$ for the hairy black holes with 
 $c_3=-c_4=2$. }
    \label{et}
\end{figure}

Remarkably, we find that these puzzling issues are not omnipresent and
the  black holes can be stable if  $\eta$ is different from $\pi/4$.  
In Fig.~\ref{et} we show $\omega^2$ 
 against $\eta$ for several values of $r_H$ for solutions with $c_3=-c_4=2$. One can see that $\omega^2(\eta)<0$ approaches zero
 and  the negative mode disappears in the  hairy Schwarzschild limit when $\eta$ approaches
 $\pi/2$.  At the same time, the bald Schwarzschild solutions for the same $r_H$ are  certainly unstable. 
 This is a very encouraging fact -- we see that adding the hair to the black hole provides  the stabilizing effect. 
 As is seen in Fig.~\ref{et},  the eigenvalue approaches zero also when $\eta$ become small,
 if only $r_H$ is also small, as seen in Fig.~\ref{et}.

 Summarizing the above discussion, for some parameter values the hairy black holes are unstable, but 
 for other parameter values they can be stable. Below we shall describe a parameter choice leading to a large 
 set of stable solutions.

\section{PARAMETER SPACE AND THE PHYSICAL SOLUTIONS\label{par}}
\setcounter{equation}{0}

In this section we give a detailed description of particular subsets  of solutions. 
Providing a complete classification of solutions depending on 4 parameters $r_H,\eta,c_3,c_4$
would be  a very difficult task.  We therefore adopt the following strategy: choosing 
 the particular values 
 \be                \label{ch1}
 c_3=-c_4=5/2
 \ee
 which fulfill condition II in \eqref{typeI}, we study the solutions for all possible $r_H,\eta$. Performing next the duality transformation 
 gives us all possible solutions for 
  \be                 \label{ch2}
 c_3=1/2,~~~c_4=3/2, 
 \ee
 which values fulfill condition I in \eqref{typeI}.  This approach reveals interesting and rather complex features which are presumably 
 generic for any $c_3,c_4$.

\begin{figure}
    \centering
    \includegraphics[scale=0.92]{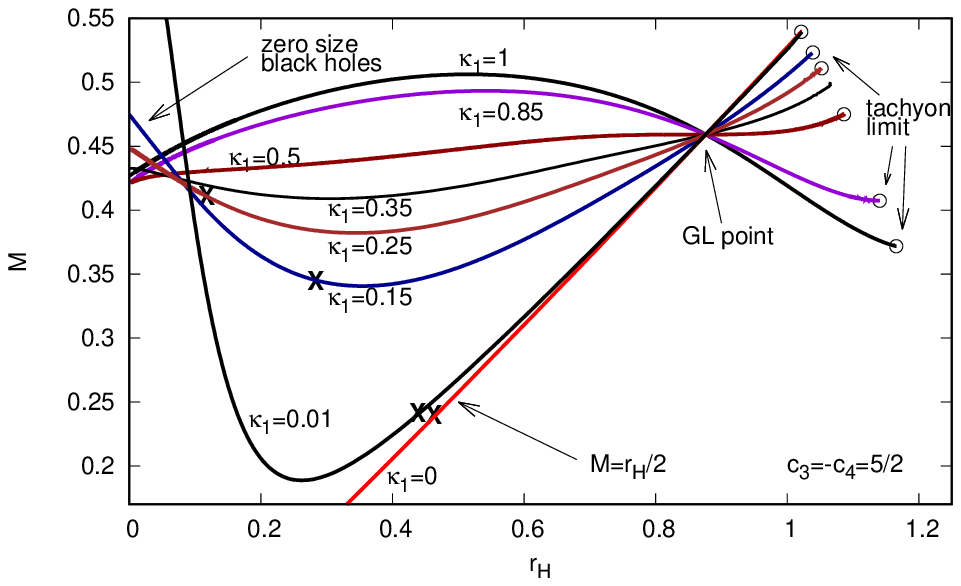}
    \includegraphics[scale=0.92]{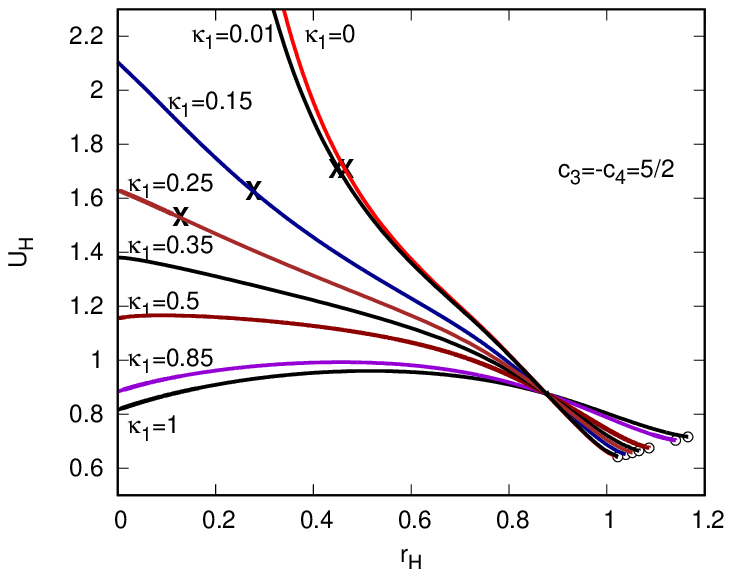}
    \caption{The mass $M(r_H)$ (left) and the functions $U_H(r_H)$ (right) for the hairy solutions with $ c_3=-c_4=5/2$. 
    \bbl{The crosses mark the points on the left of which  the f metric becomes singular.  The hollow circles mark the 
    termination points 
    beyond which the solutions would become complex valued. 
    When $\kappa_1=\cos^2\eta \to 0$, 
    the mass $M(r_H)$ develops a more and more profound minimum, 
while   the values of  $M(0)$ and $U_H(0)$ grow without bounds. }
    }
    \label{FA}
\end{figure}

 Figure \ref{FA} shows the ADM mass $M(r_H)$ and the function $U_H(r_H)$ 
 for several values of $\eta\in[0,\pi/2]$. As one can see, all curves $M(r_H)$ 
 intersect at the GL point, $(r_H,M_H)=(0.86,0.43)$, where all solutions bifurcate  with the bald Schwarzschild solution, 
 \be
 N^2=Q^2=Y^2=q^2=1-\frac{0.86}{r},~~~~U=r,
 \ee
 whereas all curves $U_H(r_H)$ pass through the point $(r_H,U_H)=(0.86,0.86)$. 
 Away from the bifurcation  point, the g metric still remains Schwarzschild if $\eta=\pi/2$, in which case $M(r_H)$ is a linear function, 
 \be 
 \eta=\frac{\pi}{2}:~~~~~N^2=Q^2=1-\frac{r_H}{r}~~~\Rightarrow M=\frac{r_H}{2},
 \ee
 but the f metric for these solutions is not Schwarzschild, even though both metrics have the same mass; 
 \bbl{as explained above, we call such solutions  hairy Schwarzschild}. 
 For $\eta\neq \pi/2$ the 
 mass  depends nonlinearly on $r_H$. 
 
 Introducing the mass function $M(r)$ via $N^2(r)=1-2M(r)/r$,  Eq.\eqref{e1}
assumes the form 
\be                   \label{ADM1}
\bbl{M^\prime(r)=\kappa_1\,\frac{r^2}{2}\left({\cal P}_0+U^\prime{\cal P}_1\frac{N}{Y}\right)
\equiv 
\kappa_1\,\rho,}
\ee
from where the ADM mass
\be                         \label{ADM2}
M=M(\infty)=\frac{r_H}{2}+\kappa_1\int_{r_H}^\infty \rho\, dr\equiv M_{\rm bare}+M_{\rm hair}.
\ee
Here the ``bare" mass $M_{\rm bare}=r_H/2$ is determined only by the horizon radius and coincides with the mass of the Schwarzschild solution of radius $r_H$, 
whereas  the mass $M_{\rm hair}$ expressed by the integral 
is the contribution of the massive hair distributed outside the horizon. 
As one can see in Fig.~\ref{FA}, one has $M>r_H/2$ if $r_H<0.86$, hence the  ``hair mass" is positive and the  hairy 
solutions are more energetic than the bare Schwarzschild black hole. However, the mass of the hair becomes negative above the GL point, where $r_H>0.86$,
and the hairy solutions are then less energetic than the bare one. Therefore the energy density $\rho(r)$ can be negative. In fact,
there are no reasons for which  the standard energy conditions should be respected within the bigravity theory.

\bbl{Each curve in Fig.~\ref{FA} is defined only in a finite interval $r_H\in[0,r_H^{\rm max}(\eta)]$. 
It is very instructive to understand what happens at the boundaries of this interval. }

\subsection{ The lower limit $r_H\to 0$}

 \bbl{All the solutions extend  down to arbitrarily small values of $r_H$.}
 Remarkably, as seen in Fig.~\ref{FA},  except  for $\eta=\pi/2$ the mass $M$ does not vanish when $r_H\to 0$ 
 but approaches a finite value, even though 
 the bare mass 
 $M_{\rm bare}=r_H/2\to 0$. Therefore, all mass is contained in the  hair mass in this limit, hence  something remains even when 
 the horizon size $r_H$ shrinks to zero. A similar phenomenon is actually well known, 
 since in many nonlinear field theories there are solutions describing a small black hole inside a soliton 
 (for example, inside the magnetic monopole) \cite{Volkov:1998cc}. Sending the horizon size to zero the black hole disappears, 
 but its external nonlinear matter fields remain and become a gravitating soliton containing a regular origin  in its center  instead of the horizon. 
 Therefore, the $r_H\to 0$ limit of a hairy black hole may correspond to a regular soliton. 
 
\begin{figure}
    \centering
    \includegraphics[scale=0.82]{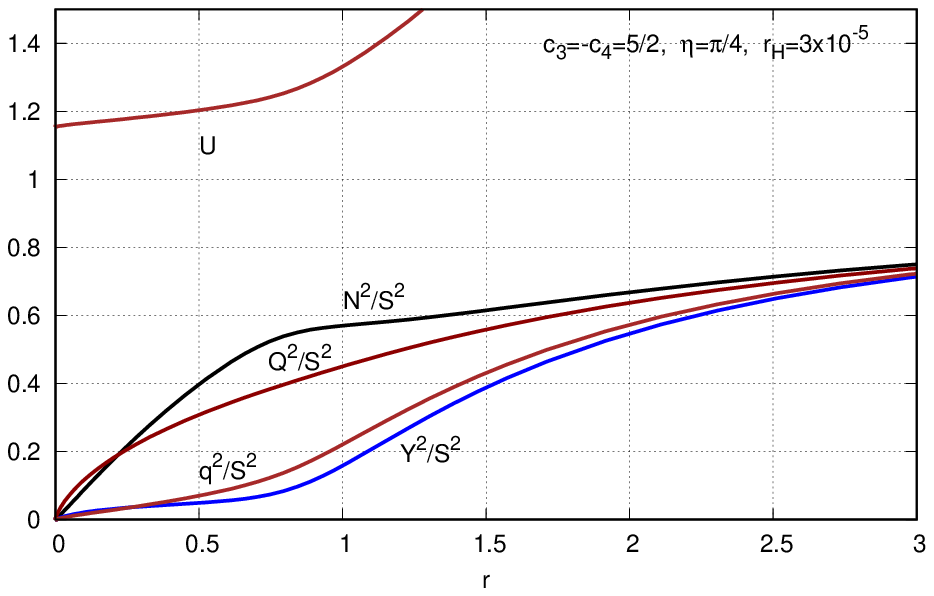}
      \includegraphics[scale=0.82]{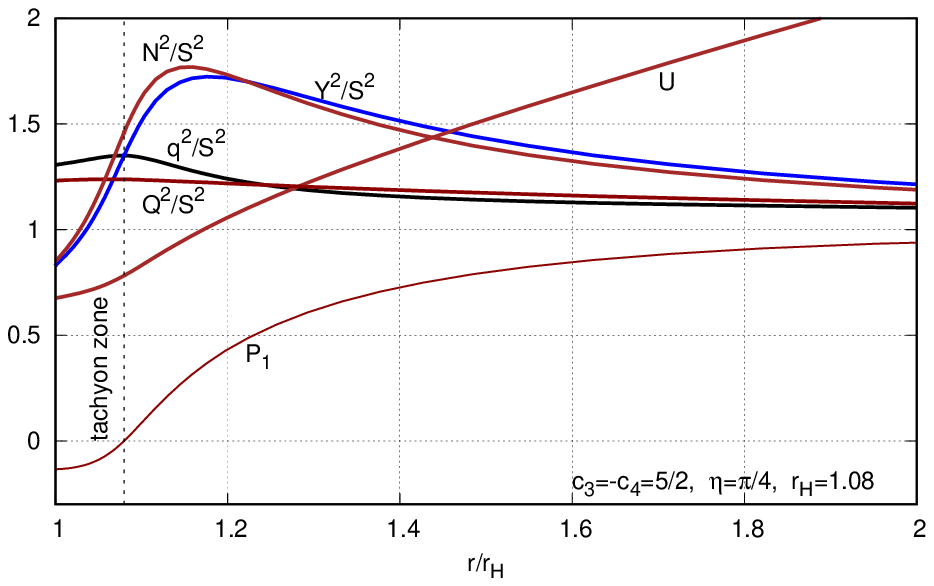}
    \caption{Profiles of the solution with $r_H\sim 10^{-5}$ that is close to the zero size black hole (left), and 
of that close to the tachyon limit, with   $D\sim 10^{-6}$ (right). One has 
  $S^2={1-r_H/r}$.  The amplitude ${\cal P}_1$ determines the graviton mass  via \eqref{FP-mass-1}
and the gravitons behave as tachyons if ${\cal P}_1<0$. 
}
    \label{lim}
\end{figure}

 One may expect the situation to be similar also in our case and that there is a limiting  configuration  to which the black hole
 solutions approach pointwise when $r_H\to 0$. Such a limiting configuration indeed exists; however it seems to be singular and not of the 
 regular soliton type. 
 First, as seen in Fig.~\ref{FA}, the value of $U_H$ which determines the size of the f horizon remains finite when $r_H\to 0$, 
 hence the f geometry remains a black hole even in the limit. Secondly, as seen in 
 Fig.~\ref{lim}, one has $N^2/S^2\sim r$ for $r\leq 0.5$ for a solution with a very small $r_H$. 
 However, one has $S=\sqrt{1-r_H/r}\to 1$ as $r_H\to 0$,
 hence one has in this limit $N^2\sim r$ and the limiting form of the g metric is something like a ``zero size black hole". 
 The numerical profiles shown in 
 Fig.~\ref{lim} suggest this limiting configuration to have  the following structure at small $r$:
 \be
 N^2\sim Y^2\sim Q^2\sim q^2\sim  r,~~~~~~U=U_{\rm min}+{\cal O}(r). 
 \ee
 The g geometry is singular since its Ricci invariant $R(g)=2/r^2+{\cal O}(1/r)$ at small $r$, 
 but the f geometry remains of the regular black hole type because $U$ does not vanish.
 Curiously, the temperature remains finite for  $r_H\to 0$ and is always the same for both metrics. 
 The limiting g temperature can be  formally computed by assuming $N^2=\alpha r$, $Q^2=\beta r$ with   $\alpha\approx 0.7$ and $\beta\approx 6$ 
 from Fig.~\ref{lim}. 
 Equation \eqref{TT} then yields 
  $T=\sqrt{\alpha\beta}/(4\pi)\approx 0.163$, which  is  very close 
 to the value $T=0.16$  for  the solution with $r_H\sim 10^{-5}$ shown in Fig.~\ref{lim}. However, these considerations 
 are of course purely formal since 
 the zero size black hole cannot evaporate and further reduce its size, and the standard WKB 
 arguments for the black hole evaporation do not apply because the geometry is singular at the horizon. 
 
 \bbl{One should say that 
the f metric can become singular for small $r_H$  because 
 the $q,Y$ amplitudes develop additional zeros outside the horizon. This happens along the parts 
 of the curves  on the left of the points marked by the crosses in Fig.\ref{FA}. 
 We have already discussed this phenomenon and said that 
 we do not exclude  such solutions from consideration 
 because  the f geometry is not observable and its singularities are invisible, while 
 the g geometry that can be directly  probed   remains always regular. 
The physical parameters of the solutions such as the ADM mass also do not show anything special 
when the $q,Y$ amplitudes starts to oscillate. The potential $V$ in the perturbation equation \eqref{eqpert} 
also remains regular. 
We therefore have no reason to exclude such solutions from consideration, 
and in fact they are necessary in order that the theory could describe 
black holes within  a broad mass spectrum.  }

 \subsection{ The upper ``tachyon" limit $r_H\to r_H^{\rm max}(\eta)$}
 
 In this limit the solutions always remain regular and disappear after a  fusion of roots of the algebraic equation \eqref{qqq} [or \eqref{alg}].
 As explained above, this equation determines the horizon values of the solutions. 
 Its two roots determine  two solution branches, 
 but only the root with $\sigma=+1$ gives rise to 
 asymptotically flat solutions, the  other branch showing    a singularity of the g metric 
 outside the horizon.  When $r_H$ increases, the determinant of \eqref{qqq} decreases and vanishes for some $r_H=r_H^{\rm tach}(\eta)$, 
 then it becomes positive again, decreases again and vanishes for the  second time  for  $r_H=r_H^{\rm max}(\eta)>r_H^{\rm tach}(\eta)$,
 after which it becomes negative and the procedure stops. 
 Specifically,  it turns out that the determinant of \eqref{qqq} factorizes, 
  \be                      \label{D}
  {\cal D}\equiv {\cal B}^2-4{\cal AC}={\cal P}^2_1(r_H)\, D~~~~\Rightarrow~~~~\sqrt{\cal D}={\cal P}_1(r_H) \sqrt{D},
  \ee 
  where ${\cal P}_1(r_H)$ is defined by \eqref{e5} with ${\bf u}=U/r$ replaced by $u=U_H/r_H$
  while $D$ is a complicated function of $r_H,U_H,\eta,c_3,c_4$.
  When $r_H$ increases, then 
  ${\cal P}_1(r_H)$ crosses zero at some $r_H=r_H^{\rm tach}(\eta)$ while $D$ remains positive, hence the 
  square root $\sqrt{\cal D}$ changes sign.  When  $r_H$  continues to increase, then $D$ approaches zero and 
  vanishes as $r_H\to r_H^{\rm max}(\eta)$. No further increase of $r_H$ is possible since 
  $D$ would then be negative thus rendering the solutions complex valued. 
  
  Although the determinant ${\cal D}$ vanishes for $r_H=r_H^{\rm tach}(\eta)$ when ${\cal P}_1(r_H)=0$ 
  and also for $r_H=r_H^{\rm max}(\eta)$ when $D=0$, the two solution branches never merge. Specifically, 
  the two horizon values $\nu_H$ determined by \eqref{alg} merge when ${\cal D}=0$, but a careful inspection 
  reveals that  $y_H,U_H$ in \eqref{Up} and \eqref{yh} remain different for the two branches when ${\cal P}_1(r_H)=0$. 
  If $D=0$ then all horizon values $\nu_H,y_H,U_H$ coincide for the two branches, but  the derivatives  $y_H^\prime$ 
  defined by \eqref{fin} remain different. This is a consequence of the fact that the existence and uniqueness theorem applies 
  only to regular points of the differential equations, whereas  the event horizon $r=r_H$ is a singular point. 

  In the interval  $r_H^{\rm tach}(\eta)<r_H<r_H^{\rm max}(\eta)$ the solutions show a ``tachyon zone" near the horizon 
  where the function ${\cal P}_1(r)$ defined by \eqref{e5} is negative, as shown in the right panel in Fig. \ref{lim}. 
   \bl{Let us remember relation 
   \eqref{FP-mass} for the  Fierz-Pauli mass of gravitons obtained via linearizing 
   the field equations around the flat background. This relation  can be written as 
   ${\rm m^2_{\rm FP}}={\cal P}_1(\infty)\, {\rm m}^2$. 
   However, the equations can be similarly linearized around an arbitrary background solution,   
    which yields in the spherically symmetric case  
   the position-dependent mass term \cite{Mazuet:2018ysa}
    \be                                \label{FP-mass-1}
{\rm m^2_{\rm FP}}={\cal P}_1(r)\, {\rm m}^2.
\ee
Therefore, if ${\cal P}_1(r)<0$ then the mass effectively becomes imaginary. }
  As a result, solutions for $r_H>r_H^{\rm tach}$  show unphysical 
   features, hence we call $r_H\to r_H^{\rm max}(\eta)$
 the  ``tachyon limit".  The horizon value $y^\prime_H$ diverges in this limit, but this seems to be an integrable divergence similar to 
   $y^\prime(r)\sim 1/\sqrt{r-r_H}$ and the limiting solution itself stays regular. We were able to approach this solution 
   rather closely,  as shown in Fig.~\ref{lim} (right panel)
 which presents ``an almost limiting" solution with the horizon value of the determinant $D\sim 10^{-6}$.   
 
  To recapitulate, hairy solutions exist only  for 
   $0<r_H\leq r_H^{\rm max}(\eta)$.

\subsection{The ADM mass}
  
  \bbl{It is important that,  unless $\kappa_1=\cos^2\eta$ is very small, 
the ADM mass of all hairy solutions always varies 
   within a finite range and can  be neither very large nor very small,    as seen in  Fig.~\ref{FA}. 
  It seems this fact was not recognized in Ref.~\cite{Brito:2013xaa}, which 
 always shows only the ratio  $M/r_H$ which 
  diverges as $r_H\to 0$. However, the mass $M$ remains  finite for $r_{H}\to 0$. 
As seen in   Fig.~\ref{FA}, the mass actually does not change much when $r_H$ changes and 
always remains close to the GL value, which is the mass of the Schwarzschild solution with $r_H=0.86$,}
\be
M\sim \frac{0.86}{2}=0.43.
\ee
This means that the  dimensionful mass  (restoring for the moment the speed of light $c$ and Newton's constant $G$) 
\be
{\rm M}=\frac{c^2\, M}{G\, \rm m}
\ee
is always close to that  of the Schwarzschild black hole of size ${\rm r}_H=0.86/{\rm m}$, which is
close to the Compton length of  massive gravitons. As a result, one cannot assume the graviton mass m to be very small
and of the order of the inverse Hubble radius as in \eqref{Hubble}. Indeed, this would imply the hairy black holes to be   as heavy 
as the Schwarzschild black hole of a cosmological size -- a physically meaningless result. \bbl{However, assuming instead 
that $1/{\rm m}=\gamma\times  10^6$~km  with $\gamma\in[0,1]$ as in \eqref{m}, which is consistent with the cosmological observations if $\kappa_1$ is parametrically small 
as expressed by \eqref{m0}, yields a physically acceptable result. The masses of the hairy black holes are then close to the 
mass of the Schwarzschild black hole of  radius $\gamma\times 10^6$~km, that is ${\rm M}\sim 0.3\times 10^6 \,\gamma\times {\rm M}_\odot$. 
If $\gamma\sim 1$ this gives the value   typical for 
supermassive astrophysical  black holes observed in the center of many galaxies. }


\bbl{If $\kappa_1$ is very small then the mass can deviate considerably from the GL value and can 
 become very small or very large. As seen in Fig.~\ref{FA}, for small $\kappa_1$ the mass $M(r_H)$ 
shows a minimum:    first it decreases  with $r_H$, 
then reaches a minimal value $M_{\rm min}$, and then increases  up to some $M(r_H=0)$.  
For smaller values of  $\kappa_1$ the minimum becomes more and more profound and the  value 
$M_{\rm min}$ approaches zero while $M(r_H=0)$ becomes larger and larger. 
If $\kappa_1$ is extremely small as in \eqref{m0}, $\kappa_1\leq 10^{-34}$, then the minimum  value $M_{\rm min}$
is extremely close to zero. One has then $M(r_H)\approx r_H/2$ for $r>(r_H)_{\rm min}$ where  $(r_H)_{\rm min}$ 
is very small,  but in the region $r<(r_H)_{\rm min}$ the mass grows rapidly when $r_H\to 0$ 
up to a very large value $M(r_H=0)$. }

 \begin{figure}
    \centering
    \includegraphics[scale=0.65]{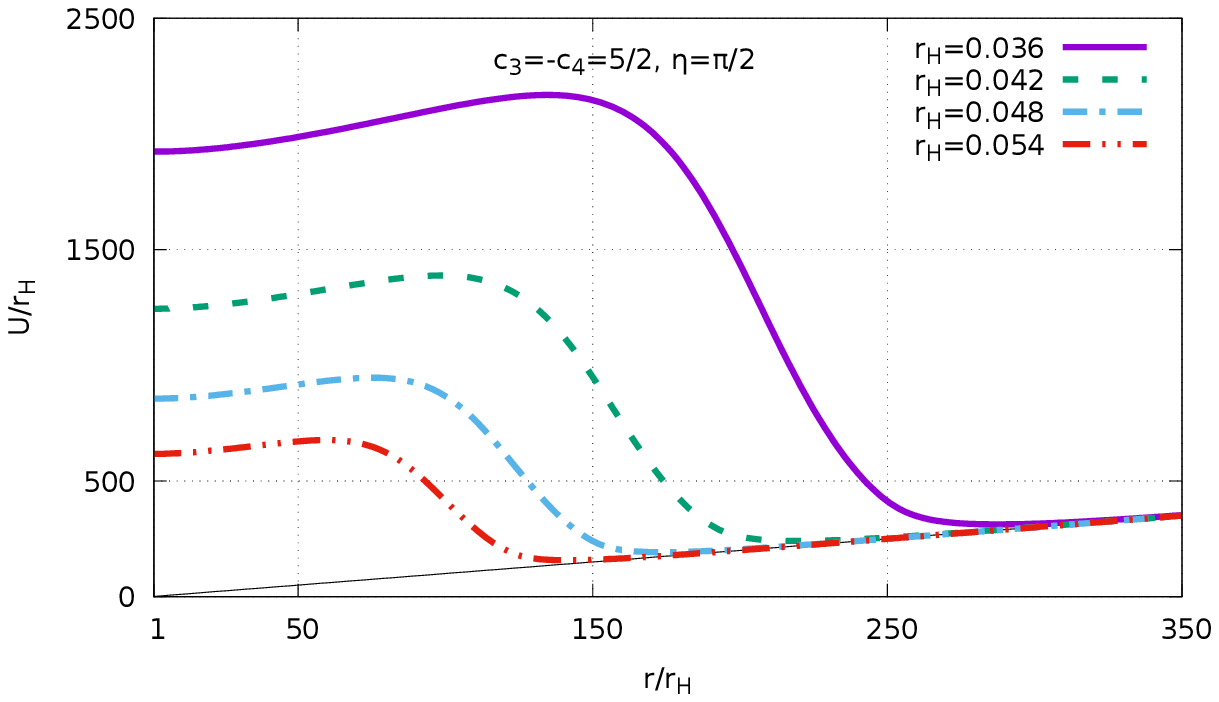}
      \includegraphics[scale=0.65]{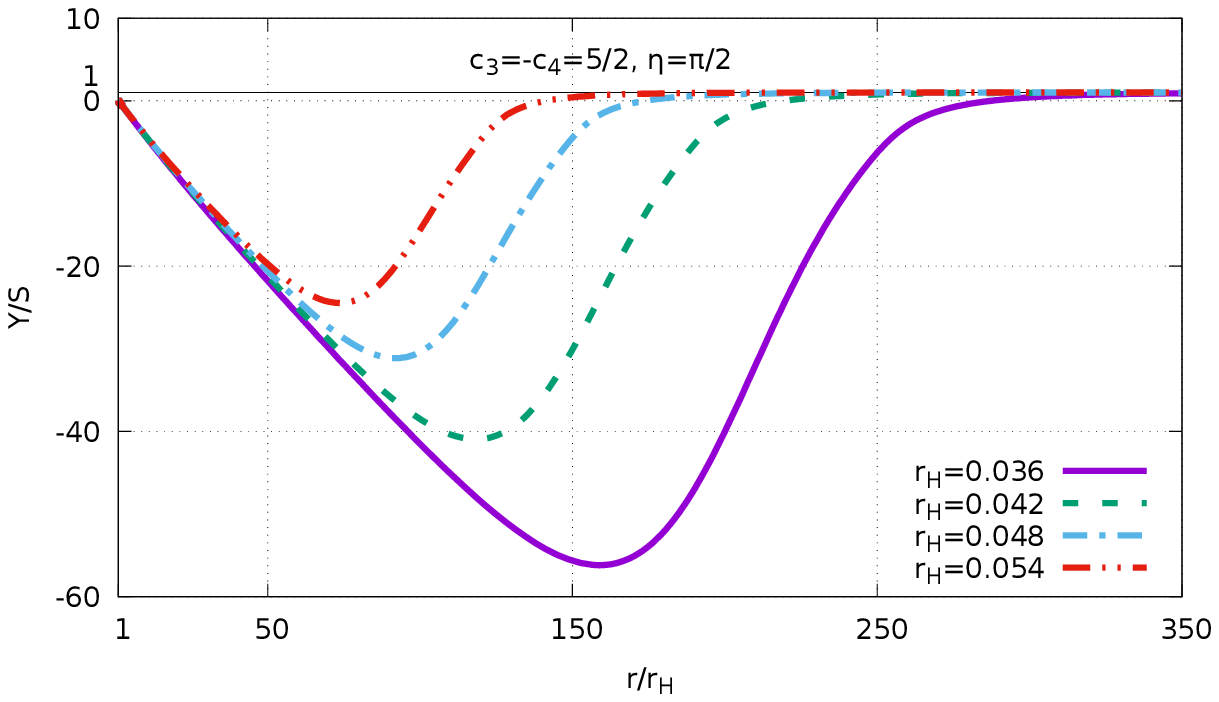}
      
         \includegraphics[scale=0.65]{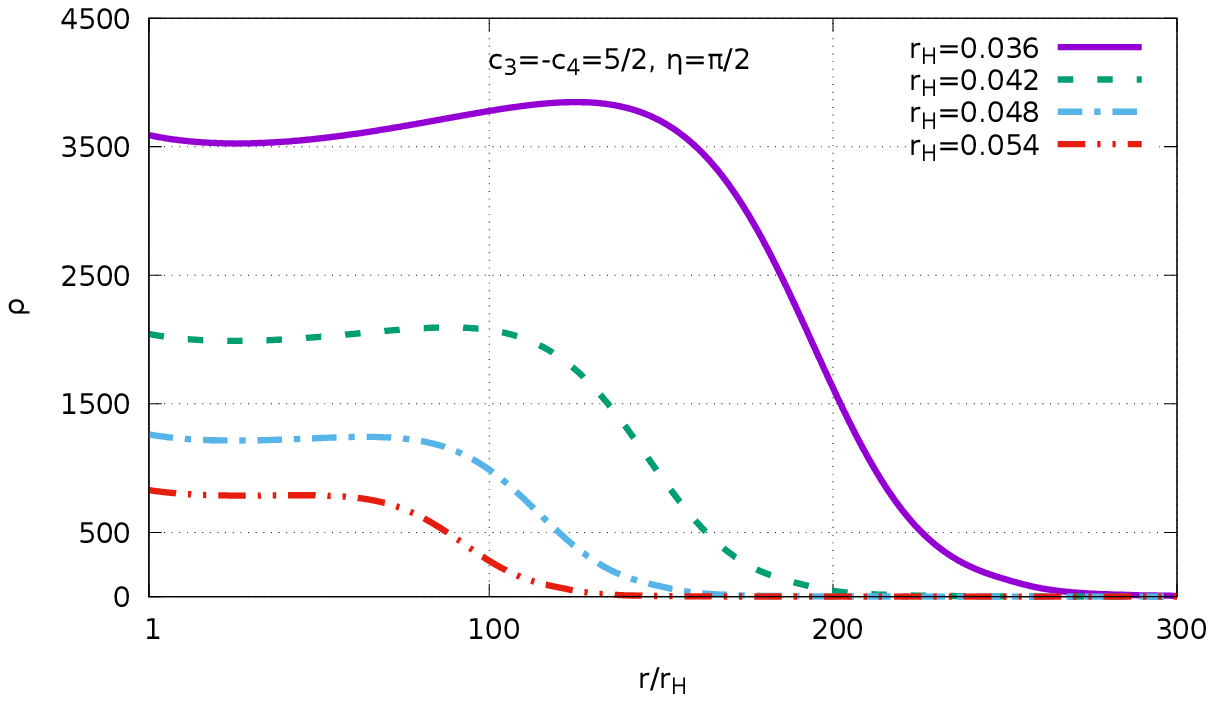}
      \includegraphics[scale=0.65]{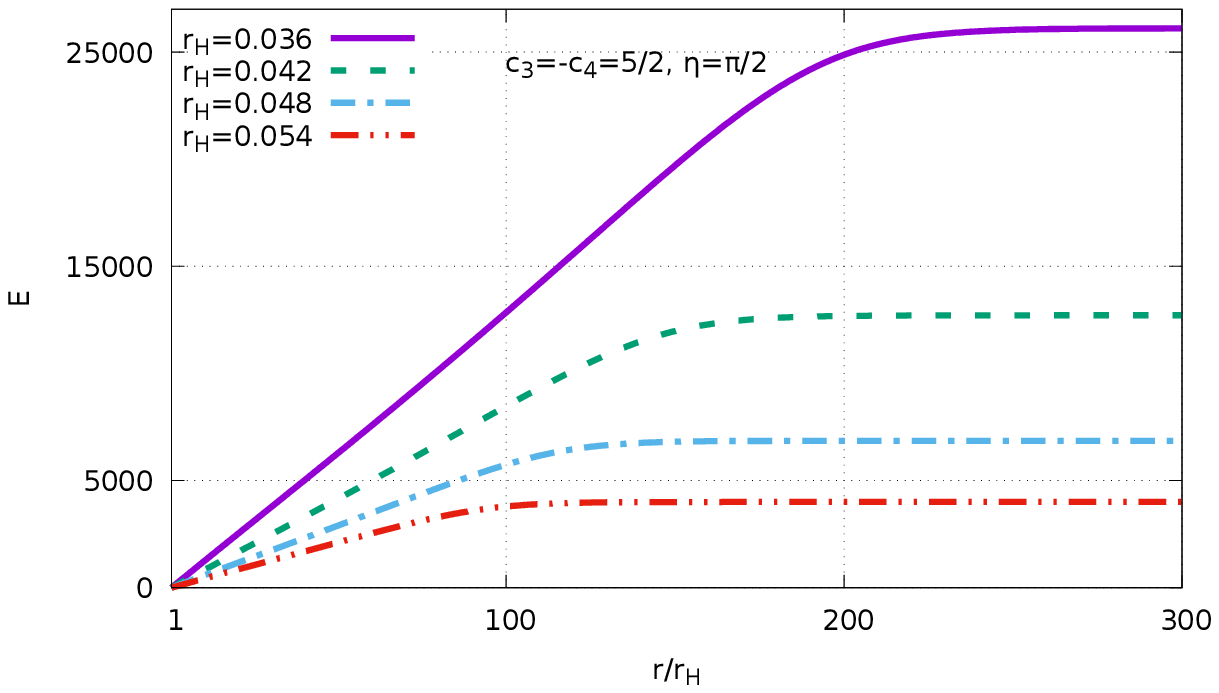}

    \caption{\bbl{The amplitudes $U(r),Y(r)$, the ``hair energy density" $\rho(r)$ and its integral $E(r)$ whose asymptotic value $E(\infty)$
    is the ``hair  energy" 
    for the hairy Schwarzschild solutions with $\kappa_1=0$ 
    and  $r_H\ll 1$.}}
    \label{pi_2}
\end{figure}

\bbl{
To get an approximation for $M(r_H)$ for very small $\kappa_1$, we consider the  hairy Schwarzschild solutions with $\kappa_1=0$. 
Their g metric is Schwarzschild with all the  hair contained in the f metric. It turns out that the $Y,U$ amplitudes 
of the f metric depend very strongly  on $r_H$ if the latter is small. As seen in Fig.~\ref{pi_2}, in the horizon vicinity 
these amplitudes  show very large values which apparently grow without bounds when $r_H\to 0$, although one always has
$Y(r)\to 1$ and $U(r)\to r$ far away from the horizon. We inject these solutions to \eqref{ADM1} and \eqref{ADM2}  to obtain the radial energy 
density $\rho(r)$ and  $E(r)=\int_{r_H}^r \rho\, dr$. They also become very large when $r_H$ decreases, as seen  in Fig.~\ref{pi_2}. 
The asymptotic value $E(\infty)$ is the ``hair energy". As seen  in Fig.~\ref{pi_2}, the hair energy is  large for small 
$r_H$, but  it does not backreact and the g metric remains Schwarzschild  if $\kappa_1=0$. However, the  hair energy starts to backreact if 
$\kappa_1\neq 0$.  If $\kappa_1\ll 1$ then one can deduce from Eq.\eqref{ADM2} that 
\be
M= \frac{r_H}{2}+\kappa_1 E(\infty)+{\cal O}(\kappa_1^2), 
\ee
where $E(\infty)$ is computed for $\kappa_1=0$. We evaluate numerically  $E(\infty)$ for various values of $r_H$ and obtain 
the following best fit approximation: 
\be                     \label{MMM} 
M\approx \frac{r_H}{2}+\kappa_1\,\frac{a}{(r_H)^s}, 
\ee
where $a=0.0056$ and $s=4.61$. 
Assuming that $\kappa_1=\gamma^2\times 10^{-34}$, this function shows an absolute minimum at 
\be
(r_H)_{\rm min}\approx 5.2 \,\gamma^{0.35}\times 10^{-7},~~~~~~~M_{\rm min}\approx 3.1\,\gamma^{0.35}\times 10^{-7}\,,
\ee
whose dimensionful versions are obtained by multiplying by ${\rm 1/m}=\gamma\times 10^6$~km 
(restoring again the speed of light and Newton's constant) 
\bea                  \label{min}
({\rm r}_H)_{\rm min}=\frac{(r_H)_{\rm min}}{\rm m} 
\approx 0.52\,\gamma^{1.35}~{\rm km},~~~~~~~
{\rm M}_{\rm min}=\frac{c^2\, M_{\rm  min}}{G\,\rm m}
\approx 0.2\,\gamma^{1.35}\times {\rm M}_\odot\,.
\eea
This determines the minimum mass for the hairy black holes.
When $r_H$ gets smaller still  then  the mass starts to grow, 
but it grows  only up to a finite although very large value as $r_H\to 0$ because 
the approximation \eqref{MMM} is not valid for however small $r_H$.}

\subsection{Parameter regions for solutions with $c_3=-c_4=5/2$ }

\begin{figure}
    \centering
    \includegraphics[scale=1.2]{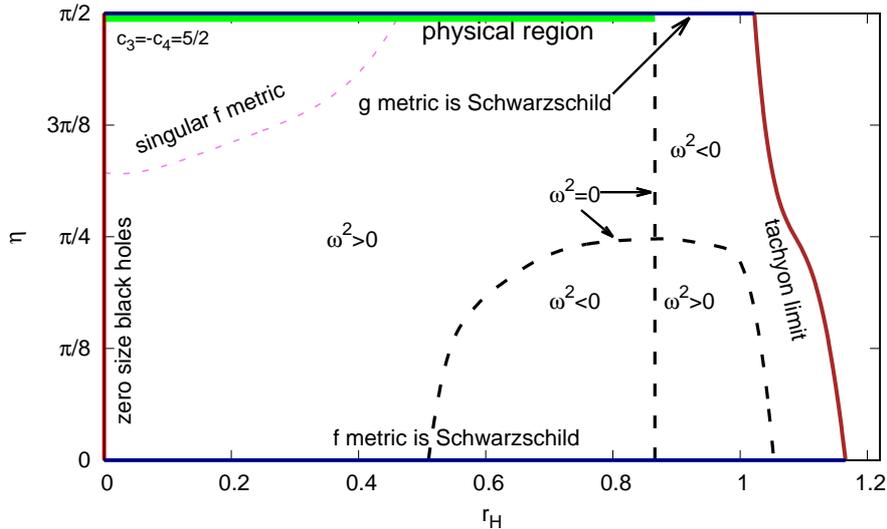}
    \caption{The parameter region  in  the $(r_H,\eta)$ plane corresponding to 
regular hairy black hole solutions with $c_3=-c_4=5/2$. The dashed black $\omega^2=0$ lines separate stable and unstable sectors. 
\bbl{The upper left corner contains solutions with a singular f metric, however, their g geometry is regular.}}
    \label{FB}
\end{figure}

Let us now collect all the facts together. The diagram in Fig.~\ref{FB} shows the region in the $(r_H,\eta)$ plane within which 
there are hairy black hole solutions.  The low boundary of this region at $\eta=0$ corresponds to solutions whose f metric is 
 Schwarzschild,  while the upper boundary at $\eta=\pi/2$ corresponds to solutions whose g metric is 
 Schwarzschild. \bbl{The left boundary corresponds to the limiting solutions with $r_H=0$: 
  the  zero size black holes. 
 The right boundary marks the ``tachyon limit" beyond which the solutions would become complex valued. 
 The upper-left corner of the diagram contains solutions with a singular f geometry, but their g geometry, 
 which is physically measurable,  is regular. }
 
 The diagram also shows lines corresponding to the zero modes, $\omega^2=0$, of the 
 perturbative eigenvalue problem \eqref{eqpert}.  The vertical line  corresponds to the GL value $r_H=0.86$. 
 The eigenvalue $\omega^2$ changes sign when 
 crossing these lines, therefore, the lines separate sectors  where $\omega^2>0$ and hence the solutions are stable, 
 from sectors  where $\omega^2<0$ and the solutions are unstable. There are altogether two stable and two unstable sectors. 
 It is worth noting that the stability region is now much larger than for solutions with $c_3=-c_4=2$ considered in the previous section. 
 One also notices  that the tachyonic solutions 
 are in the unstable sector. 
 
 Finally, the diagram shows the ``physical region" corresponding to physically acceptable solutions. 
 As explained above, for such solutions the coupling $\kappa_1=\cos^2(\eta)$ should be very small for their mass not to be 
 too large, hence $\eta$ should be very close to $\pi/2$. The solutions should be stable, hence they should 
 correspond to the sector  where $\omega^2>0$. These conditions specify the physical region to be 
 the thick (green online) line at the top of the diagram. 
 
 Physical solutions are therefore described by the g metric which is extremely close to Schwarzschild, since  
 \bea                                  \label{Enst-eq-1}
G_{\mu\nu}({g})&=&\kappa_1\, T_{\mu\nu}(g,f),~~~~~~~~\mbox{where}~~~~\kappa_1~\bbl{\leq}~ 10^{-34}.
\eea
 The ``hairy features" of the solutions hidden in the f metric should  be difficult to observe, unless in violent 
 processes like black hole collisions producing large enough  $T_{\mu\nu}(g,f)$
 to overcome the $10^{-34}$ suppression. 
 Summarizing, the static bigravity black holes should be extremely   similar to the GR black holes, but their strong field dynamics
 is expected to be different.

 \bbl{As explained above, the physical region contains stable hairy black holes whose masses range 
 from the minimal value
 $\sim 0.2\,\,\gamma^{1.35}\times  {\rm M}_\odot$ up to the maximal value $\sim 0.3\times  10^6\,\gamma^{1.35}\times  {\rm M}_\odot$
 with $\gamma\in [0,1]$.
 Yet  heavier black holes also exist in the theory but they cannot be hairy and should be described by 
 the ``bald" Schwarzschild solution \eqref{b2}, which is stable for $r_H>0.86$. 
 Stable black holes with ${\rm M}\leq 0.2\,\gamma^{1.35}\times {\rm M}_\odot$ 
 can only be of the  type \eqref{b1}.   }

 \subsection{Parameter regions for dual solutions with $c_3=1/2$, $c_4=3/2$ }
 
 Let us now see how the described above solutions look after the duality transformation \eqref{dual}. 
 This transformation converts the parameter values \eqref{ch1}
 into \eqref{ch2}, flips the sign of $\eta-\pi/4$  and swaps the $Q,N,r$ with $q,Y,U$.  
 Graphically, this amounts to relabelling  the functions 
 and plotting them 
 against $U$ instead of $r$. 
 The ADM mass and temperature are invariant under duality. The stability property also does not change since,
 for example, if a solution is unstable and admits growing in time perturbations, 
 then its dual version  will contain the same growing modes and 
 hence will be unstable as well. 
 
 \begin{figure}
    \centering
    \includegraphics[scale=0.92]{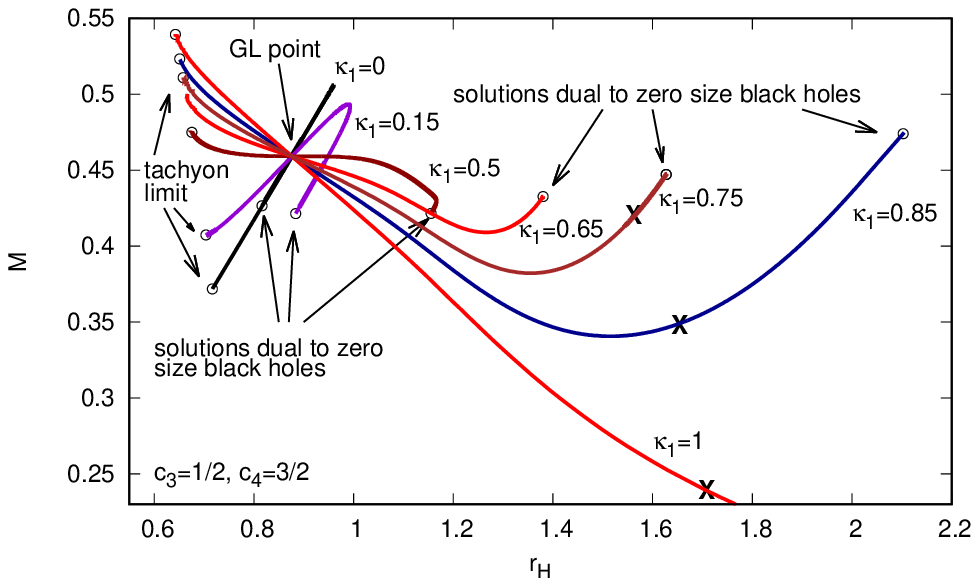}
    \includegraphics[scale=0.92]{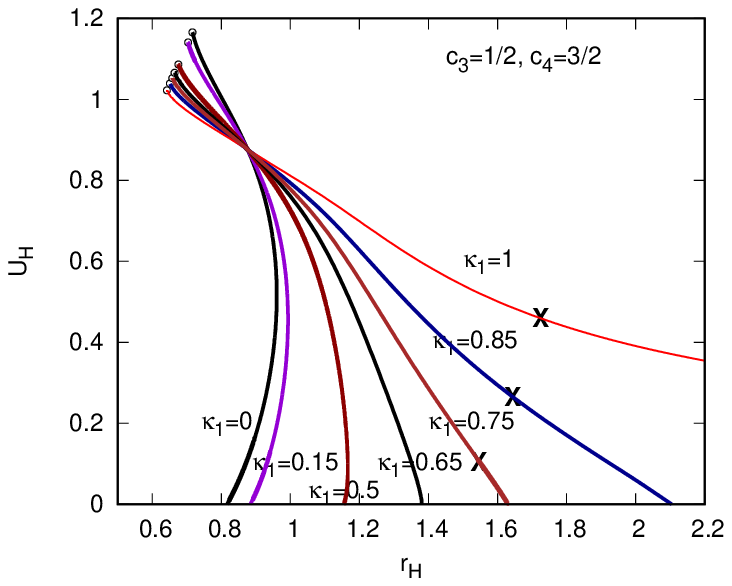}
    \caption{The mass $M(r_H)$ (left) and the functions $U_H(r_H)$ (right) for the hairy black hole solutions with $c_3=1/2$, $c_4=3/2$.
  \bbl{  The crosses mark points on the right of which the g metric becomes singular, hence these pars of the curves 
    correspond to unphysical solutions that should be excluded from consideration.}}
    \label{FAA}
\end{figure}

 Figure \ref{FAA} shows the dual version of Fig.~\ref{FA}. The mass curves $M(r_H)$ still intersect in the GL point but they 
 look quite different as compared to those in  Fig.~\ref{FA}. In particular, not all of them are single valued. 
 The reason is that the functions $U_H(r_H)$ 
 in Fig.~\ref{FA}  are not always monotone, hence their inverses shown in Fig.~\ref{FAA} are not single-valued. As a result, 
 for each $\eta$ such that $0\leq \cos^2\eta\leq 0.6$ there are two different 
 solutions with the same $r_H$ but with different $U_H$, hence the curves $M(r_H)$ are not always single valued.

The solutions now exist  for  $r_H\in[r_H^{\rm min}(\eta),r_H^{\rm max}(\eta)]$.
 The lower limit $r_H^{\rm min}(\eta)$ corresponds to what used to be the upper limit before the duality -- 
 the tachyon solutions with vanishingly small  horizon determinant $D$.
\bbl{The upper limit $r_H^{\rm max}(\eta)$  corresponds  for small $\eta$ to solutions whose g metric starts being singular. 
Before the duality these were solutions whose f metric started being singular while their g-metric was regular. 
After the duality their g-metric becomes singular, hence such solutions are no longer allowed and should be excluded.  }
For larger values of $\eta$ the right boundary 
 $r_H^{\rm max}(\eta)$ corresponds to 
points where the two different 
 solutions with the same $r_H$ but with different $U_H$ merge to each other.

\begin{figure}
    \centering
    \includegraphics[scale=1.2]{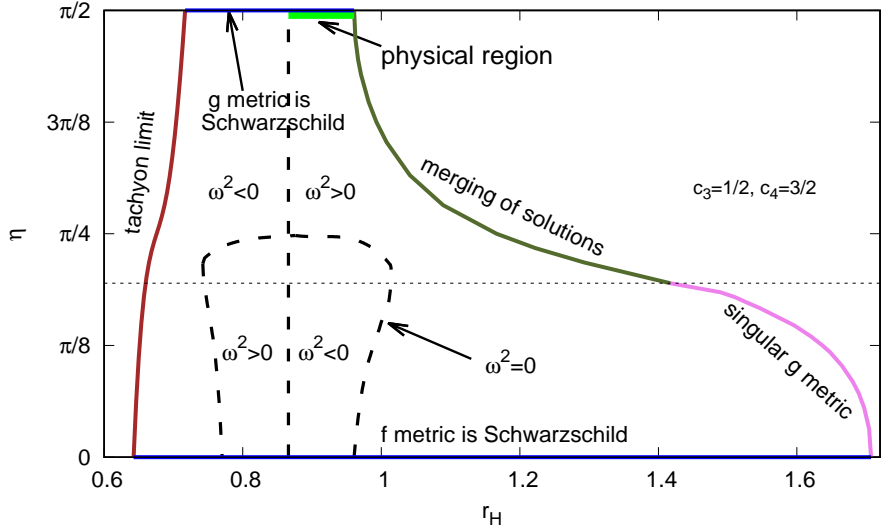}
    \caption{The parameter region  in the $(r_H,\eta)$ plane corresponding to 
regular hairy black hole solutions with $c_3=1/2$, $c_4=3/2$. The dashed black $\omega^2=0$ lines separate stable and unstable sectors.}
    \label{FC}
\end{figure}

The solutions below the GL point, for $r_H<0.86$, are still more energetic than the solution with $\eta=\pi/2$,
 hence their  hair mass  $M_{\rm hair}$ is positive, whereas  above the GL point it becomes negative.  Finally, Fig.~\ref{FC} shows
 the existence diagram in the  $(r_H,\eta)$ plane, together with the stability regions. The diagram now looks quite different as compared 
 to that in Fig.~\ref{FA}, although it corresponds to essentially the same solutions, up to the duality transformation. 
 Although the duality does not change stability, it interchanges positions of the stability sectors. Therefore,  the physical region corresponding 
 to stable solutions with $\eta$ close to $\pi/2$ is now above the GL point, where the hair mass is negative. 
 The physical solutions are again characterized by the g metric that is extremely close to Schwarzschild, but the novel feature is  that now
 for each value of $r_H$ from the physical region there are two different solutions whose g metrics are almost the same but the f metrics 
 are different. 
 
 \bbl{
 As one can see, the physical region in Fig.\ref{FAA} is rather short and corresponds only to supermassive black holes with $0.86<r_H<r_{H}^{\rm max}$. 
 All black holes of smaller masses are unstable. Therefore, the parameter choice $c_3=1/2$, $c_4=3/2$ is not physically interesting. }
 

\section{CONCLUDING REMARKS \label{CR}}

To recapitulate, we presented above a detailed analysis of static and asymptotically flat black holes in the ghost-free 
massive bigravity theory.  Extending the earlier result of \cite{Brito:2013xaa}, we find that for given values of the theory 
parameters $c_3,c_4,\eta$ and for a given event horizon size varying within a finite range, 
$r_H\in [r_H^{\rm min}, r_H^{\rm max}]$, there are one or sometimes two different black holes supporting a nonlinear massive graviton hair, 
in addition to the ``bald Schwarzschild" solution with $g_{\mu\nu}=f_{\mu\nu}$ described by \eqref{b2}. 
The hairy solutions are more energetic than the Schwarzschild one
if $r_H<0.86$ and they are less energetic otherwise. When $r_H$ approaches the limiting values $r_H^{\rm min}$ or $r_H^{\rm max}$, the solutions
either become complex valued or merge between themselves. 
For some values of $c_3,c_4$ zero-size black holes exist for which $r_H^{\rm min}=0$ but the corresponding $U_H$ remains finite. 
Depending on values of $r_H,c_3,c_4,\eta$, the hairy solutions can be either stable
of unstable. 

\bbl{To avoid the hairy black holes being unphysically heavy, one is bound to assume the 
massive graviton Compton length to be 
${\rm 1/m}= \gamma \times \, 10^6$ km where the parameter  $\gamma$ may range in the interval $[0,1]$. 
The agreement with the cosmological data is then achieved by assuming 
that $\kappa_1=\cos^2\eta=\gamma^2\times  ({\rm M}_{\rm ew}/{\rm M}_{\rm Pl})^2=\gamma^2\,\times  10^{-34}$.
The stable hairy black holes are  described by a g metric which is 
extremely close to Schwarzschild, but their f metric is quite different. These black holes have the mass and size close to those of 
ordinary  black holes with  the masses ranging  from $\sim 0.2\,\gamma^{1.35}\times  {\rm M}_\odot$  to 
$\sim 0.3\times 10^6\,\gamma^{1.35}\times  {\rm M}_\odot$, the latter being the value typical for the 
supermassive astrophysical black holes if $\gamma\sim 1$. Yet heavier black holes in the theory 
should  be  bald.  As a result, if the bigravity theory indeed applies to describe physics, the astrophysical   black holes should 
support  the  hair hidden  in the 
f metric. }

\bbl{
Since the f metric is not coupled to matter and  cannot be directly probed, 
while the deviation of the ``visible" g metric from Schwarzschild 
is suppressed by the factor of $\kappa_1= \gamma\times 10^{-34}$, the hairy black holes should 
normally be undistinguishable  from the usual GR black holes. However, in violent 
processes like black hole coalescences 
the interaction between the two metrics may produce an energy momentum tensor strong enough to overcome the $10^{-34}$ 
suppression in $G_{\mu\nu}({g})=\kappa_1\, T_{\mu\nu}(g,f)$. In this case the deviation from GR should become visible. 
Therefore, it is possible that signals from black hole mergers detected by LIGO/VIRGO  \cite{Abbott:2016blz} may  
carry  information about the hairy structure of the black holes. One could expect a ``hair imprint" in the signal to be stronger for {\it small}
black holes, since we know that for small black holes the amplitudes $U,Y$ of the f metric become very large, which should influence the 
$T_{\mu\nu}(g,f)$ of the merger. It is therefore possible that the   hair imprints  will be  visible when smaller mass mergers 
(see \cite{Abbott:2020niy} for a recent review) are detected. 
However, to actually determine   the  hair imprint in the signal 
would require calculations  going beyond the scope of the present paper. We therefore 
leave this problem for a separate  project and for the time being 
simply refer to 
the  recent preprint \cite{Dong} where calculations of this type are performed 
within the context of the ghost-gree massive gravity \cite{deRham:2010kj}
[where  the only static black holes are those described by \eqref{b1}]. 
 }

 Finally, we should discuss the paper \cite{Torsello:2017cmz} that also considers  black holes in the 
ghost-free  massive bigravity theory. \bl{This paper presents  essentially the same classification of different types of black holes as 
 the one previously given in  \cite{Volkov2012} but in a more refined way, extending it and paying attention  to some subtle points. } 
 The paper addresses in particular the issue of  convergence of the solutions  to the flat background
 in the asymptotic region. 
 Among other things, it claims that the Schwarzschild solution is the only asymptotically flat black hole in the  theory. 
 \bl{At the same time, the paper does not contain a rigorous proof of this statement but gives just a number of plausibility arguments,
 so that the claim  should rather be viewed  as a conjecture, as actually  explicitly stated   in some places of  \cite{Torsello:2017cmz}. }
 These arguments are  as follows. 
 
 First of all, it was emphasized in  \cite{Torsello:2017cmz} that the usual practice of starting the numerical integration not at the horizon $r=r_H$,
 which is a singular point of the differential equations, but at a regular nearby point  $r=r_H+\epsilon$, as was done in   \cite{Brito:2013xaa}, 
 could in principle  lead to numerical instabilities. We agree with this, and it is for this reason that we use the desingularization 
 procedure (described in Appendix \ref{Des} below) which allows us to start the numerical integration exactly at $r=r_H$
 \bl{(initial conditions exactly at $r=r_H$ were described  also in  \cite{Torsello:2017cmz})}. 
 
 The paper \cite{Torsello:2017cmz} makes also another  remark concerning the behavior at the horizon. 
 It is known that 
 in order to be able to {\it cross} the horizon, for example when studying geodesics, one cannot use the Schwarzschild coordinates 
 and one should introduce instead regular at the horizon coordinates. These can  be, for example,  
 Eddington-Finkelshtein (EF) coordinates in which $g_{00}=g_{11}=0$, $g_{01}=g_{10}\neq 0$. 
 \bl{It was noticed in \cite{Torsello:2017cmz} 
  that the f metric, when expressed in the same coordinates, generically does not have the same form, since it has $f_{11}\neq 0$, hence 
 the two metrics cannot be simultaneously EF. 
 We understand  this, but this does not invalidate  the background solutions (Ref.~\cite{Torsello:2017cmz} agrees on this). }
 The horizon geometries are regular, and if one wishes, 
 one can use the same boundary conditions at the horizon to integrate  inside the horizon to recover the interior solutions. 
 Within the parametrization described  in Appendix \ref{Des}, this is achieved by simply changing the sign of the numerical integration step.

Next, small initial deviations from the Schwarzschild solution via setting at the horizon 
$u=U_H/r_H=1+\epsilon$ were considered in \cite{Torsello:2017cmz}.
Integrating the equations toward large $r$ then yields 
metrics whose components diverge as $r\to\infty$ instead of approaching finite values. 
\bl{This observation, made  already in \cite{Volkov2012},   shows that 
there are no regular and asymptotically flat solutions {\it in a small vicinity} of the Schwarzschild solution.
However, there  can be regular solutions corresponding 
to $u$ considerably deviating from unit value. }

\bl{Finally, the paper \cite{Torsello:2017cmz}  reproduces and analyzes   (in  Appendix A) one of the asymptotically flat solution 
(with a singular f metric) 
found in  \cite{Brito:2013xaa}. 
It obtains a pathological result, and the reason is the following. 
Appendix D of \cite{Torsello:2017cmz}  describes the numerical method used 
-- a straightforward integration starting from the horizon with  the standard routine 
of {\it Mathematica}. This adequately produces  the solution with a given precision, but only within a finite range 
of the radial coordinate $r$. If one integrates farther on trying to approach flat space, then the growing  
$C e^{+r}$ mode generically present in the solution leads to a rapid accumulation of numerical errors triggering 
a numerical instability.  Trying to suppress this mode by adjusting the horizon boundary conditions, 
one typically observes the derivatives of some functions in the solution growing without bounds  at a some  finite $r$.
Precisely this type of behavior at the end of the integration interval is seen  in Fig.~11 in  \cite{Torsello:2017cmz}. }

\bl{
One cannot get asymptotically flat solutions within the numerical scheme adopted in \cite{Torsello:2017cmz},
since pathological  features  arise in this way inevitably. This must be the reason behind the conviction 
that such solutions do not exist.  However, all pathologies can be eliminated within the more elaborate 
numerical scheme described above -- via suppressing the growing  
$C e^{+r}$ mode from the very beginning. }

\section*{ACKNOWLEDGEMENTS} 

\bl{We thank Francesco Torsello, Mikica Kocic and Edvard Mörtsell for clarifying  discussions and useful remarks. }
The work of  M.S.V. was partly supported by the  French National Center of Scientific Research 
within the collaborative French-Russian research program, Grant No. 289860, 
as well as by the Institute of Theoretical and Mathematical Physics at  the Moscow University during the visit 
in early 2020, and also by the Russian Government Program of Competitive Growth of the Kazan Federal University.




\section*{APPENDIX A: DESINGULARIZATION AT THE HORIZON \label{Des}}

\renewcommand{\theequation}{A.\arabic{equation}}

The horizon  $r=r_H$ is a singular point of the differential equations -- the derivatives $N^\prime$ and $Y^\prime$ 
expressed by Eqs.\eqref{eqs} are not defined at this point. The usual practice to handle this difficulty is to use the local power series expansions 
\eqref{l1} and \eqref{l2} to start the numerical integration not exactly at $r=r_H$ but at a nearby point with $r=r_H+\epsilon$ where $\epsilon$ 
is a small number. One may then hope that the results will not be very sensitive to the value of $\epsilon$. However, in such an approach 
$\epsilon$ remains an arbitrary parameter not defined by any prescription. This inevitably affects the stability of the numerical procedure,
which becomes evident when one studies the dependence of the solutions on the parameters.

At the same time, it is possible to reformulate the problem in such a way that the numerical integration starts 
exactly at $r=r_H$.  Let us make the change of variables 
\be
N={S}\,\nu,~~~~~Y={S}\,y~~~~~\mbox{with}~~S=\sqrt{1-\frac{r_H}{r}}. 
\ee  
The  functions  $\nu,y$ and their derivatives are defined also  at $r=r_H$. 
Equations \eqref{e1} and \eqref{e2} then yield 
\be                    \label{desing}
\nu^\prime=-\frac{\nu}{2r}+\frac{{\cal C}_1}{2\nu y\, r^2 S^2},~~~~~~~y^\prime=-\frac{y U^\prime}{2U}+\frac{{\cal C}_2}{2\nu y\, r^2 U S^2},
\ee
where 
\bea             \label{CCC}
{\cal C}_1&=&(r-r_H\nu^2 -\kappa_1\,r^3 {\cal P}_0)\,y-\kappa_1\,r^3\,{\cal P}_1 U^\prime \,\nu\,, \nonumber \\
{\cal C}_2&=&\nu\,r^2 (1-\kappa_2\,r^2 {\cal P}_2)\,U^\prime -\kappa_2\,r^4\,{\cal P}_1\,y-r_H U\nu\,y^2\,.
\eea
At the horizon the derivatives $\nu^\prime$ and $y^\prime$ are finite, which requires that 
\be
{\cal C}_{1|_{r_H}}=0,~~~~~{\cal C}_{2|_{r_H}}=0,
\ee
from where one obtains the horizon values
\bea
U^\prime_H&=&\frac{(1-\nu^2-\kappa_1\,r^2{\cal P}_0)\,y }{\kappa_1\, r^2 {\cal P}_1\,\nu}_{|_{r_H}}\,, \label{Up} \\
y_H&=&\left.\frac{1+(\kappa_2\,r^2{\cal P}_2-1)\nu^2
+\kappa_1\kappa_2({\cal P}_0{\cal P}_2-{\cal P}_1^2) \,r^4-(\kappa_1{\cal P}_0+\kappa_2{\cal P}_2)\,r^2}{\kappa_1 r{\cal P}_1 U\nu}\right|_{_{r_H}}.  \label{yh}
\eea
At the same time, the horizon value of $U^\prime$ can be obtained from \eqref{eqs},
\bea       \label{Up1} 
U^\prime_H=\lim_{r\to r_H}{\cal D}_U(r,U,{S}\nu,{S}y)\equiv {\cal D}_{UH}(r_H,U_H,\nu_H,y_H).
\eea
This value must agree with the one given by \eqref{Up}, which yields a condition on $\nu_H$, and using \eqref{yh},
this condition reduces  [if $b_k$ are chosen according to \eqref{bbb}] to a biquadratic equation
\be               \label{alg}
{\cal A}\,(\nu_H^2)^2+{\cal B}\,\nu_H^2+{\cal C}=0,
\ee 
where the coefficients ${\cal A}$, ${\cal B}$, ${\cal C}$ are (rather complicated) functions of $r_H,U_H$. As a result,
for given $r_H,U_H$ there are two possible horizon values $\nu_H^{(1)}$ and $\nu_H^{(2)}$.  Injecting to \eqref{Up} and \eqref{yh}, 
this determines the horizon values $y_H$ and $U^\prime_H$. Finally, the horizon values of $\nu^\prime$ and $y^\prime$ are obtained 
from \eqref{desing} by taking the $S\to 0$ limit and using  l'Hopital's rule, which yields 
\be                    \label{des}
\nu^\prime_H=-\frac{\nu_H}{2r_H}+\frac{{\cal C}_{1|_{r_H}}^\prime }{2r_H\nu_H y_H },~~~~~~~
y^\prime_H=-\frac{y_H U^\prime_H}{2U_H}+\frac{{\cal C}^\prime_{2|_{r_H}}}{2r_H \nu_H y_H U_H }. 
\ee
There remains to compute the derivatives here. One has, for example, 
\be
{\cal C}_{1|_{r_H}}^\prime=\left.\left(\frac{\partial}{\partial r} +\nu^\prime_H\frac{\partial}{\partial \nu} 
 +y_H^\prime\frac{\partial}{\partial y}  +
 U'_H \frac{\partial}{\partial U}
 +U^{\prime\prime}_H\frac{\partial}{\partial U^\prime}\right){{\cal C}_1}(r,U,\nu,y,U^\prime)\right|_{r=r_H,U=U_H,\nu=\nu_H,y=y_H}
\ee
where the second derivative is similarly obtained from \eqref{Up1}, 
\be
U^{\prime\prime}_H=\left.\left(\frac{\partial}{\partial r}+\nu^\prime_H\frac{\partial}{\partial \nu} 
 +y_H^\prime\frac{\partial}{\partial y} +
 U'_H \frac{\partial}{\partial U}
\right){\cal D}_U(r,U,{S}\nu,{S} y)\right|_{r=r_H,U=U_H,\nu=\nu_H,y=y_H},
\ee
and similar expressions for ${\cal C}_{2|_{r_H}}^\prime$.
Injecting this to \eqref{des} yields {\it linear} in $\nu_H^\prime$ and $y_H^\prime$ relations, which can be 
resolved to give (we do not show explicit formulas in view of their complexity)
\be              \label{fin}
\nu_H^\prime=\nu_H^\prime(r_H,U_H,\nu_H,y_H),~~~~~y_H^\prime=y_H^\prime(r_H,U_H,\nu_H,y_H). 
\ee

Summarizing the above discussion, the equations in the desingularized form read 
\bea                    \label{desint}
\nu^\prime&=&-\frac{\nu}{2r}+\frac{{\cal C}_1}{2\nu y r^2 S^2}\equiv {\cal F}_\nu(r,U,\nu,y),  \nonumber \\
y^\prime&=&-\frac{y U^\prime}{2U}+\frac{{\cal C}_2}{2\nu y r^2 U S^2}\equiv {\cal F}_y(r,U,\nu,y),\nonumber \\
U^\prime&=&{\cal D}_U(r,U,{S}\nu,{S}y)\equiv {\cal F}_U(r,U,\nu,y),
\eea
where ${\cal C}_1$ and ${\cal C}_2$ are defined by \eqref{CCC} while ${\cal D}_U$ is the same as in \eqref{eqs}. 
These equations apply for $r>r_H$, while at $r=r_H$ they should be replaced by 
\bea
\nu^\prime&=&\nu_H^\prime(r_H,U_H,\nu_H,y_H),~\nonumber \\
y^\prime&=&y_H^\prime(r_H,U_H,\nu_H,y_H), \nonumber \\
U^\prime&=&U_H^\prime(r_H,U_H,\nu_H,y_H),
\eea
where $\nu^\prime_H$, $y^\prime_H$, $U^\prime_H$ are defined by Eqs.\eqref{Up} and \eqref{fin}. 
The horizon values $r_H$ and $U_H\equiv ur_H$ can be
arbitrary, while $\nu_H$ is not arbitrary but must fulfil the algebraic equation \eqref{alg}, whereas  $y_H$ is determined by \eqref{yh}. 
This formulation allows one to start  the  integration exactly at the horizon $r=r_H$ and then continue  to the $r>r_H$ region.

\section*{APPENDIX B: FIELD EQUATIONS WITH TIME DEPENDENCE \label{time}}
\renewcommand{\theequation}{B.\arabic{equation}}
\label{completefieldeq}

Let us allow  both metrics to depend on time, assuming that they are still 
 spherically symmetric. The gauge freedom of reparametrizations  of the $t,r$ coordinates can be used to make 
  the g metric  diagonal, but the f metric will in general contain an off-diagonal term.  The two metrics can be written as
  \cite{Volkov:2011an} 
\begin{align*}
    ds^2_g&=-Q^2 dt^2+\frac{dr^2}{\Delta^2}+R^2 d\Omega^2,\\
    \label{anz}
    ds^2_f&=-\big(q^2-\alpha^2Q^2\Delta^2\big)dt^2-2\alpha\bigg(q+\frac{Q\Delta}{W}\bigg)dtdr+\bigg(\frac{1}{W^2}-\alpha^2\bigg)dr^2+U^2 d\Omega^2,\numberthis{}
\end{align*}
where $d\Omega^2=d\theta^2+\sin^2\theta\;d\phi^2$ and $Q$, $q$, $\Delta$, $W$, $\alpha$, $U$, $R$ are functions of $r$ and $t$.

One can check that the tensor 
\begin{equation}
    \tensor{\gamma}{^\mu_\nu}=\begin{pmatrix}
    q/Q & \alpha/Q & 0 & 0\\
    -\alpha Q \Delta^2 & \Delta/W & 0 & 0\\
    0 & 0 & U/R & 0\\
    0 & 0 & 0 & U/R
    \end{pmatrix}
\end{equation}
has the property $\gamma^\mu_{~\sigma}\gamma^\sigma_{~\nu}=g^\mu_{~\sigma}f^\sigma_{~\nu}$. This tensor is used to 
compute the energy-momentum tensors $T^\mu_{~\nu}$ and ${\cal T}^\mu_{~\nu}$ in \eqref{T}. 

One can redefine the two amplitudes similarly to \eqref{NY}
\be                                 \label{NYa}
N=\Delta R^\prime\,,~~~~Y=WU^\prime\,,
\ee
where the prime denotes the derivative with respect to $r$, and one can impose the gauge condition 
\be
R=r. 
\ee
As a result, the independent field equations  \eqref{Enst-eq} become
\bea                                  \label{Ein00}
G^0_0(g)&=&\kappa_1\, T^{0}_{~0}, ~~~~
G^1_1(g)=\kappa_1\, T^{1}_{~1}, ~~~~
G^0_1(g)=\kappa_1\, T^{0}_{~1}, ~~~~
\nonumber \\
G^0_0(f)&=&\kappa_2\, {\cal T}^{0}_{~0},~~~~
G^1_1(f)=\kappa_2\, {\cal T}^{1}_{~1},~~~~~
G^0_1(f)=\kappa_2\, {\cal T}^{0}_{~1},
\eea
plus two nontrivial components of the the conservation condition$ \stackrel{(g)}{\nabla}_\mu T^\mu_{~\nu}=0\,$, 
\be                        \label{cons00}
\stackrel{(g)}{\nabla}_\mu T^\mu_{~0}=0\,,~~~~~~~~\stackrel{(g)}{\nabla}_\mu T^\mu_{~1}=0. 
\ee
Here one has explicitly 
\begin{align*}
    G(g)^0_0=\frac{N^2-1}{r^2}+\frac{2NN'}{r},~~~~~~~
     G^1_1(g)=\frac{N^2-1}{r^2}+\frac{2N^2Q'}{rQ},~~~~~~~~~
    G^0_1(g)=\frac{2\dot{N}}{rNQ^2},~~
    \numberthis
\end{align*}
where the dot denotes the partial derivative with respect to $t$, while 
\begin{align}
    T^0_{~0}=-\mathcal{P}_0-\mathcal{P}_1\frac{NU'}{Y},~~~~~~~~
     T^1_{~1}=-\mathcal{P}_0-\mathcal{P}_1\frac{q}{Q},~~~~~~~~
    T^0_{~1}=\mathcal{P}_1\frac{\alpha}{Q},~
 \end{align}
where $\mathcal{P}_m$ are defined in \eqref{e5}. The components of the second stress-energy tensor are
\begin{align*}
{\cal T}^0_{~0}=&-\frac{r^2}{N U^2\mathcal{A}}\bigg(\mathcal{P}_1 q Y+\mathcal{P}_2\big(\alpha^2 N^2 QY+qNU'\big)\bigg),\nn \\
 {\cal T}^1_{~1}=&-\frac{r^2}{U^2\mathcal{A}}\bigg(\mathcal{P}_1 QU'+\mathcal{P}_2\big(\alpha^2NQY+qU'\big)\bigg),\nn\\
 {\cal T}^0_{~1}=&-\frac{r^2}{N U^2\mathcal{A}}\mathcal{P}_1 Y\alpha, \numberthis 
\end{align*}
where $\mathcal{A}=N Q Y \alpha^2+qU'$. The components of the Einstein tensor for $f_{\mu\nu}$, are 
complicated:
\begin{align*}
   \tensor{G(f)}{^0_0}=&-\frac{1}{U^2 Y \mathcal{A}^3}\bigg(N^3 Q^3 Y^4 \alpha^6+\big(-N Q \dot{U}^2 Y^4+N^3 Q^3 U'^2 Y^4+2 N^3 Q^3 U U''Y^4\\
   &+3 N^2 q Q^2 U' Y^3\big) \alpha^4+\big(-2 N Q U \dot{U} \dot{\alpha} Y^4-2 q Q U
   \dot{U} N' Y^4+2 N Q U \dot{U} q' Y^4\\
   &-2 N q U \dot{U} Q' Y^4+2 N
   q Q \dot{U} U' Y^4-2 N^3 Q^3 U U' \alpha ' Y^4+2 N q Q U \dot{U}' Y^4\\
   &+2
   N^2 Q^2 \dot{U} U'^2 Y^3+2 N^2 Q^2 U U' \dot{U}' Y^3+2 N^2 Q^2 U \dot{U} U''
   Y^3-2 N^2 Q^2 U \dot{U} U' Y' Y^2\big) \alpha^3\\
   &+\big(-N q^2 Q U'^2 Y^4+2 q^2 Q U
   N' U' Y^4-2 N q Q U q' U' Y^4+2 N q^2 U Q' U' Y^4\\
   &-2
   N q Q U \dot{U} \alpha ' Y^4-2 N q^2 Q U U'' Y^4+N^2 q Q^2 U'^3 Y^3+2
   N q Q^2 U N' U'^2 Y^3\\
   &-2 N^2 Q^2 U q' U'^2 Y^3+2 N^2 q Q U
   Q' U'^2 Y^3-q \dot{U}^2 U' Y^3-2 N^2 Q^2 U \dot{U} U' \alpha ' Y^3\\
   &+N Q
   \dot{U}^2 U'^2 Y^2+3 N q^2 Q U'^2 Y^2+2 N^2 q Q^2 U U'^2 Y' Y^2+2 N Q
   U \dot{U} U' \dot{U}' Y^2\\
   &-2 N Q U \dot{U} \dot{Y} U'^2 Y\big) \alpha^2+\big(4 N q^2 Q
   U U' \alpha ' Y^4+2 q^2 \dot{U} U'^2 Y^3-2 q U \dot{U} \dot{\alpha} U' Y^3\\
   &+2 N^2 q
   Q^2 U U'^2 \alpha ' Y^3+2 q^2 U U' \dot{U}' Y^3-2 q^2 U \dot{U} U'' Y^3+2 N q Q
   \dot{U} U'^3 Y^2\\
   &+2 q Q U \dot{U} N' U'^2 Y^2-2 N Q U \dot{U} q' U'^2
   Y^2+2 N q U \dot{U} Q' U'^2 Y^2+2 q^2 U \dot{U} U' Y' Y^2\\
   &+2 N q Q U
   U'^2 \dot{U}' Y^2\big) \alpha-q^3 Y^3 U'^3+q Y \dot{U}^2 U'^3+q^3 Y U'^3-2 q U \dot{U}
   \dot{Y} U'^3\\
   &-2 q^3 U Y^2 U'^2 Y'+2 N q Q U Y^2 \dot{U} U'^2 \alpha '+2 q^2 U
   Y^3 \dot{U} U' \alpha '+2 q U Y \dot{U} U'^2 \dot{U}'\bigg),
   \end{align*}
   \begin{align*}
   \tensor{G(f)}{^0_1}=&-\frac{2}{U Y\mathcal{A}^3}\bigg(\big(-Q \dot{U} N' Y^4-N \dot{U} Q' Y^4+N Q \dot{U}' Y^4\big) \alpha^4+\big(-N Q\dot{\alpha} U' Y^4\\
   &+q Q N' U' Y^4+N q Q' U' Y^4-N Q \dot{U}
   \alpha ' Y^4-N q Q U'' Y^4+N Q^2 N' U'^2 Y^3\\
   &+N^2 Q Q'
   U'^2 Y^3-N^2 Q^2 U' U'' Y^3\big) \alpha^3+\big(2 N q Q U' \alpha ' Y^4-\dot{U}
   q' U' Y^3+q U' \dot{U}' Y^3\\
   &+2 N^2 Q^2 U'^2 \alpha ' Y^3-q \dot{U} U''
   Y^3+Q \dot{U} N' U'^2 Y^2+N \dot{U} Q' U'^2 Y^2+q \dot{U} U' Y'
   Y^2\\
   &-N Q \dot{U} U' U'' Y^2-N Q \dot{Y} U'^3 Y+N Q \dot{U} U'^2
   Y' Y\big) \alpha^2+\big(-q \dot{\alpha} U'^2 Y^3+q q' U'^2 Y^3\\
   &+q \dot{U} U' \alpha ' Y^3+N Q q' U'^3 Y^2-q^2 U'^2 Y' Y^2+2 N Q \dot{U} U'^2 \alpha ' Y^2-N q Q U'^3 Y' Y\big) \alpha\\
   &-q \dot{Y} U'^4+Y \dot{U} q' U'^3\bigg),
\end{align*}
\begin{align*}
   \tensor{G(f)}{^1_1}=&-\frac{1}{U^2\mathcal{A}^3}\bigg(N^3 Q^3 Y^3 \alpha^6+\big(-N Q \dot{U}^2 Y^3+N^3 Q^3 U'^2 Y^3+2 Q U \dot{N} \dot{U}
   Y^3+2 N U \dot{Q} \dot{U} Y^3\\
   &-2 N Q U \ddot{U} Y^3+2 N^2 Q^3 U N' U'
   Y^3+2 N^3 Q^2 U Q' U' Y^3+3 N^2 q Q^2 U' Y^2\big) \alpha^4\\
   &+\big(2 N Q U
   \dot{U} \dot{\alpha} Y^3-2 q Q U \dot{N} U' Y^3+2 N Q U \dot{q} U' Y^3-2 N q
   U \dot{Q} U' Y^3\\
   &+2 N q Q \dot{U} U' Y^3+2 N^3 Q^3 U U' \alpha ' Y^3+2
   N q Q U \dot{U}' Y^3+2 N^2 Q^2 \dot{U} U'^2 Y^2\\
   &+4 N^2 Q^2 U U' \dot{U}' Y^2-2
   N^2 Q^2 U \dot{Y} U'^2 Y\big) \alpha^3+\big(-N q^2 Q U'^2 Y^3-2 N q Q U \dot{\alpha} U' Y^3\\
   &-2 N q Q U q' U' Y^3+N^2 q Q^2 U'^3 Y^2-2 N^2 Q^2 U
   \dot{\alpha} U'^2 Y^2+2 N q Q^2 U N' U'^2 Y^2\\
   &+2 N^2 q Q U Q' U'^2
   Y^2-q \dot{U}^2 U' Y^2+2 U \dot{q} \dot{U} U' Y^2-2 q U \ddot{U} U' Y^2+2 q U \dot{U}
   \dot{U}' Y^2\\
   &+N Q \dot{U}^2 U'^2 Y+3 N q^2 Q U'^2 Y-2 Q U \dot{N} \dot{U}
   U'^2 Y-2 N U \dot{Q} \dot{U} U'^2 Y+2 N Q U \ddot{U} U'^2 Y\\
   &-2 q U \dot{U}
   \dot{Y} U' Y+2 N Q U \dot{U} U' \dot{U}' Y-2 N Q U \dot{U} \dot{Y} U'^2\big)
   \alpha^2+\big(2 q Q U Y \dot{N} U'^3\\
   &-2 N Q U Y \dot{q} U'^3+2 N q U Y
   \dot{Q} U'^3+2 N q Q Y \dot{U} U'^3+2 q^2 Y^2 \dot{U} U'^2+2 q^2 U Y \dot{Y}
   U'^2\\
   &-4 N Q U Y \dot{U} \dot{\alpha} U'^2+2 N^2 q Q^2 U Y^2 \alpha ' U'^2+2
   N q Q U Y \dot{U}' U'^2-2 q U Y^2 \dot{U} \dot{\alpha} U'\big) \alpha\\
   &+q^3 U'^3-q^3
   Y^2 U'^3+q \dot{U}^2 U'^3-2 U \dot{q} \dot{U} U'^3+2 N q Q U Y \dot{\alpha} U'^3+2
   q U \ddot{U} U'^3\\
   &+2 q^2 U Y^2 \dot{\alpha} U'^2-2 q^2 U Y^2 q' U'^2\bigg). \numberthis 
 \end{align*}

Finally, there are two nontrivial components of the conservation law,
\begin{align*}
    \stackrel{(g)}{\nabla}_\mu T^\mu_{~0}=&-\mathcal{P}_1\bigg(\alpha N'NQ+2\alpha N^2Q+\alpha'N^2Q+\frac{q\dot{N}}{NQ}+\frac{N\dot{U}'}{Y}-\frac{NU'\dot{Y}}{Y^2}\bigg)\\
    &-\frac{d\mathcal{P}_0}{r}\Big(\alpha N^2Q+\dot{U}\Big)-\frac{d\mathcal{P}_1}{r}\bigg(\alpha N^2QU'+\frac{N\dot{U}U'}{Y}\bigg),  \\
     \stackrel{(g)}{\nabla}_\mu T^\mu_{~1}=&\mathcal{P}_1\bigg(\frac{\dot{\alpha}}{Q}-\frac{\alpha\dot{N}}{NQ}-\frac{q'}{Q}+\frac{NQ'U'}{QY}\bigg)+\frac{d\mathcal{P}_1}{r}\bigg(\alpha^2N^2+\frac{\alpha\dot{U}}{Q}+\frac{qNU'}{QY}-\frac{qU'}{Q}\bigg)\\
    &+\frac{d\mathcal{P}_0}{r}\bigg(\frac{NU'}{Y}-U'\bigg),\numberthis
\end{align*}
where $d\mathcal{P}_m$ are defined in \eqref{e5a}. 
Equations \eqref{Ein00}, \eqref{cons00} comprise a system of 8 equations for 6 functions  $Q$, $q$, $\Delta$, $W$, $\alpha$, $U$.
For this system not to be overdetermined, only 6 equations out of 8 should be independent. As shown in 
Sec. \ref{pert}, this indeed happens at least for small $\alpha$, when the perturbative analysis of the equations 
shows that some of them coincide.



\providecommand{\href}[2]{#2}\begingroup\raggedright\endgroup

\end{document}